%% file: Planck2011-5.1b.tex
\documentclass[a4paper,traditabstract,longauth]{aa}  
\usepackage{graphicx}
\usepackage{lscape}
\usepackage{txfonts}
\usepackage{natbib}
\usepackage{color}
\usepackage{amssymb}
\usepackage{booktabs}
\usepackage{subfig}

\def\ergscm{\rm ergs\,s^{-1}\,cm^{-2}}

\def\msol{{M$_{\odot}$}}
\def\xmm{XMM-{\it Newton} }

\def\xmm{{\it XMM-Newton}}
\def\planck{{\it Planck}}
\def\rosat{{\it ROSAT}}
\def\rass{{\rm RASS}}

\def \xspec {\hbox{\tt xspec}}

\def \sas {\hbox{SAS}}
 
\def \epic {\hbox{\sc EPIC}} 
\def \mos {\hbox{\sc EMOS}} 
\def \pn {\hbox{\sc EPN}} 
\def \mekal {\hbox{\sc mekal}}

\newfont{\gwpfont}{cmssq8 scaled 1000}
\newcommand{\rexcess}{{\gwpfont REXCESS}}
\newcommand{\excpres}{{\gwpfont EXCPRES}}
\newcommand{\reflex}{{REFLEX}}
\newcommand{\noras}{{NORAS}}
\newcommand{\bcs}{{BCS}}
\newcommand{\ebcs}{{eBCS}}
\newcommand{\macs}{{MACS}}
\def\M500{M_{500}}
\def\R500{R_{500}}
\def\Mgv{M_{\rm g,500}}

\def\YX {Y_{\rm X}}
\def\TX {T_{\rm X}}
\def\kT {{\rm k}T}
\def\YSZ {Y_{\rm SZ}}
\def\YSZ {Y_{500}}

\def\kT {{\rm k}T}
\def\Mv {M_{\rm 500}}
\def \Rv {R_{500}}
\def\keV {\rm keV}
\def\Yv {Y_{500}}
\def\LX {L_{500}}
\def\NH {N_{\rm H}}

\def\MYX {$M_{500}$--$Y_{\rm X}$}

\def\Lxz{$L_{\rm X}$--$z$}

\def\YSZYX {$\YSZ$--$\YX$}

\def\msol {{\rm M_{\odot}}}

\def\lesssim{\mathrel{\hbox{\rlap{\hbox{\lower4pt\hbox{$\sim$}}}\hbox{$<$}}}}
\def\gtrsim{\mathrel{\hbox{\rlap{\hbox{\lower4pt\hbox{$\sim$}}}\hbox{$>$}}}}

\newcommand{\propsim}{\lower 3pt \hbox{$\, \buildrel {\textstyle
       \propto}\over {\textstyle \sim}\,$}}

\begin{document}
%

\input{Proj_Ref_5_1b_authors_and_institutes.tex}

\title{\textit{Planck} Early Results. IX. \textit{XMM-Newton}  follow-up for validation of \textit{Planck}  cluster candidates}

  \abstract
  {We present the \xmm\ follow-up for confirmation of \planck\ cluster candidates. Twenty-five candidates have been observed to date using  snapshot ($\sim\!\!10\,{\rm ksec}$) exposures, ten as part of a pilot programme to sample a low range of signal-to-noise ratios ($4<\textrm{S/N}<6$), and a further 15 in a programme to observe a sample of $\textrm{S/N}>5$ candidates. The sensitivity and spatial resolution of \xmm\  allows unambiguous discrimination between clusters and false candidates.  {The 4 false candidates have $\textrm{S/N}\leq 4.1$.}  A total of 21 candidates are confirmed as extended X-ray sources. Seventeen are single clusters, the majority of which are found to have highly irregular and disturbed morphologies {(about $\sim 70\%$)}. The remaining four sources are multiple systems, including the unexpected discovery of a supercluster at $z=0.45$. For {20} sources we are able to derive a redshift estimate from the X-ray Fe~K line (albeit of variable quality). The new clusters span the redshift range $0.09 \lesssim z \lesssim 0.54$, with a median redshift of $z \sim 0.37$. A first determination is made of their X--ray properties including the characteristic size, which is used to improve the estimate of the SZ Compton parameter, $\YSZ$.  The follow--up validation programme has helped to optimise the \planck\  candidate selection process.  It has also provided a preview of the X--ray properties of these newly-discovered clusters, allowing comparison with their SZ properties, and to the X-ray and SZ properties of known clusters observed in the \planck\ survey.  Our results suggest that \planck\ may have started to reveal a non-negligible population of massive dynamically perturbed objects that is under-represented in X-ray surveys. However, despite their particular properties, these new clusters appear to follow the $\YSZ$--$\YX$ relation established for X--ray selected objects, where $\YX$  is the product of the gas mass and temperature.}

\authorrunning{Planck Collaboration}
\titlerunning{Planck early results. IX}
  \maketitle

\input{Planck.tex}

\allearlypapers

\section{Introduction}

The \Planck\footnote{\Planck\ (http://www.esa.int/Planck) is a project of the European Space Agency (ESA) with instruments provided by two scientific consortia funded by ESA member states (in particular the lead countries: France and Italy) with contributions from NASA (USA), and telescope reflectors provided in a collaboration between ESA and a scientific consortium led and funded by Denmark.
} satellite has been surveying the sky across nine frequencies in the microwave band since August 2009. The resulting data set allows the detection of galaxy clusters through the Sunyaev-Zeldovich (SZ) effect \citep{sun72}, the spectral distortion of the cosmic microwave background (CMB) generated via inverse Compton scattering of CMB photons by the hot electrons in the Intra-Cluster Medium (ICM).  
The total SZ signal is expected to be closely related to the cluster mass \citep[e.g.,][]{das04} and its brightness is insensitive to redshift dimming.  As a result, SZ surveys can potentially  provide unbiased cluster samples, covering a wide range of redshifts, that are expected to be close to mass-selected.  As compared to other SZ instruments, \planck\ brings a unique nine-band coverage from 30 to 857 GHz and  a relatively high, band-dependent spatial resolution of 5--10 arcmin. 
Most crucially, the \planck\ SZ survey covers an exceptionally large volume,  being the first all-sky survey capable of blindly detecting clusters since the \rosat\ All-Sky Survey (\rass) in the X-ray domain.  
As a consequence, \planck\  is detecting previously unknown, massive clusters that do not appear in other SZ surveys. Its all-sky coverage allows detection of the rarest clusters, the most massive objects lying in the exponential tail of the mass function. 
These are the best clusters for precision cosmology:  their abundance evolution is the most sensitive to the cosmological parameters \citep[][]{voi05}, and their gas mass fractions can be used as  distance indicators \citep{all08}.  In addition, clusters in this high-mass regime are X--ray bright, making their observation easier, and their ICM is expected to be the least affected by non-gravitational processes. These newly-discovered \planck\ clusters will thus also be ideal targets for studying the physics of the gravitational collapse that drives all cluster formation. 

The \planck\ survey is providing a sample  of  cluster {\it candidates}.  Any such survey sample is expected to include a fraction of false detections, due for example to fluctuations in the complex microwave astrophysical sky.   In addition, as a result of \planck's moderate spatial resolution at SZ frequencies with respect to typical cluster sizes, a \planck\ cluster candidate SZ measurement essentially provides only coordinates and total SZ flux estimates; these estimates are further hampered by the flux-size degeneracy discussed extensively in \citet[][]{planck2011-5.1a}. 
A vigourous follow-up programme is therefore required to scientifically exploit \planck\ cluster candidate data. Such a programme includes candidate confirmation, which is the final part of the catalogue validation, in addition to redshift measurements, estimation of relevant physical parameters (including cluster size, allowing precise SZ flux estimates),  and investigation of scaling properties.  In particular, measurement of the relation between the SZ `luminosity' and the mass as a function of redshift, $z$, is essential for calculation of the survey selection function and for related cosmological applications.  

The all-sky nature of the \planck\ survey means that confirmation and redshift measurement of cluster candidates is not a trivial task. In the optical domain, the only publicly available large survey is the Sloan Digital Sky Survey (SDSS)\footnote{http://www.sdss.org/}. Although cross-correlation with this survey can be used to confirm candidates up to $z\sim0.6$, it covers only part of the northern sky. Furthermore, optical confirmation is hampered by the relatively large \planck\ source position uncertainty, which can be up to 5\arcm\ \citep{planck2011-5.1a}.  To discriminate between a true counterpart and a chance association with low-mass systems at various redshifts within the \planck\ error box, optical mass and {spectroscopic redshift or} photometric redshift estimates with a wide-field, multi-band, instrument are required.

In contrast, confirmation in X--rays offers definite advantages. Above the Galactic Plane, the detection of extended X--ray emission is an unambiguous signature of a cluster, and the density-squared dependence of the X--ray emission reduces projection effects nearly to zero.  Furthermore, the low space density of  groups and clusters in a typical X--ray exposure makes spurious association with a \planck\ candidate unlikely.  For instance, the XMM-LSS survey found 29 systems in 5 deg$^2$ using 10 ksec \xmm\ exposures \citep{pac07}.  Such a detection rate corresponds to only a $10$ per cent probability of finding a cluster by chance within a 5\arcm\ aperture, the conservative \planck\ error box. Finally, as both X--ray and SZ observations probe the same medium, spurious associations can be readily assessed from a consistency check between the X--ray and SZ flux, assuming a reasonable redshift range (as illustrated in Sec.~\ref{sec:valfalse}). 

In this context, and because of its superior sensitivity, \xmm\ \citep{jan01} is the best instrument for following up newly-detected \planck\ clusters up to high redshift.  A short snapshot  \xmm\  exposure is  sufficient to confirm any \planck\ cluster candidate  at least up to $z\,=\,1.5$ (Sec.~\ref{sec:xsnr}), and for the X-ray brightest objects, provides the source redshift from the iron K line  (Sec~\ref{sec:zx}).  Because of their high mass, clusters are expected to be larger than $1\arcm$ and the \xmm\ spatial resolution is sufficient to discriminate between a point source and extended emission.

In order to assess the galaxy cluster nature of the \planck\ SZ sources and to help guarantee the integrity of the final \planck\ SZ legacy catalogue to be released in 2012, we have thus proposed to use \xmm\ to confirm the highest
significance cluster candidates discovered by \planck. This validation programme consists of snapshot ($\sim10\,{\rm ksec}$) observations and  is undertaken via an agreement between the ESA \xmm\ and \planck\ project scientists. In this paper we present the definition and results of this programme. To date, 25 \planck\ SZ sources have been observed, making use of \xmm\ Director's Discretionary Time.  Of these,  21 sources have been confirmed.  In compliance with \planck\ policies for follow-up, the \xmm\ data of the 25 \planck\ sources are made public along with the publication of the Early Release Compact Source Catalogue (ERCSC).  

The \xmm\ follow-up for validation is the backbone of a larger programme for the confirmation and redshift measurement of \planck\ SZ cluster candidates. The \planck\ collaboration has also been granted time on the following facilities: the ENO, the ESO/MPG $2.2\,$m and the Palomar telescopes. Observations with these facilities are ongoing or being processed. Some optical results from the ENO observations are presented together with the \xmm\ results in this paper (Sect.~\ref{sec:zopt} and~\ref{sec:apa}). Other early astrophysics results on clusters of galaxies are presented in \citet{planck2011-5.1a, planck2011-5.2a, planck2011-5.2b, planck2011-5.2c}.

We adopt a $\Lambda$CDM cosmology with $H_0=70\,\kmsMpc$, $\Omega_{\rm M}=0.3$ and $\Omega_\Lambda=0.7$. The factor $E(z)= \sqrt{\Omega_{\rm M} (1+z)^3+\Omega_\Lambda}$ is the ratio of the Hubble constant at redshift $z$ to its present day value. The quantities $\Mv$ and $\Rv$ are the total mass and radius corresponding to a total density contrast $\delta=500$, as compared to $\rho_{\rm c}(z)$,   the critical density of the Universe at the cluster redshift; thus $\Mv = (4\pi/3)\,500\,\rho_c(z)\,\Rv^3$. The quantity $\YX$ is defined as the product  of  $\Mgv$, the gas mass within $\Rv$, and $\TX$, the spectroscopic temperature measured in the $[0.15$--$0.75]~\Rv$ aperture. The SZ signal is characterised by  $Y_{500}$ throughout. This quantity is defined as $ Y_{500} D_{\rm A}^2 = (\sigma_{\rm T}/m_{\rm e} c^2)\int P dV$. Here $D_{\rm A}$ is the angular--diameter distance to the system, $\sigma_{\rm T}$ is the Thomson cross-section, $c$ is the speed of light, $m_{\rm e}$ is the electron rest mass; $P=n_{\rm e} T$ is the pressure, the product of the electron number density and temperature, and the integration is performed over a sphere of radius  $\Rv$. 
The quantity $Y_{500}\,D_{\rm A}^2$ is the spherically integrated Compton parameter and $Y_{500}$ is proportional to the apparent magnitude of the SZ signal from within $\Rv$.

%

\begin{table*}[t]
\begingroup
\caption{ {\footnotesize Observation log of the \xmm\ validation follow-up. The 10 targets of the pilot programme are listed first.  Column (1):  \planck\ source name. Asterisked objects denote sources that were found to be multiple systems in X-rays.  Column (2): Signal-to-noise ratio of the detection  of the \planck\ cluster candidate in the version of the \planck-HFI maps available for each programme. Columns (3) and (4): Right ascension and declination  of the \planck\ source (J2000). Columns (5)--(8):  \xmm\ observation identification number, filter used, on--source exposure time with the \pn\ camera and fraction of useful time after cleaning for periods of high background due to soft proton flares (\mos\ and \pn\ camera, see also Sec.~\ref{sec:xred}). Column (9): Confirmed clusters are flagged. Those included in the ESZ sample  \citep{planck2011-5.1a} are also identified.  
}\label{tab:obs}}
\centering  
\begin{tabular}{lrrrrcccl}
\toprule
\multicolumn{1}{l}{Name} & \multicolumn{1}{c}{$\textrm{S/N}$} &  \multicolumn{1}{c}{RA$_{\rm SZ}$} & 
\multicolumn{1}{c}{DEC$_{\rm SZ}$}  & \multicolumn{1}{c}{OBSID} & 
\multicolumn{1}{c}{filter} & \multicolumn{1}{c}{$t_{\rm exp}$} &
\multicolumn{1}{c}{Clean fraction} & \multicolumn{1}{c}{Confirmed}\\

\noalign{\smallskip}

\multicolumn{1}{c}{} & \multicolumn{1}{c}{} & \multicolumn{1}{c}{(deg)} & 
\multicolumn{1}{c}{(deg)}  & \multicolumn{1}{c}{}  &  
\multicolumn{1}{c}{} & \multicolumn{1}{c}{(ks EPN)} &
\multicolumn{1}{c}{(MOS/EPN)} & \multicolumn{1}{c}{} \\

\midrule

PLCK~G277.8$-$51.7 & 6.1 & 43.596 & $$-$$58.964 &  0656200301 & THIN & 16.5 & 0.5/0.2 &Y  ~ESZ \\
PLCK~G334.8$-$38.0* &4.9 &313.177 & $-$61.202 & 0656200701 & THIN & 21.2 & 0.8/0.5 &Y \\
PLCK~G250.0+24.1 & 4.9&143.042 & $-$17.647 &  0656200401 & THIN & 10.4 & 0/0 &Y \\
PLCK~G286.3$-$38.4 & 4.7&59.800 & $-$72.067 &  0656200501 & THIN & 13.6 & 0.7/0       &Y \\
PLCK~G004.5$-$19.5 & 4.6&289.226 & $-$33.509 &  0656201001 & MED &  10.0 & 1/0.6 &Y \\
PLCK~G214.6+36.9* &4.2& 137.206 & 14.642 & 0656200101 & THIN & 17.6 & 0.7/0.7 &Y \\
PLCK~G070.8$-$21.5 & 4.1&321.410& 19.941& 0656200201 & MED & 25.4 & 0.4/0.1 & \ldots \\
PLCK~G317.4$-$54.1 & 4.1&355.247 & $-$61.038 &  0656200801 & THIN & 12.0 & 0.9/0.7 & \ldots \\
PLCK~G226.1$-$16.9 & 4.0&93.139 & $-$19.040 &  0656200601 & THIN & 10.0 & 0.7/0.4 & \ldots \\
PLCK~G343.4$-$43.4 & 3.9& 320.145 & $-$53.631 & 0656200901 & MED  & 10.0 & 0.9/0.9 & \ldots \\

\midrule

PLCK~G287.0+32.9  & 10.2& 177.714 & $-$28.074&  0656201201 & THIN & 10.0 & 0.7/0.4 &Y ~ESZ \\
PLCK~G171.9$-$40.7  &10.7 & 48.231 &8.380 & 0656201101 & THIN & 10.0 & 1/0.8 &Y  ~ESZ \\
PLCK~G285.0$-$23.7  & 8.3 & 110.805 & $-$73.457 &  0656201401 & THIN & 10.0 & 0.9/0.6 &Y  ~ESZ \\
PLCK~G271.2$-$31.0  &8.3 & 87.315 & $-$62.087 &  0656201301 & THIN & 10.0 & 1/1 &Y  ~ESZ \\
PLCK~G262.7$-$40.9  & 7.4 & 69.624 & $-$54.309 &  0656201601 & THIN & 14.7 & 1/0.9 &Y  ~ESZ \\
PLCK~G308.3$-$20.2* & 7.4 & 229.588 & $-$81.523 & 0656201501 & THIN & 10.0 & 1/1 &Y  ~ESZ \\
PLCK~G337.1$-$26.0* & 6.4 & 288.583 & $-$59.513 & 0656201701 & THIN & 13.7 & 1/1 &Y  ~ESZ \\
PLCK~G292.5+22.0  & 6.2 & 180.241 & $-$39.889 &  0656201801 & MED  & 13.2 & 0.6/0.5 &Y  ~ESZ \\
PLCK~G205.0$-$63.0  & 6.1 & 41.593& $-$20.527 &  0656201901 & THIN & 11.7 & 1/1   &Y \\
PLCK~G241.2$-$28.7  & 5.9 & 85.768 & $-$36.022 &  0656202001 & THIN & 10.0 & 1/1&Y \\
PLCK~G286.6$-$31.3  & 5.9 & 82.8430 & $-$75.164 &  0656202101 & THIN & 10.0 &0.7/0.3 &Y  ~ESZ  \\
PLCK~G018.7+23.6  & 5.6 & 255.553 & $-$1.004 & 0656202201 & THIN & 7.2 & 1/1   &Y \\
PLCK~G100.2$-$30.4  & 5.5 & 350.589 & 28.563& 0656202301 & THIN & 10.0 & 0.9/0.7 &Y \\
PLCK~G272.9+48.8  & 5.1 & 173.310 & $-$9.479 &  0656202601 & THIN & 11.7 & 0/0  &Y \\
PLCK~G285.6$-$17.2  & 5.2 & 130.956 & $-$71.190&  0656202501 & THIN & 10.0 & 1/1  &Y  ~ESZ \\

\bottomrule

\end{tabular}

\endPlancktable 
 \endgroup

\end{table*}

\section{The \textit{XMM-Newton}  validation follow-up of \textit{Planck}  cluster candidates}

\subsection{The \planck\ survey}

\planck\ \citep{tauber2010a,planck2011-1.1} is the third generation space mission to measure the anisotropy of the CMB.  It observes the sky in nine frequency bands covering 30--857\,GHz with high sensitivity and angular resolution
from 31\arcm\ to 5\arcm.  The Low Frequency Instrument (LFI; \citealt{Mandolesi2010,Bersanelli2010,planck2011-1.4} covers the 30, 44, and 70\,GHz bands with amplifiers cooled to 20\,\hbox{K}.  The High Frequency Instrument (HFI; \citealt{Lamarre2010, planck2011-1.5}) covers the 100, 143, 217, 353, 545, and 857\,GHz bands with bolometers cooled to 0.1\,\hbox{K}.  Polarisation is measured in all but the highest two bands \citep{Leahy2010, Rosset2010}.  A combination of radiative cooling and three mechanical coolers produces the temperatures needed for the detectors and optics \citep{planck2011-1.3}.  Two data processing centres (DPCs) check and calibrate the data and make maps of the sky \citep{planck2011-1.7,   planck2011-1.6}.  \planck's sensitivity, angular resolution, and frequency coverage make it a powerful instrument for Galactic and extragalactic astrophysics as well as cosmology.  Early astrophysics results are given in Planck Collaboration, 2011h--z.

\subsection{Blind detection of SZ clusters in  \planck}

The blind search for clusters in \planck\ data relies on a Multi-Matched Filter (MMF) approach \citep{mel06}\footnote{Results from other methods have been cross-compared to those from the MMF search, including from the PowellSnakes-based algorithm \citep{car09b}.}.  This detection algorithm operates on all-sky maps divided into a set of overlapping square patches, using simultaneously the 6 frequency maps of the HFI instrument  \citep[][]{planck2011-5.1a}.  Within the algorithm, the SZ spectral signature and the universal pressure profile derived by \citet{arn10} are used as spectral and spatial templates, respectively. In such a blind search, the position, the characteristic scale of the profile ($\propto R_{500}$) and the amplitude ($\propto \YSZ$) are left free, being optimised by the MMF algorithm. In practice the algorithm is run in an iterative way: after a first detection run to locate candidates, consecutive runs on sky patches centred on the candidate positions  refine the estimated signal-to-noise ratio ($\textrm{S/N}$) and other properties.

Cluster candidates then undergo a validation process, extensively described in \citet{planck2011-5.1a}. This process includes  internal quality checks (e.g., map artefacts, cross-comparison between detection algorithms, SZ spectral signature, astrophysical contamination by Galactic dust, point sources or {structures in the CMB}) and  cross-correlation with ancillary data and catalogues allowing known clusters to be identified. This process produces a list of new \planck\ SZ cluster candidates above a given $\textrm{S/N}$ threshold ($\textrm{S/N}=4$ in this work).

\subsection{\xmm\ target selection}

From the list of new potential clusters detected as SZ sources in the \planck\ survey, we selected 25 targets in a two step process:

\begin{enumerate}
\item Pilot programme: 10 targets were selected on the basis of the \planck\ survey as it stood at the end of October 2009, i.e., $\sim 62\%$ sky coverage. These targets were explicitly chosen to sample the lower range of signal-to-noise ($4<\textrm{S/N}<6$) in order to better characterise the nature and quality of the SZ signal.

\item High $\textrm{S/N}$ programme: a further 15 targets were chosen  in the spring of 2010 when the first full-sky coverage was close to completion ($99.5\%$  sky coverage). In contrast to the pilot programme, here we focused on high-significance SZ sources ($\textrm{S/N}>5$) and selected candidates starting from the highest $\textrm{S/N}$. 
\end{enumerate}

In both cases the selection process was intimately linked to the \planck-HFI data Time Ordered Information processing status, calibration, attitude and map versions (as of  Dec 7, 2009 and April 19, 2010 for the two programmes, respectively).  The choice of targets was also constrained by their \xmm\ visibility in a period of 2--3 months following their submission to the Science Operations Centre. For both programmes, maps and spectra of each potential target were visually inspected, including reprocessing with aperture photometry methods. Cross--correlation with the \rass\  Bright Source Catalogue \citep[RASS-BSC,][]{vog99} and Faint Source Catalogue \citep[RASS-FSC,][]{vog00} was undertaken.  For the two targets of the pilot programme falling in the SDSS area, we ran a dedicated algorithm to search for galaxy overdensities \citep{fro11}, allowing us to track significant concentrations of matter down to $z\sim 0.6$. These two targets were chosen to test the SDSS based confirmation at high $z$. The first candidate, {PLCK~G070.8-21.5}, was not confirmed \citep[see][for discussion]{fro11}; the other candidate, {PLCK~G214.6+37.0}, is discussed in Sec.~\ref{sec:scn}.  Detailed searches in \xmm, {\it Chandra} and {\it Suzaku} observatory logs were also undertaken in order to avoid duplication of performed or accepted observations with similar facilities. 
\begin{figure}[t]
\center
\includegraphics[scale=1.,angle=0,keepaspectratio,width=0.9\columnwidth]{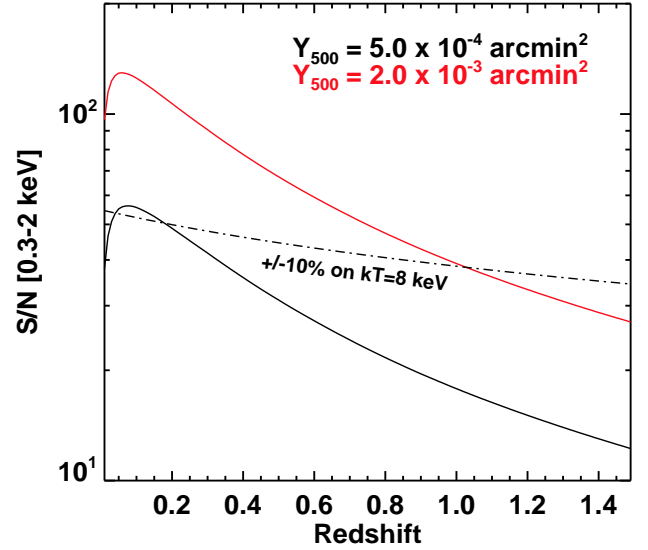}
\caption{{\footnotesize Sensitivity of \xmm\ observations to typical \planck\ SZ sources. The expected signal-to-noise ratio ($\textrm{S/N}$) of the cluster detection  in the $[0.3$--$2]\,\keV$ energy band with the \epic\ camera is plotted as a function of redshift for an exposure time of 10 ksec assuming two typical SZ fluxes. See text for details of the model assumed to convert SZ to X--ray flux and count rate. The dash-dotted line indicates the $\textrm{S/N}$ required for $10\%$ uncertainty on the temperature measurement of an 8~keV cluster. }}\label{fig:SNR}
\end{figure}

Six of the ten pilot programme targets were confirmed (see Table~\ref{tab:obs}); two of these are multiple systems. Taking into account the result of the pilot project, for the second programme we set a lower $\textrm{S/N}$ threshold of $\textrm{S/N}=5$ and refined and strengthened the selection criteria.  In particular, we required that the source be independently  detected by at least two of the three blind detection methods, and more quality flags were considered. The internal checks were very similar to those defined for constructing the Early SZ (ESZ) sample \citep{planck2011-5.1a}, which benefit from the result of the \xmm\ Pilot programme.  We also performed a search for emission in the \rass\  hard band images, looking for X-ray signatures beyond those recorded in the \rass\ source catalogues.  However, \rass\ information never took precedence over the internal \planck\ quality flags.  Note that two of the false candidates of the Pilot programme (PLCK~G343.4--43.4 and PLCK~G226.1--16.9)  were associated with a \rass-FSC source that \xmm\ subsequently revealed to be several point sources  (see Sec.~\ref{sec:valfalse}). Thus association with an \rass\ source alone is not sufficient for cluster candidate confirmation. 

The ESZ sample \citep{planck2011-5.1a} consists of  a high signal-to-noise{, i.e. primarily $\textrm{S/N}\geq6$,} list of 189 clusters and cluster candidates based on data from the first 10 months of the \planck\ survey. Ten of the 21 objects presented in the present paper passed the {$\textrm{S/N}$}  ESZ selection criteria and are thus part of the ESZ sample released to the community in January 2011. The original $\textrm{S/N}$ of their detection in the \planck\ maps is given in Table~\ref{tab:obs}, whereas the S/N values provided in Table~\ref{tab:xray} are derived from the 10 month \planck\ maps on the basis of which the ESZ sample was constructed.

\subsection{Observation setup}
\label{sec:xsnr}

As discussed by \citet{sch02}, the optimum energy band for \xmm\ cluster detection is the soft energy band {(energy below 2~keV)}, for which the signal-to-noise ratio reaches a maximum.  We calculated expectations for \xmm\ sensitivity in that band for two representative values of the SZ flux from within $R_{500}$: $Y_{\rm 500} =5 \times 10^{-4}$ arcmin$^{-2}$ and $Y_{\rm 500} = 2 \times10^{-3}$ arcmin$^{-2}$. In each case, the expected cluster luminosity $L_{500}$ for various redshifts was estimated  using the $L_{500}$--$D_{\rm A}^2 Y_{{\rm 500}}$ relation of \citet{arn10}, assuming self-similar evolution.  We then derived the corresponding total \xmm\ count rates, $R$, in the $[0.3-2]\,\keV$ band for the \epic\ MOS--CCD (herafter \mos) and pn--CCD (herafter \pn) camera \citep{tur01,str01}. We used the \xspec\  software \citep{arn96} to simulate an absorbed thermal model (assuming $\kT = 7\,\keV$, $N_{\rm H} = 2 \times 10^{20}$ cm$^{-2}$), convolved with the instrument response.  The corresponding angular extent $\theta_{500}$  was estimated from the $L_{500}$--$M_{500}$ relation of \citet{pra09}. The signal--to--noise ratio of the detection is then given by $\textrm{S/N} = (R\, \sqrt{t_{\rm exp}}) / (\sqrt{R + (2\, R_{\rm bkg}\,A)})$, where $t_{\rm exp}$ is the exposure time, $R_{\rm bkg}$ is the background count rate, and $A=4\,\pi\, \theta_{500}^2 $ is the integration area in square arc minutes. We assumed a $[0.3$--$2]\,\keV$ band background count rate of $R_{\rm bkg} = 4.5 \times 10^{-3}$ counts s$^{-1}$ arcmin$^{-2}$, as estimated from the blank sky backgrounds of \citet{rea03}. Figure~\ref{fig:SNR} shows the resulting $\textrm{S/N}$ of an \xmm\ detection as a function of redshift. 

Since the goal of the \xmm\ observations is confirmation of new \planck\ SZ cluster candidates, the nominal observing time was set to 10 ksec (net \pn\  camera time) per target. Such a snapshot observation is sufficient to detect the cluster -- if real -- at better than $10 \sigma$ up to $z=1.5$ (Fig~\ref{fig:SNR}). The nominal setup used the THIN filters (unless optical loading had to be avoided) and EFF  mode for the \pn\  camera.  The boresight was optimised to avoid camera gaps.

\subsection{\xmm\ data reduction}
\label{sec:xred}

We produced calibrated event lists using v10.0 of the \xmm\ Science Analysis System (\sas). Observations were cleaned for periods of high background due to soft proton flares, {\sc pattern}-selected and corrected for vignetting as described in \citet{pra07}. Point sources were  identified from the small scales of wavelet-decomposed images in the $[0.3$--$2]$ and $[2$--$5]\,\keV$ bands. These sources were excluded in the analysis of confirmed clusters,  with the exclusion radius matched to the PSF size (which varies across the detector).  

Above $\sim\!\!2$ keV the \xmm\ background is dominated by particle events. We subtracted this background using a stacked event list built from observations obtained with the filter wheel in the {\sc closed} position, recast to the pointing position and renormalised using the count rate in the high energy band free of cluster emission. The remaining background (due to the cosmic X-ray background of unresolved AGN and Galactic emission) was estimated from the particle-background subtracted emission from an annulus beyond the cluster emission. For the spectral analysis, we modeled this background emission as arising from two thermal sources and a power-law source with index $\Gamma = 1.4$, taking into account the absorbing Galactic column density in the direction of the object \citep[see, e.g.,][]{del04}. 

As Table~\ref{tab:obs} shows, the observations are of variable quality. In three cases the \pn\  data were completely contaminated by soft proton flares and formally had no useful observing time. For two of these observations,  the \mos\ data were completely contaminated too. In these instances, we used \mos\  data only (uncleaned in the last two cases). The power-law index in the background model was left free, which empirically produces a relatively good fit to the background spectrum. The spectroscopic results for these objects should be treated with caution.

Spectral fits were undertaken in the $[0.3$--$10]\,\keV$ energy range,
after binning the spectra to a minimum of 25 counts per bin and
excluding background fluorescence line regions. The cluster component
was modelled with an absorbed \mekal\ model with the reference solar
abundances from the data of \citet{and89}.  {The  hydrogen column
density $\NH$ was fixed at the 21 cm value from \citet{dic90},  except for
PLCK~G286.3$-$38.4, PLCK~G308.3$-$20.2 and PLCK~G018.7+23.6. Their best fit  $\NH$ values were  found to be  significantly higher by a factor 1.8, 1.8 and  2.4, respectively.   
These clusters are located at low latitude, in regions of high IR dust emission \citep[][Fig.~11]{sno97}. The 21 cm value may underestimate the
total  $\NH$, measured from X--ray data,  due to a non-negligible H$_{2}$ contribution.  To check this hypothesis, we used
 the IRIS maps \citep{miv05}
  as tracers of the dust emission and the correlation between
  the Galactic dust emission and the total hydrogen column density
  \citep{bou96}   to estimate the  $\NH$
  values   at the cluster locations \citep[see][]{poi04}. A better agreement was found with X--ray values, with 
 ratios of 1.30, 1.06 and 1.48. It must also be noted that the PLCK~G286.3$-$38.4 observation
is highly contaminated by solar flares and only \mos\ data are used.  
Some residual background  may also affect the low energy part of the spectrum and thus the $\NH$ estimate}


\begin{figure}[t]
\center
\includegraphics[scale=1.,angle=0,keepaspectratio,width=0.9\columnwidth]{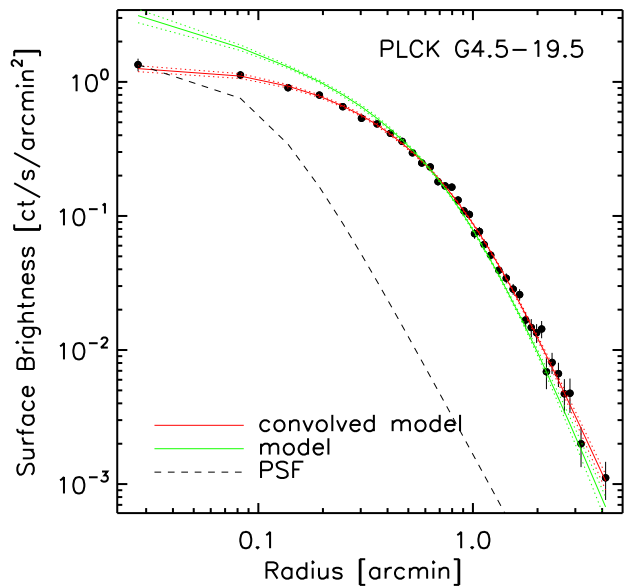}
\caption{{\footnotesize Surface brightness profile of PLCK~G4.5-19.5 {at $z=0.54$}, the highest-$z$ cluster of the sample, as measured with \xmm. The data of the \mos1$\&$2 and \pn\ camera in the $[0.3$--$2.]\,\keV$ energy band are combined.  The green line indicates the best fitting model (see text); the red line is the best fitting model  convolved with the Point Spread Function (PSF) of \xmm\ and the dashed line is the on axis XMM/PSF, normalised to the central intensity. The source is clearly extended.}}\label{fig:cfPSF}
\end{figure}

\begin{figure*}[tbp]
\center
\includegraphics[scale=1.,angle=0,keepaspectratio,height=0.77\textheight]{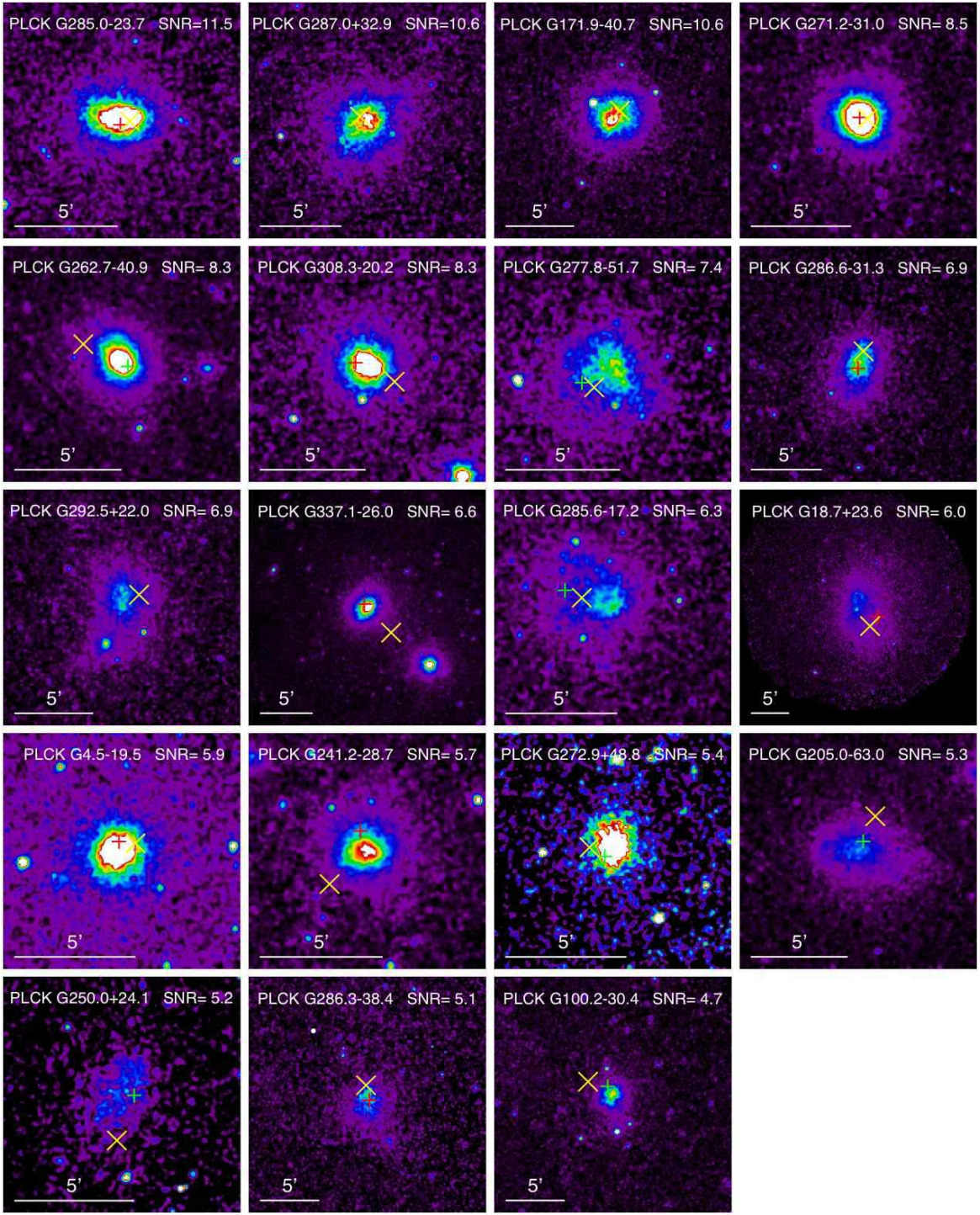}
\caption{{\footnotesize \xmm\ images of all confirmed cluster candidates, except for the two triple systems {which are shown on Fig.~\ref{fig:sc} and discussed in Sec,~\ref{sec:multi}}, in the $[0.3$--$2]\,\keV$ energy band. The observations of PLCK~G272.9+48.8 and PLCK~G250.0+24.1 suffer from high background that has only been crudely subtracted.  Image sizes are $3 \theta_{500}$ on a side, where $\theta_{500}$ is estimated from the $\Mv$--$\YX$ relation (see Sec.~\ref{sec:xquan}). Images are corrected for surface brightness dimming with $z$, divided by the emissivity in the energy band, taking into account galactic absorption and instrument response, and scaled according to the self-similar model. The colour table is the same for all clusters, so that the images would be identical if clusters obeyed strict self-similarity. The majority of the objects show evidence for significant morphological disturbance.  A yellow cross indicates the \planck\ position and a red/green plus sign the position of a \rass-BSC/FSC source.}}
 \label{fig:gal}
\end{figure*}

\section{\xmm\ validation: methods and outcome}

The observations were completed by the end of October 2010.  The median clean  \pn\ observation time is 7 ksec (Table~\ref{tab:obs}).  Of 25 targets, 21 are confirmed as X-ray extended sources.{Only four targets with $\textrm{S/N}\leq 4.1$  were not confirmed}.  The confirmation status of each \xmm\ observation is given in Table~\ref{tab:obs}. 

\subsection{Confirmed cluster candidates}

Our procedure for candidate cluster confirmation  consists of identifying an extended X-ray source coincident with the \planck\ SZ source and checking that the SZ and X--ray properties are consistent. Generally, a candidate cluster (or supercluster) is clearly visible within $5\arcm$ of the \planck\ candidate position, in which case we simply have to  confirm the X--ray source extent. This is achieved by comparing the  surface brightness profile extracted in the [0.3-2.0] keV band with the \xmm\ Point Spread Function. A typical cluster $\beta$--model with a cusp \citep[Eq.~2 in ][]{pra02} is also fitted to the data. Figure~\ref{fig:cfPSF} shows this comparison for the highest-$z$ confirmed extended source.  

17 systems show extended emission from a single source and are confirmed as new clusters of galaxies. Using the Fe K line in the X--ray spectrum we have estimated a redshift for all these objects, albeit with large uncertainties in some cases (see Sec.~\ref{sec:zx}). We have also calculated the  $\YX$ parameter (Sec.~\ref{sec:xquan}). A final check of the candidate confirmation is the good agreement found between the measured SZ signal and that expected from the $\YX$ value (Sec.~\ref{sec:xquan}).  A further two confirmed candidates were revealed to be double systems, one of which is a projection of two independent clusters at different redshifts.  More unexpected are two additional newly-discovered triple systems. All of the  confirmed candidates revealed by \xmm\ to be multiple clusters are discussed in more detail in Sec.~\ref{sec:multi}. 

The \xmm\ images of confirmed single and double systems are shown in Fig.~\ref{fig:gal}. In each panel, the \planck\ source centre position is marked with a cross; in addition, when relevant a red/green plus sign shows the associated RASS-FSC/BSC source. 

\begin{figure}[t]
\begin{centering}
\begin{minipage}[t]{\columnwidth}
\resizebox{\hsize}{!} {
\includegraphics{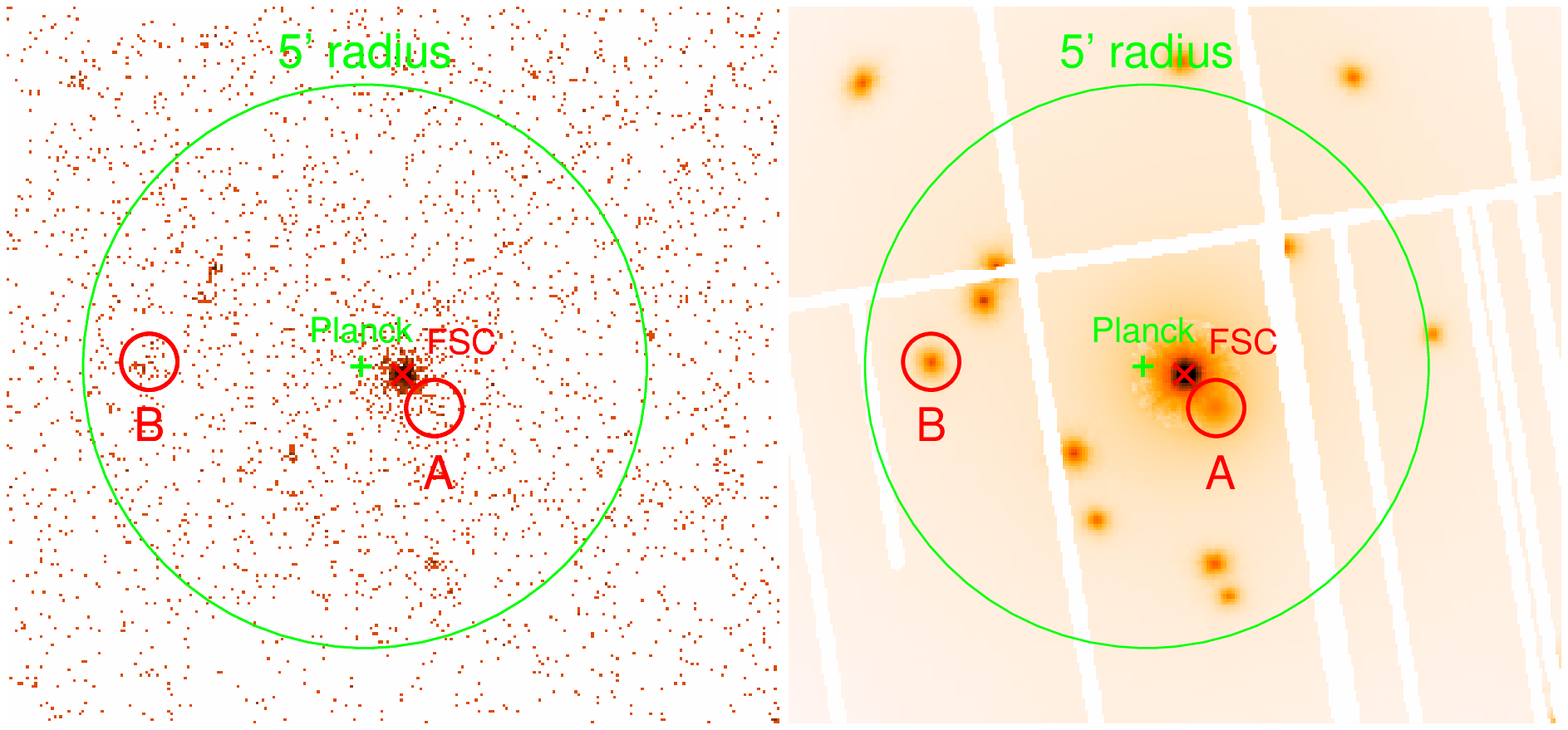}
} 
\end{minipage}\\[2mm]
\begin{minipage}[t]{\columnwidth}
\resizebox{\hsize}{!} {
\includegraphics{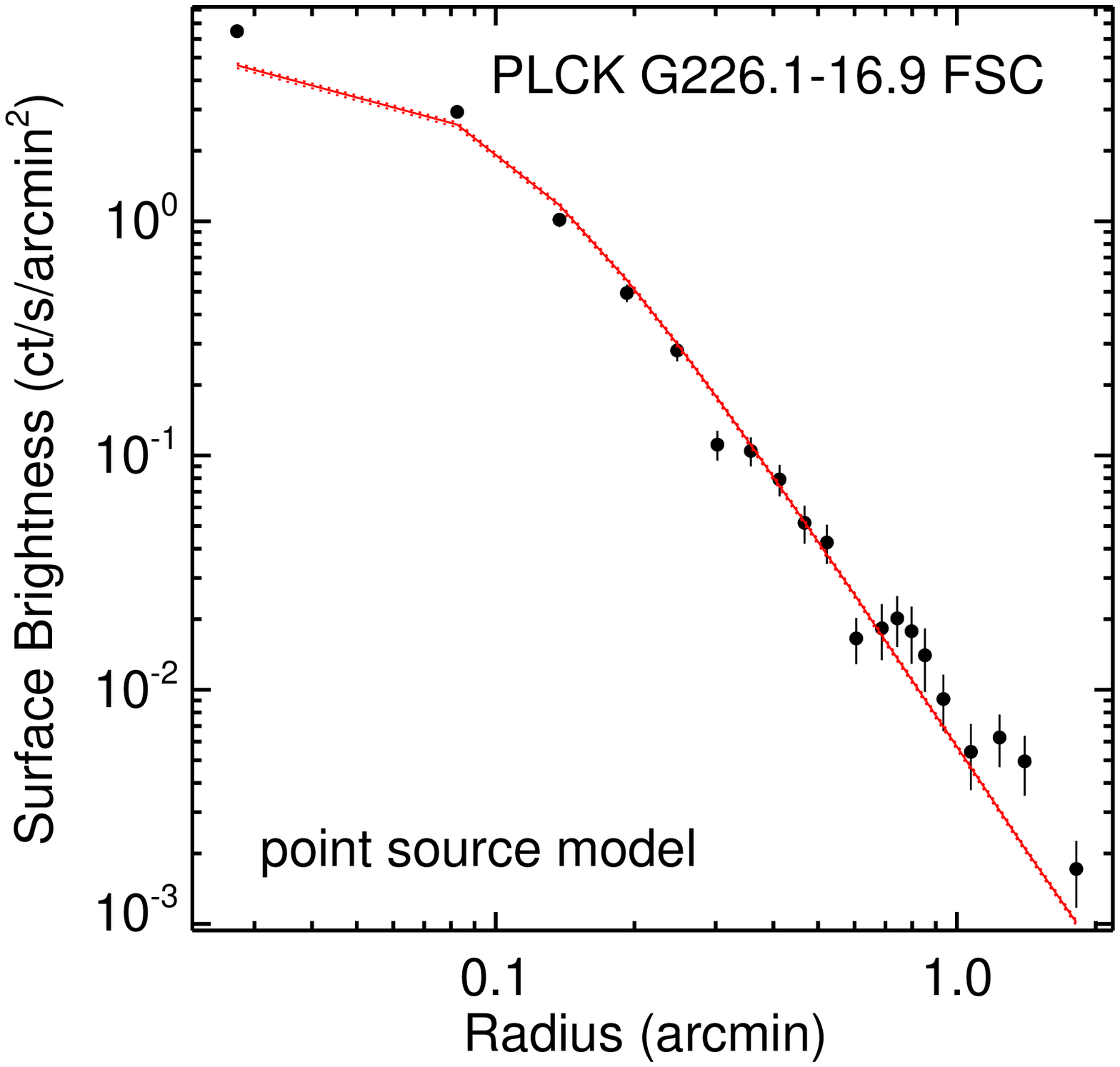}
\hspace{4mm}
\includegraphics{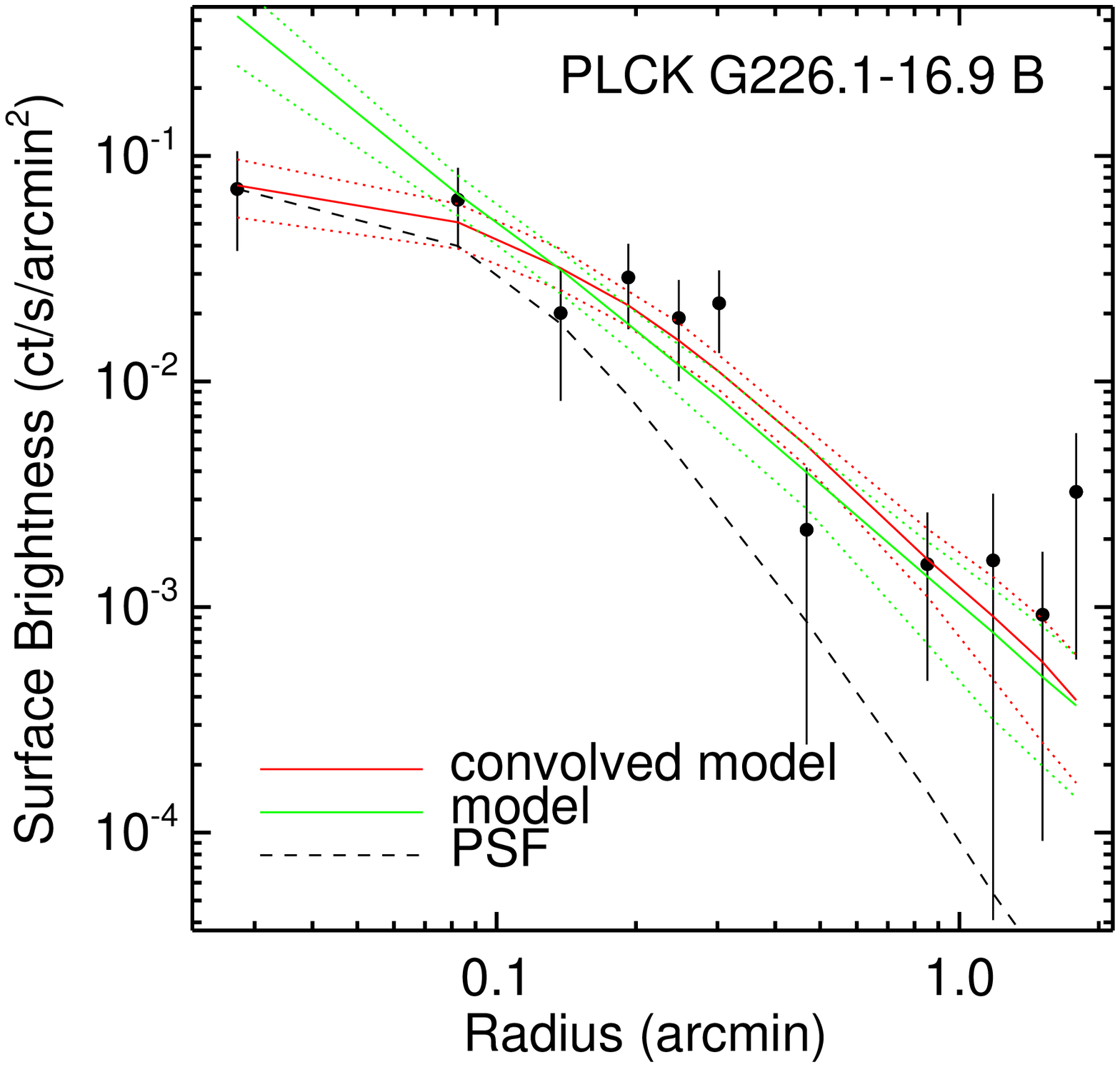}
} 
\end{minipage}
\end{centering}
\caption{{\footnotesize Illustration of the \xmm\ validation procedure results for  a false candidate,  PLCK~G226.1$-$16.9. Top panels: Sum of the \mos\ and \pn\ raw  image (left) and reconstructed \pn\ image (right) in the soft band. The circle of $5\arcm$ radius centred on the \planck\ position (green cross) corresponds to the conservative position error box. The \rass-FSC source is clearly detected (red cross) as a point source: its surface brightness profile (black points in left bottom panel) is well fitted by the \xmm\ PSF (red line).  Two extended sources labelled A and B are also detected. Bottom right panel: same as Fig.~\ref{fig:cfPSF} for source B. }}
\label{fig:false}
\end{figure}

\subsection{False cluster candidates}
\label{sec:valfalse}

In some cases a source is not clearly visible in the image and then the relatively large FWHMs of the HFI beams \citep[$\sim
4\farcm5-9\farcm5$; ][]{planck2011-1.7} complicate source search and confirmation. For these observations we employ the approach described in \citet{suh10}, applying the \xmm--\sas\ source detection algorithms {\tt eboxdetect} and~{\tt emldetect} to the images to determine whether an extended source lies within the \planck\ beam. In brief, images are produced in the $[0.35$--$2.4]\,\keV$ band and {\tt   eboxdetect} is first run in local mode, where the background is estimated locally for each source. Sources found in this first step are then excised, leaving an image suitable for background estimation. The background image is modeled with two components, a vignetted component to represent the X-ray background, and a non-vignetted component to represent the particle and instrumental background. The model is based on a linear combination of two templates based on vignetted and non-vignetted exposure maps, and is fit to the source-subtracted image. We then re-run {\tt eboxdetect} with this model background. All sources found in this step are then analysed with the maximum likelihood (ML)  task {\tt emldetect}, that analyses each source by fitting with a 2D King function convolved with the PSF. The  log of the detection likelihood of each source as defined in the code is ${\tt det\_ml}$ $= - \ln{P_{\rm rand}}$, where the latter is the probability of the observed counts arising from Poisson fluctuations. We set the minimum  ${\tt det\_ml} =6$, corresponding to a $\gtrsim 3\sigma$ detection\footnote{see {\tt http://xmm.esac.esa.int/\\sas/current/doc/emldetect/node3.html} for more details}. In addition to the above, we also searched for possible extended sources using visual inspection of a wavelet-smoothed image.

Figure~\ref{fig:false} illustrates application of the method for the false source PLCK~G226.1$-$16.9. This candidate was the lowest $\textrm{S/N}$  candidate of the Pilot sample ($\textrm{S/N}=4.0$) and located close to a \rass-FSC source, which may have been the cluster counterpart. The top panel shows the raw \xmm\ image and the reconstructed  \pn\ ML source image.  The \rass-FSC source located at 0.8\arcm\  from the \planck\ source is clearly detected with \xmm\ (red plus sign in the top panels). The surface brightness profile is well fitted by a point source convolved with the \xmm\ PSF (bottom left panel). The source spectrum is clearly a power law, and thermal emission from a 0.3 solar abundance ICM is rejected at high confidence at all redshifts and temperatures. This source is most likely an AGN and is definitively not the \planck\ counterpart. 

\begin{figure*}[t]
\centering
\begin{minipage}[t]{0.95\textwidth}
\resizebox{\hsize}{!} {
\includegraphics[scale=1.,angle=0,keepaspectratio]{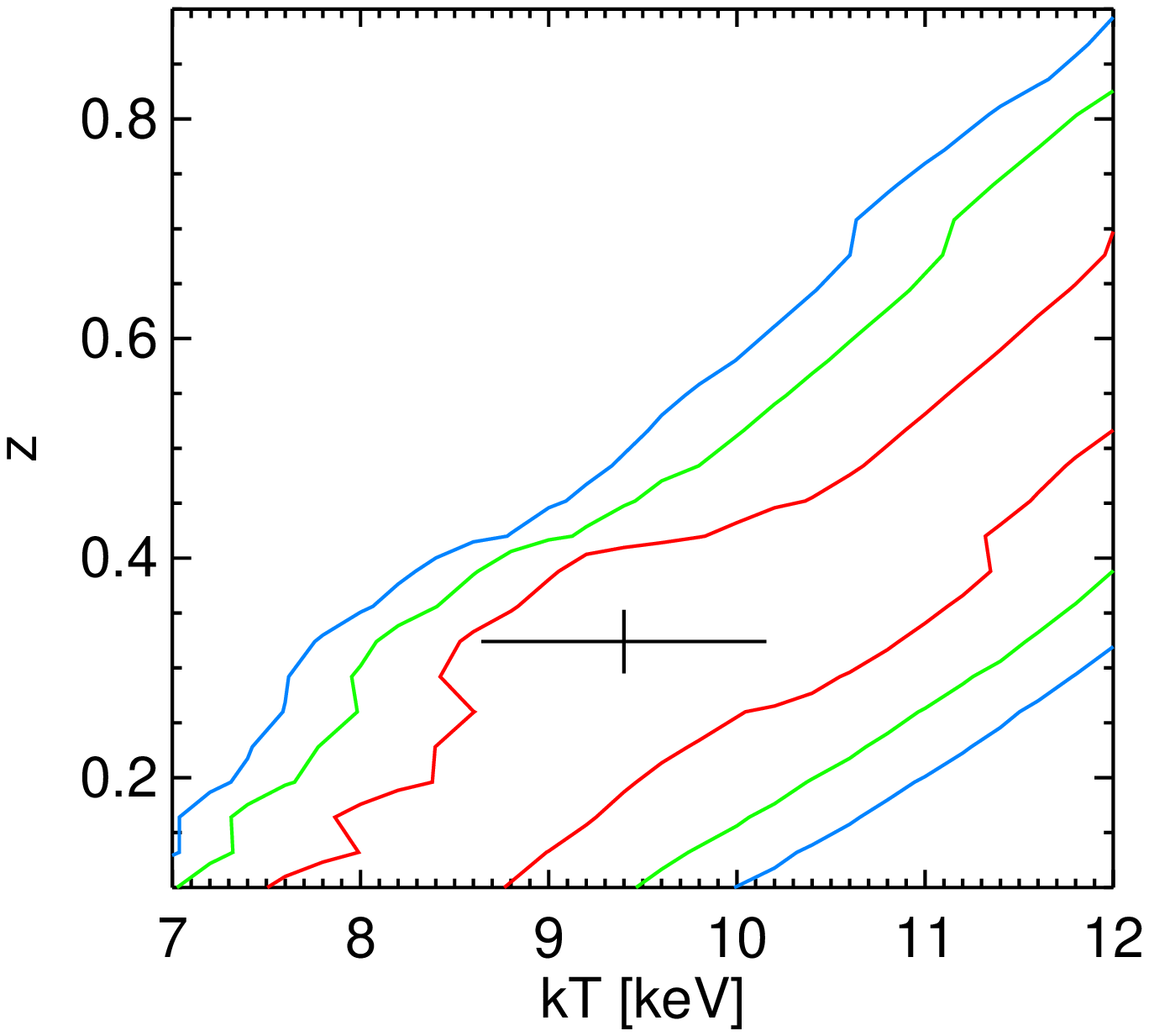}%
\hspace{14mm}
\includegraphics[scale=1.,angle=0,keepaspectratio]{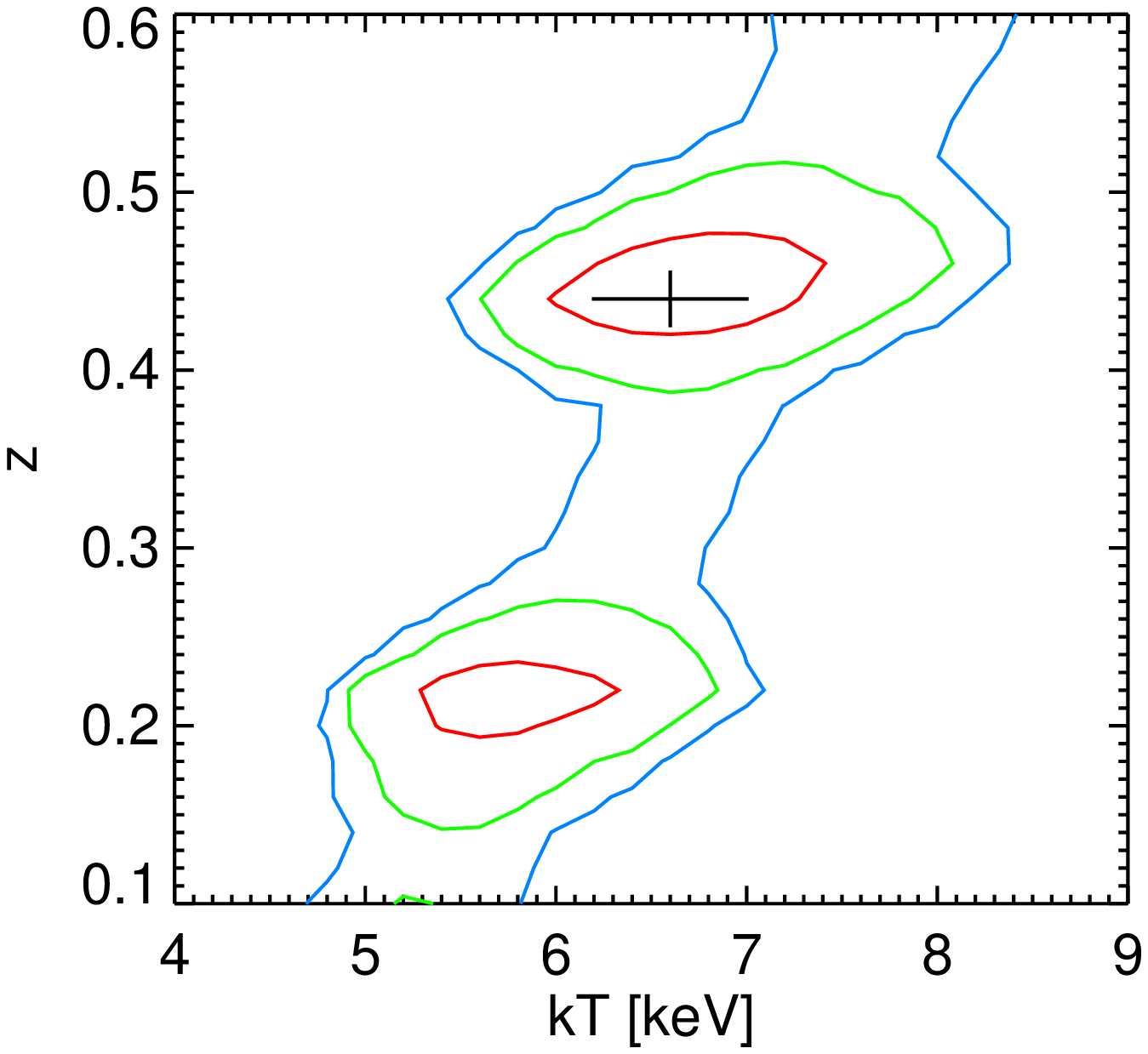}%
\hspace{14mm}
\includegraphics[scale=1.,angle=0,keepaspectratio]{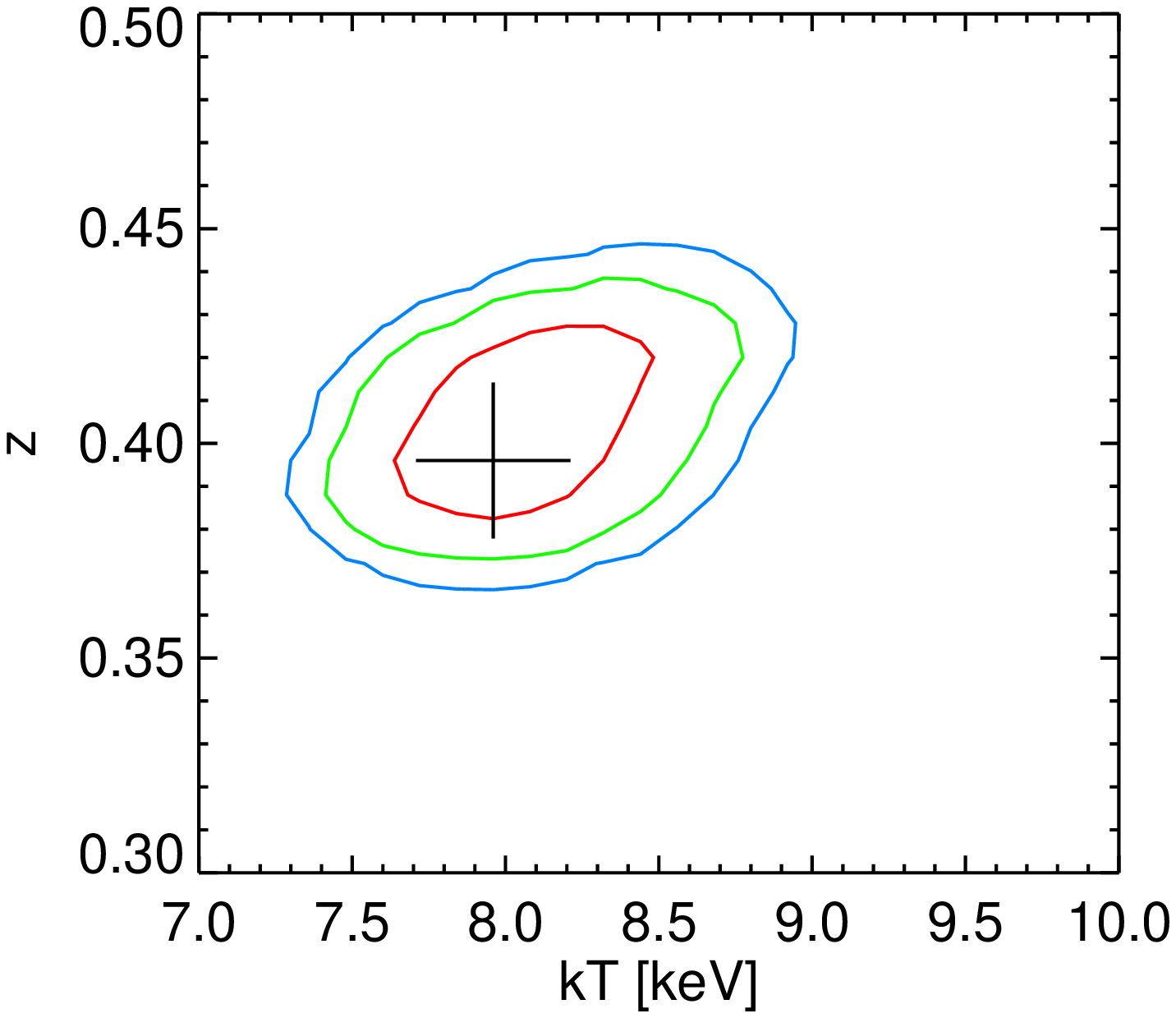}
} 
\end{minipage}
\begin{minipage}[t]{0.95\textwidth}
\resizebox{\hsize}{!} {
\includegraphics[scale=1.,angle=270,keepaspectratio,width=0.33\textwidth]{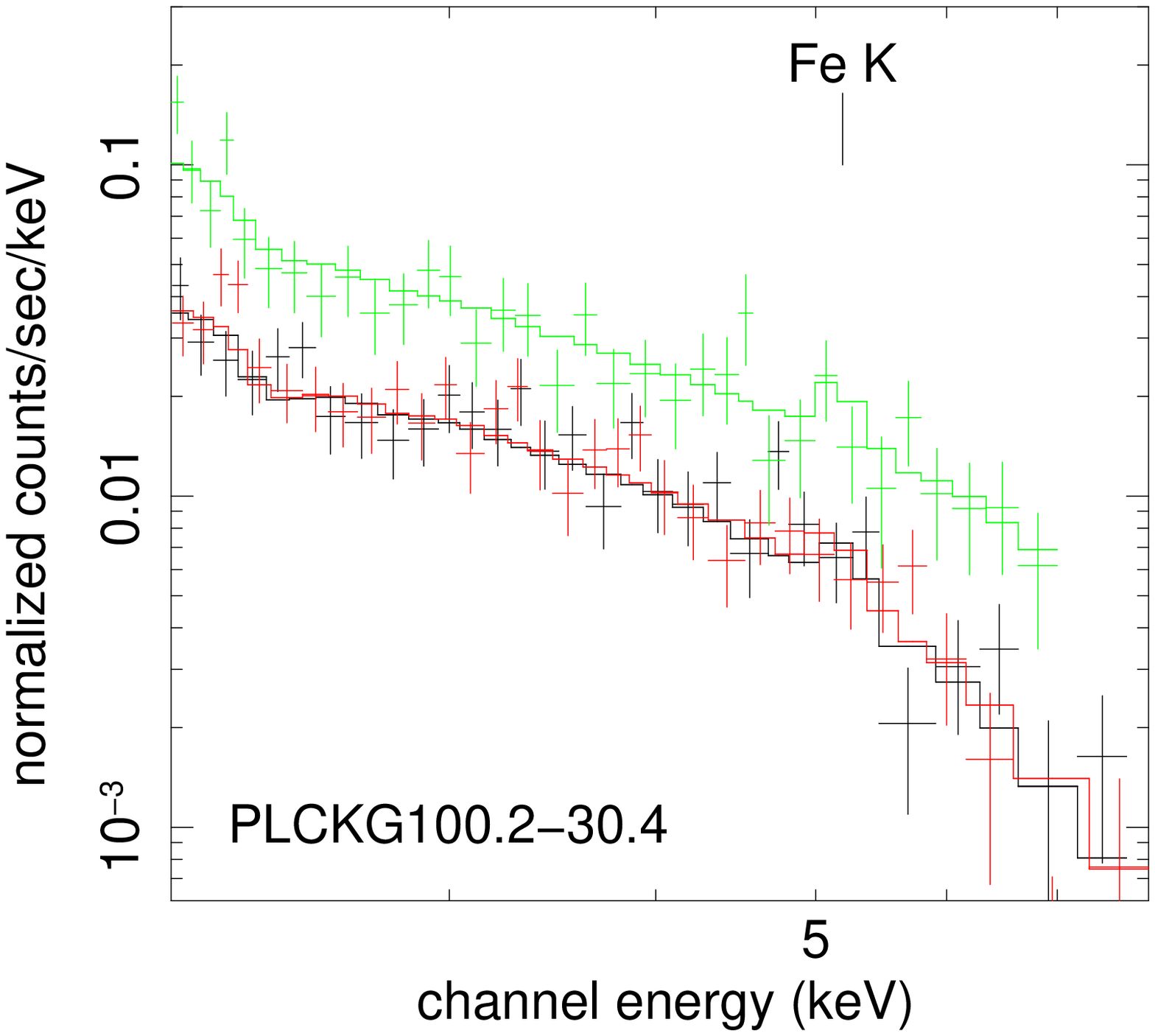}%
\hspace{2mm}
\includegraphics[scale=1.,angle=270,keepaspectratio,width=0.33\textwidth]{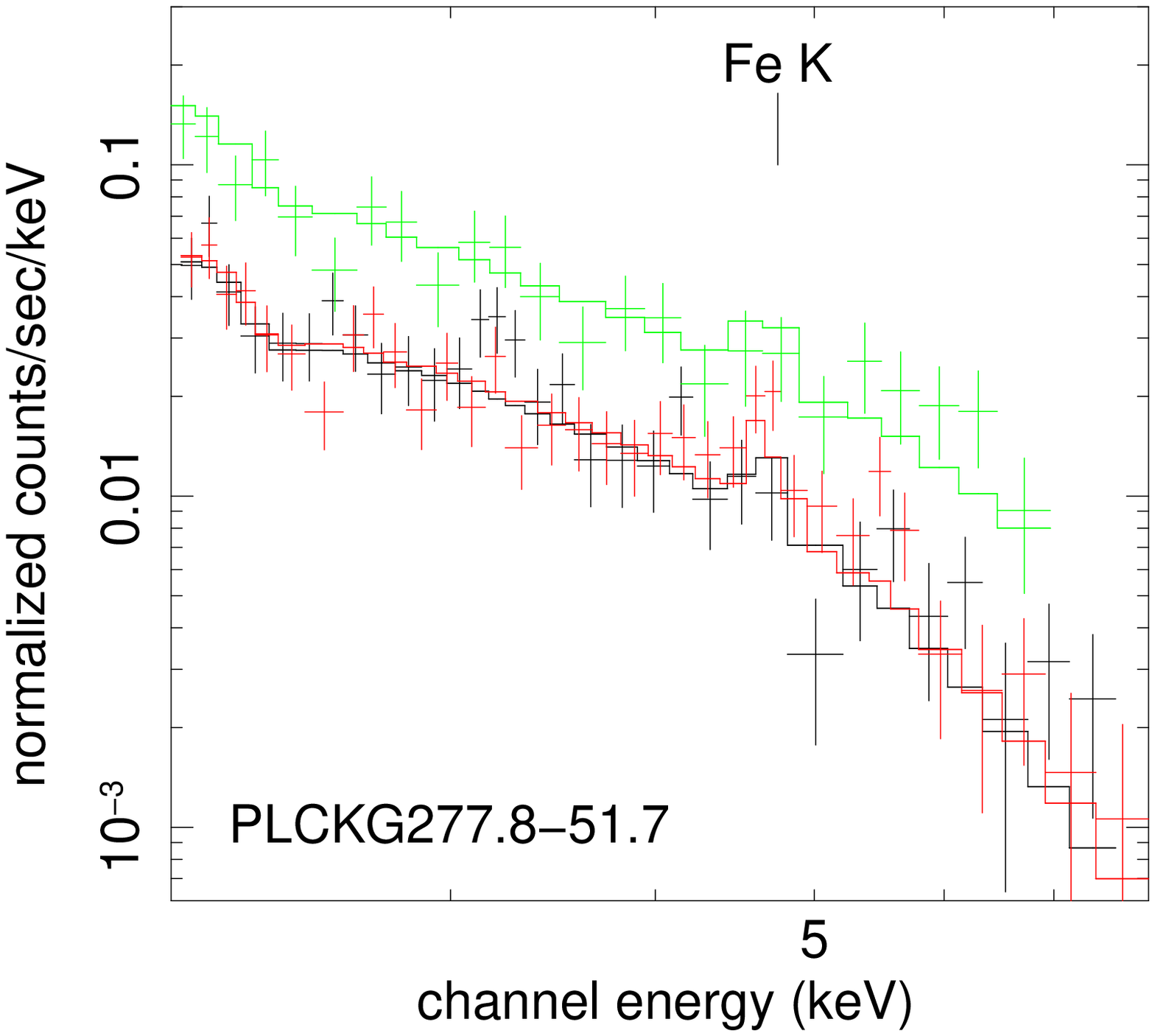}%
\hspace{2mm}
\includegraphics[scale=1.,angle=270,keepaspectratio,width=0.33\textwidth]{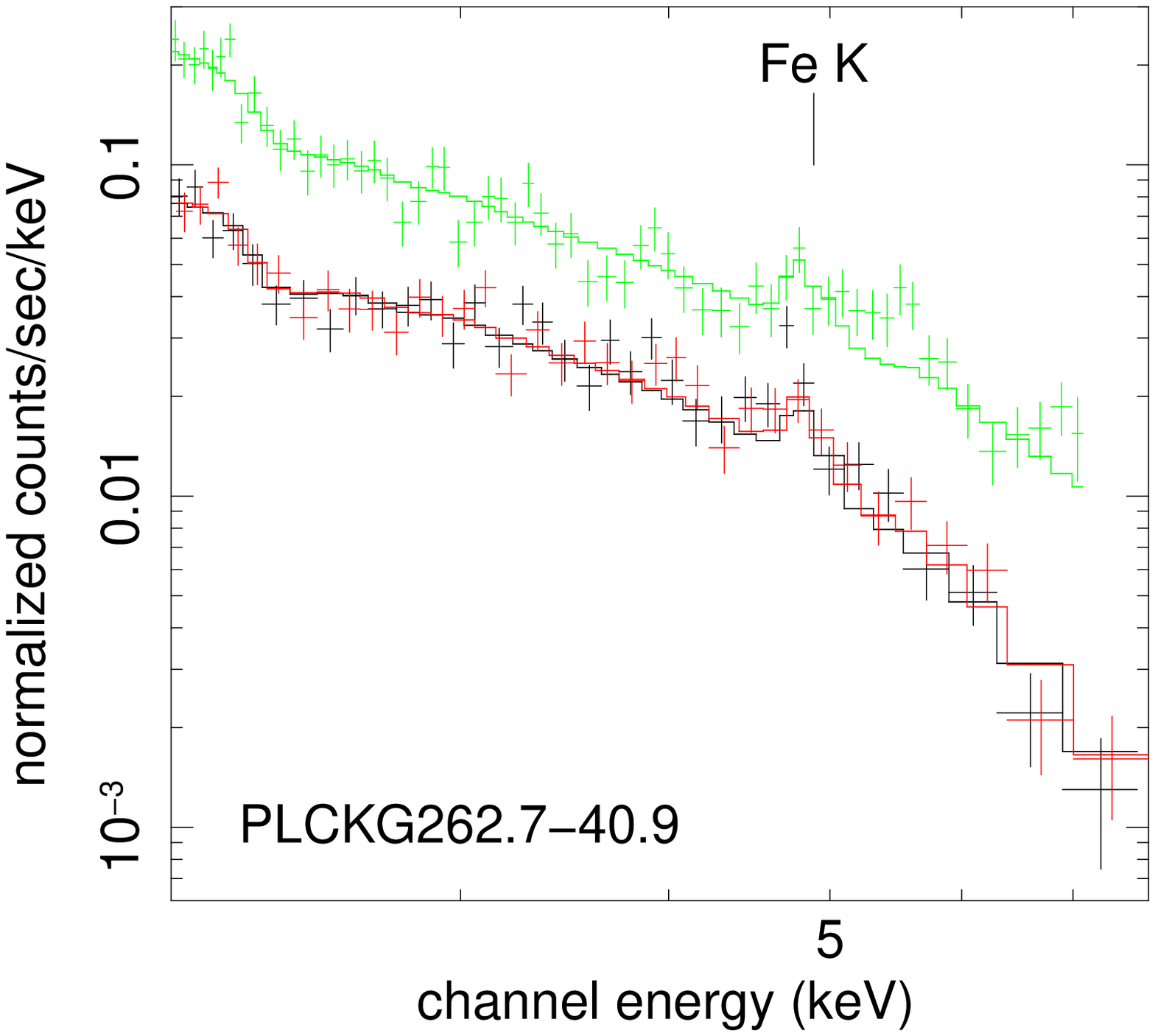}
}
\end{minipage}
\caption{{\footnotesize Top row: Redshift determination from \xmm\ spectroscopy in the $\kT$--$z$ plane. Red, green and blue contours trace  68, 95 and 99.9 per cent confidence levels, respectively. The black error point shows the final best-fitting spectral results with associated statistical errors. Bottom row: \epic\ \mos1$\&$2 (red and black points) and \pn\ (green points) spectra.  Only the data points above $2\,\keV$ are shown for clarity but data down to $0.3\,\keV$ are used in the spectral fitting. The line is the thermal model for the best-fitting redshift. The position of the redshifted Fe K line is marked. From left to right the figures are for sources PLCK~G100.2$-$30.4, PLCK~G277.8$-$51.7 and PLCK~G241.2$-$28.7.  }\label{fig:z}}
\end{figure*}


\begin{table*}[t]
\caption{{\footnotesize X-ray and SZ properties of the confirmed \planck\ sources. 
Column (2): Signal-to-noise ratio derived from the 10-month \planck\ maps on the basis of which the ESZ sample was constructed. Columns (3)--(8): Right ascension and declination of the peak of the X--ray image (J2000). Column (9): redshift from X-ray spectral fitting. Column (10): optical redshift. Column (11): Quality flag for the X-ray redshift measurement (see Sec.~\ref{sec:zx}). Column (12): Total \epic\  count rates  in the $[0.3$--$2]\,\keV$ band,  within the maximum radius of detection given in column (13).  Columns (14)--(20): $\Rv$ is the radius corresponding to a density contrast of 500,  estimated iteratively from the \MYX\ relation (Eq.~\ref{eqn:Yx}), where $Y_{\rm X}=\Mgv\TX$ is the product of the gas mass within $\Rv$ and the spectroscopic temperature $\TX$, and $\Mv$ is the total mass within $\Rv$.  $\LX$ is the luminosity in the $[0.1-2.4]$ keV band and in the  $\Rv$ aperture.  $\Yv$ is the spherically integrated Compton parameter measured with \planck, centred on the X--ray peak, interior to  $\Rv$ as estimated with \xmm.}
\label{tab:xray}} 
\resizebox{\textwidth}{!} {
\begin{tabular}{llrrlcrrrrrrrrrrrrrrrrrrrrr}
\toprule
\multicolumn{1}{l}{Name} & \multicolumn{1}{c}{$\textrm{S/N}$} & 
\multicolumn{1}{c}{RA$_{\rm X}$} & \multicolumn{1}{c}{DEC$_{\rm X}$} & 
\multicolumn{1}{c}{$z_{\rm Fe}$} & \multicolumn{1}{c}{$z_{\rm opt}$} & \multicolumn{1}{c}{$Q_{\rm z}$} &
\multicolumn{1}{c}{$R$} & \multicolumn{1}{c}{$\theta_{\rm det}$} &
\multicolumn{1}{c}{$R_{500}$} & \multicolumn{1}{c}{$\TX$} &
\multicolumn{1}{c}{$M_{\rm gas,500}$} & \multicolumn{1}{c}{$\YX$} &
\multicolumn{1}{c}{$Y_{500}$} & \multicolumn{1}{c}{$M_{500}$} &
\multicolumn{1}{c}{$\LX$} \\

\noalign{\smallskip}
\multicolumn{1}{c}{} & \multicolumn{1}{c}{} & 
\multicolumn{1}{c}{[h:m:s]} &\multicolumn{1}{c}{[d:m:s]} & 
\multicolumn{1}{c}{} & \multicolumn{1}{c}{} &
\multicolumn{1}{c}{} & \multicolumn{1}{c}{$[{\rm cts\,s^{-1}}]$} & \multicolumn{1}{c}{$[\arcmin]$} &
\multicolumn{1}{c}{[kpc]} & \multicolumn{1}{c}{[keV]} &
\multicolumn{1}{c}{$[10^{14}\,{\rm M_{\odot}}]$} & \multicolumn{1}{c}{$[10^{14}\,{\rm M_{\odot}\,\keV}]$} &
\multicolumn{1}{c}{$[10^{-4}\,{\rm Mpc^2}]$} & \multicolumn{1}{c}{$[10^{14}\,{\rm M_{\odot}}]$} &
\multicolumn{1}{c}{$[10^{44}\,{\rm erg\,s^{-1}}]$} \\

\midrule

 PLCK~G285.0$-$23.7&11.5&{07:23:18.4}&{$-$73:27:20.6}   & 0.39 &  0.37$^{a}$ & 2 & 1.85$\pm$ 0.02 & 4.1 & 1216 & 6.98$\pm$ 0.74 & 1.23$\pm$ 0.04 & 8.62$\pm$ 1.28 & 1.27$\pm$ 0.35 & 7.71$\pm$ 0.50 & 16.91$\pm$ 0.27 \\
  PLCK~G287.0+32.9&10.6&{11:50:49.2}&{$-$28:04:36.5} & 0.39 & \ldots & 1 & 2.68$\pm$ 0.01 & 6.8 & 1541 & 12.86$\pm$ 0.42 & 2.39$\pm$ 0.03 & 30.69$\pm$ 0.36 & 3.30$\pm$ 0.16 & 15.72$\pm$ 0.27 & 17.20$\pm$ 0.11 \\
  PLCK~G171.9$-$40.7&10.6&{03:12:57.4}&{ 08:22:10.3}  & 0.27 & \ldots & 2 & 2.19$\pm$ 0.03 & 5.3 & 1428 & 10.65$\pm$ 0.42 & 1.43$\pm$ 0.04 & 15.26$\pm$ 0.72 & 2.05$\pm$ 0.21 & 10.92$\pm$ 0.37 & 11.28$\pm$ 0.19 \\
  PLCK~G271.2$-$31.0& 8.5&{05:49:19.5}&{$-$62:05:16.0}  & 0.37 & \ldots & 2 & 3.32$\pm$ 0.01 & 5.2 & 1212 & 7.94$\pm$ 1.23 & 1.02$\pm$ 0.04 & 8.08$\pm$ 1.03 & 1.02$\pm$ 0.17 & 7.47$\pm$ 0.70 & 18.95$\pm$ 0.16 \\
  PLCK~G262.7$-$40.9& 8.3&{04:38:17.2}&{$-$54:19:25.1}  & 0.39 & 0.54$^{b}$ & 2 & 1.72$\pm$ 0.02 & 5.6 & 1169 & 7.77$\pm$ 0.87 & 0.90$\pm$ 0.06 & 7.01$\pm$ 0.98 & 1.14$\pm$ 0.22 & 6.87$\pm$ 0.68 & 9.94$\pm$ 0.47 \\
  PLCK~G277.8$-$51.7& 7.4&{02:54:16.7}&{$-$58:56:44.0} & 0.44 & \ldots & 1 & 1.33$\pm$ 0.02 & 5.8 & 1172 & 6.37$\pm$ 0.84 & 1.26$\pm$ 0.02 & 8.01$\pm$ 0.57 & 1.70$\pm$ 0.20 & 7.32$\pm$ 0.38 & 9.46$\pm$ 0.07 \\
  PLCK~G286.6$-$31.3& 6.9&{05:31:27.5}&{$-$75:10:41.2}  & 0.21 & \ldots & 2 & 1.82$\pm$ 0.04 & 4.6 & 1149 & 6.85$\pm$ 0.94 & 0.60$\pm$ 0.07 & 4.14$\pm$ 3.01 & 0.61$\pm$ 0.32 & 5.32$\pm$ 0.86 & 3.72$\pm$ 0.23 \\
  PLCK~G292.5+22.0& 6.9&{12:01:05.3}&{$-$39:52:26.2}  & 0.30 & \ldots & 2 & 1.42$\pm$ 0.03 & 6.8 & 1336 & 9.82$\pm$ 0.84 & 1.17$\pm$ 0.04 & 11.49$\pm$ 1.33 & 1.45$\pm$ 0.24 & 9.25$\pm$ 0.60 & 5.46$\pm$ 0.09 \\
  PLCK~G285.6$-$17.2& 6.3&{08:43:44.4}&{$-$71:13:13.7} & 0.35 & \ldots & 1 & 0.69$\pm$ 0.02 & 4.4 & 1044 & 4.87$\pm$ 0.35 & 0.71$\pm$ 0.01 & 3.46$\pm$ 0.40 & 0.79$\pm$ 0.13 & 4.67$\pm$ 0.22 & 4.45$\pm$ 0.08 \\
   PLCK~G18.7+23.6& 6.0$^{c}$&{17:02:21.3}&{$-$00:59:58.9} &0.09 & \ldots & 2 & 3.94$\pm$ 0.01 & 10.5 & 1034 & 4.63$\pm$ 0.32 & 0.39$\pm$ 0.02 & 1.80$\pm$ 0.32 & 0.32$\pm$ 0.17 & 3.42$\pm$ 0.22 & 1.21$\pm$ 0.05 \\
    PLCK~G4.5$-$19.5& 5.9&{19:17:04.6}&{$-$33:31:21.9}& 0.54 & \ldots & 2 & 1.29$\pm$ 0.02 & 4.8 & 1245 & 10.39$\pm$ 0.52 & 1.37$\pm$ 0.03 & 14.27$\pm$ 0.87 & 1.99$\pm$ 0.20 & 9.88$\pm$ 0.35 & 17.78$\pm$ 0.11 \\
  PLCK~G241.2$-$28.7& 5.7&{05:42:56.8}&{$-$35:59:54.8}& 0.42 & \ldots & 2 & 0.92$\pm$ 0.02 & 3.6 & 1065 & 6.08$\pm$ 0.32 & 0.75$\pm$ 0.02 & 4.58$\pm$ 0.39 & 0.79$\pm$ 0.11 & 5.37$\pm$ 0.20 & 6.72$\pm$ 0.12 \\
  PLCK~G272.9+48.8& 5.4&{11:33:10.5}&{$-$09:28:52.2}& 0.40 & \ldots & 2 & 0.66$\pm$ 0.01 & 3.7 & 1053 & 4.67$\pm$ 0.32 & 0.88$\pm$ 0.02 & 4.09$\pm$ 0.32 & 0.87$\pm$ 0.18 & 5.07$\pm$ 0.21 & 12.36$\pm$ 0.09 \\
  PLCK~G205.0$-$63.0& 5.3&{02:46:25.8}&{$-$20:33:16.9}& 0.31 & \ldots & 2 & 1.12$\pm$ 0.03 & 5.5 & 1111 & 6.06$\pm$ 0.39 & 0.72$\pm$ 0.02 & 4.38$\pm$ 0.30 & 0.83$\pm$ 0.09 & 5.37$\pm$ 0.21 & 3.89$\pm$ 0.08 \\
  PLCK~G250.0+24.1& 5.2& {09:32:13.8}&{$-$17:38:06.7}& 0.40 & \ldots & 0 & 0.18$\pm$ 0.02 & 3.8 & 1061 & 6.75$\pm$ 0.28 & 0.63$\pm$ 0.03 & 4.27$\pm$ 0.42 & 0.72$\pm$ 0.12 & 5.19$\pm$ 0.21 & 3.37$\pm$ 0.19 \\
  PLCK~G286.3$-$38.4& 5.1&{03:59:10.2}&{$-$72:04:46.1}& 0.31 & 0.307$\pm$0.003$^{d}$ & 2 & 0.32$\pm$ 0.05 & 4.5 & 1064 & 5.60$\pm$ 0.14 & 0.62$\pm$ 0.01 & 3.48$\pm$ 0.08 & 0.80$\pm$ 0.06 & 4.72$\pm$ 0.08 & 4.07$\pm$ 0.02 \\
  PLCK~G100.2$-$30.4& 4.7&{23:22:14.9}&{ 28:31:13.5} & 0.31 & 0.38$\pm$0.04$^{a}$ & 0 & 0.76$\pm$ 0.01 & 4.7 & 1128 & 9.03$\pm$ 0.30 & 0.53$\pm$ 0.02 & 4.76$\pm$ 0.29 & 0.45$\pm$ 0.14 & 5.63$\pm$ 0.22 & 3.36$\pm$ 0.08 \\
\midrule
PLCKG308.3$-$20.2 & 8.3 \\
{\it A} & \ldots & {15:18:55.5} & {$-$81:30:30.1} & 0.48 & \ldots & 2 & 1.06$\pm$ 0.01 & 3.9 & 1250 & 9.55$\pm$ 0.56 & 1.31$\pm$ 0.04 & 12.55$\pm$ 0.92 &  \ldots& 9.32$\pm$ 0.38 & 15.65$\pm$ 0.19 \\
{\it B} & \ldots & {15:16:52.6} & {$-$81:35:50.0} & 0.48$^{f}$ & \ldots & $-$1 & 0.42$\pm$ 0.01 & 2.9 & 894 & 3.79$\pm$ 0.35 & 0.55$\pm$ 0.03 & 2.09$\pm$ 0.25 & \ldots & 3.41$\pm$ 0.23 & 8.67$\pm$ 0.26 \\
PLCKG337.1$-$26.0 & 6.6 \\
{\it A} & \ldots & {19:14:37.7} & {$-$59:28:16.7} & 0.26 & \ldots & 2 & 3.28$\pm$ 0.03 & 7.5 & 1177 & 6.16$\pm$ 0.23 & 0.86$\pm$ 0.01 & 5.30$\pm$ 0.25 & \ldots & 6.05$\pm$ 0.16 & 8.95$\pm$ 0.07 \\
{\it b} & \ldots & {19:13:51.4} &{ $-$59:33:51.6} & 0.12 & \ldots & 2 & 1.79$\pm$ 0.02 & 4.5 & 749 & 2.84$\pm$ 0.20 & 0.12$\pm$ 0.01 & 0.34$\pm$ 0.03 & \ldots & 1.34$\pm$ 0.07 & 1.02$\pm$ 0.01 \\
PLCKG214.6+36.9 & \textcolor{red}{3.6} \\
{\it A} & \ldots &  {09:08:49.5 }&{14:38:29.4} & 0.45  & 0.45$^{d}$ &2 & 0.38$\pm$ 0.01 & 4.3 & 767 & 3.46$\pm$ 0.26 & 0.25$\pm$ 0.01 & 0.85$\pm$ 0.09 &  \ldots& 2.08$\pm$ 0.12 & 3.03$\pm$ 0.07 \\
{\it B} & \ldots & {09:09:02.1} & {14:39:41.7} & 0.45 & 0.46$^{a}$ & 1 & 0.11$\pm$ 0.01 & 2.3 & 750 & 4.16$\pm$ 0.75 & 0.18$\pm$ 0.01 & 0.76$\pm$ 0.18 & \ldots & 1.94$\pm$ 0.26 & 1.03$\pm$ 0.06 \\
{\it C} & \ldots & {09:08:51.4 }& {14:45:55.3} & 0.45$^{f}$ & 0.45$^{d}$ & $-$1 & 0.25$\pm$ 0.01 & 3.1 & 809 & 3.72$\pm$ 0.50 & 0.30$\pm$ 0.01 & 1.13$\pm$ 0.19 & \ldots & 2.44$\pm$ 0.23 & 2.28$\pm$ 0.08 \\
PLCKG334.8$-$38.0 & \textcolor{red}{3.4} \\
{\it A} & \ldots & {20:52:16.8} & {$-$61:12:29.4} & 0.35$^{e}$ & \ldots & 2 & 0.13$\pm$ 0.01 & 1.8 & 722 & 3.14$\pm$ 0.33 & 0.15$\pm$ 0.01 & 0.48$\pm$ 0.07 &  \ldots  & 1.55$\pm$ 0.13 & 0.77$\pm$ 0.06 \\
{\it B} & \ldots & {20:53:08.0} & {$-$61:10:35.3} & 0.35$^{f}$ & \ldots & $-$1 & 0.08$\pm$ 0.01 & 3.2 & 605 & 2.02$\pm$ 0.31 & 0.09$\pm$ 0.01 & 0.19$\pm$ 0.05 &  \ldots  & 0.91$\pm$ 0.12 & 0.47$\pm$ 0.33 \\
{\it C} & \ldots & {20:52:44.3} & {$-$61:17:24.5} & 0.35$^{f}$ & \ldots & $-$1 & 0.03$\pm$ 0.00 & 1.2 & 607 & 3.13$\pm$ 1.71 & 0.06$\pm$ 0.02 & 0.19$\pm$ 0.18 &  \ldots  & 0.92$\pm$ 0.44 & 0.21$\pm$ 0.06 \\
\bottomrule
\end{tabular}
}
{\footnotesize
$^{\rm a}$ Photometric redshift. See Appendix~\ref{sec:apa}. 
$^{\rm b}$ Photometric redshift for ACT-CL~J0438-5419 in \citet{men10}. 
{$^{\rm c}$ Rounded from 5.99 to 6.0, therefore not included in ESZ \citep{planck2011-5.1a}.}
$^{\rm d}$ Spectroscopic redshift. 
$^{\rm e}$ Redshift constrained from Fe L complex.
$^{\rm f}$ Redshift assumed to be identical to that of component $A$.
}
\normalsize
 \end{table*}

The source list produced by the ML method includes two potentially extended sources, only one of which is within 5\arcm\ of the \planck\ source position (source labelled A in the Figure). It is located $\sim0.8\arcm$ from the RASS-FSC source position and is much fainter, showing the capability of \xmm\ to separate sources. The source has an estimated  $[0.35$--$2.4]\,\keV$ flux of $\sim 2\times10^{-14}\,{\rm erg\,cm^{2}\,s^{-1}}$, which is  more than 5 times lower than that expected from the SZ source even if located at $z\sim1.5$.  Furthermore its extent (although not well constrained) is small and it is perfectly coincident with a 2MASS galaxy. This source again could not  be the \planck\ counterpart.  Finally, from a wavelet-smoothed image, there was a hint that another source, located $3.5\arcm$ away from the \planck\ candidate position, was extended, although it was not classified as such with the ML method (source labelled B).  We extracted its profile and confirmed it as extended,  although the extent was not very significant (bottom right panel of Fig.~\ref{fig:false}). However, its flux was half that expected from the observed \planck\ flux, even for a cluster at $z=1.5$. Nevertheless, in view of possible errors in the \planck\ position, we re-analysed the \planck\ data by re-extracting the signal exactly at the position of the source. The SZ detection was no longer significant, leading us to conclude that source B was definitively not the counterpart to the \planck\ candidate.  From the \xmm\ observation we thus concluded that this \planck\ candidate was a false detection.
~\\


\section{Redshift estimate}

\subsection{\xmm\ estimates}
\label{sec:zx}

The ICM has a typical abundance of $0.3$ times solar, implying that metals are present in large amounts (\citealp[see][for recent
work on metals in the cluster context]{bal07,lec08,mau08}). The spectroscopic sensitivity of \xmm\ allows the measurement of the intensity and centroid energy of the strongest line emission, namely the  {Fe~K and Fe~L} line complexes (respectively found at $E\sim6.4$ and $E\sim 1$~keV at $z=0$). As a consequence the Fe line emission can be used to constrain the cluster redshift. We have thus searched for their signature in the \xmm\ observations, focusing mainly on the Fe~K complex, which is about 10 times as strong as any other line emission in the ICM. A clear detection then provides an estimate of the X--ray redshift $z_{\rm Fe}$. 

The intrinsic spectral resolution of \xmm\  is $\Delta E \sim 150$~eV at  $6.4$~keV and  $\Delta E  \sim 100$~eV at $3.2$~keV; the energies here correspond to  the Fe~K complex centroid energy  for a cluster at a redshift of $z=0$ and $z=1$, respectively.  Such resolution allows centroid determination to typically $10$--$15\,$eV for high quality spectra, of the same magnitude as the systematic uncertainty of the calibration of the energy reconstruction (about $5\,$eV and $10\,$eV in the central CCD of the \mos\ and \pn\ camera, respectively). The overall energy uncertainty would yield a typical corresponding redshift uncertainty of  $\Delta z\sim0.002$.  In practice,  the limiting factor affecting the accuracy of the redshift determination is the statistical uncertainty in the spectrum, which is linked to the observation duration and overall quality (background conditions). Furthermore, \planck-detected clusters are mostly massive, hot objects with low {Fe~K} line equivalent widths \citep{rot85}. This makes $z_{\rm Fe}$ determination more difficult than for cooler objects.   

To estimate $z_{\rm Fe}$ using \xspec\ we first performed a spectral fit of the region corresponding to the maximum significance of the detection (defined from the surface brightness profile in the soft band), with the redshift as one of the free parameters. The abundance was left free to fit within a typical cluster range ($0.2$--$0.6$ times solar).  From this starting point we investigated the $\chi^2$ in the $\kT$--$z_{\rm Fe}$ plane using a regular grid. The best fitting $\kT$ and $z_{\rm Fe}$ values were recovered from a simple maximum likelihood analysis, whereupon these best fitting values were used as input for a final spectral fit. When a two- or three--peak degeneracy appeared in the $\kT$--$z_{\rm Fe}$ plane, we checked the various potential $z_{\rm Fe}$ values and chose the redshift giving the best spectral fit as defined by the $\chi^2$ and the null probability hypothesis.

This redshift estimation process is illustrated by three cases in Fig.~\ref{fig:z} with the left panel showing a fully degenerate case, the middle panel a double-peaked case and the right panel a well-constrained case. These redshifts  are flagged with quality values $Q_{\rm z}=0,1,2$, respectively, in Table~\ref{tab:xray}. The few cases where no redshift estimate was possible are flagged with $Q_{\rm z}=-1$.
\subsection{Optical estimates}
\label{sec:zopt}

For three clusters, we have estimated the redshift either from existing optical archive observations or dedicated follow-up observations as part of the overall \planck\ cluster candidate validation programme. The most recent corresponds to telescope time acquired by the \planck\ consortium at the ENO telescopes, Observatorio del Teide (Tenerife, Spain -- AO 2010A
and 2010B). The details of the observation setup and data processing can be found in Appendix~\ref{sec:apa}.
\begin{itemize}
\item PLCK~G100.2$-$30.4. The source was observed in 4 bands ({\it griz}) with the CAMELOT camera at the 0.82-m IAC80 telescope. After data reduction, we derived a photometric redshift of $z_{\rm phot}=0.38 \pm 0.04$, using the {\sc   bpz} code \citep{bpz00}. This estimate is compatible within $3 \sigma$, with the {$z_{\rm Fe}=0.31$} derived from the X-ray spectroscopy.
\item  PLCK~G285.0$-$23.7. We reduced the ESO NTT/SUSI2 archive images for this object, deriving a red-sequence redshift of $z_{\rm phot}=0.37$. This estimate is in good agreement with the X-ray spectroscopic redshift $z_{\rm Fe}=0.39$.
\item PLCK286.3$-$38.4. ESO NTT/SUSI2 images and NTT/EMMI spectroscopic archive data targetting the X-ray source RX\,J0359.1$-$7205 were available. From a poor quality NTT/EMMI spectrum, we extracted a redshift of $z_{\rm spec}=0.307 \pm 0.003$, backed-up by the presence of two absorption line features (H$\beta$ and \ion{Mg}{I}). Again this value agrees well with the X-ray spectroscopic redshift of $z_{\rm Fe}=0.31$.

\end{itemize}

Finally, the source PLCK~G262.7$-$40.9 appeared to be one of the  ACT SZ optically-confirmed clusters \citep{men10}, accepted for observation by {\it Chandra} after it was scheduled for observation with \xmm.  The reported photometric redshift is $z_{\rm   phot}=0.54\pm 0.05$, in disagreement with our X-ray-derived value of $z_{\rm Fe}=0.38$ at the $3\sigma$ level. Although slightly weak, the Fe~K line is  clearly seen in the X-ray spectrum (see Fig.~\ref{fig:z} right panels). We thus adopt the X--ray estimate.  However, optical spectroscopic observations are clearly needed to confirm the cluster redshift.  All compiled and derived optical redshifts are reported in column 6 ($z_{\rm opt}$) in Table~\ref{tab:xray}.

\section{Physical parameter estimates  of confirmed clusters}
\subsection{\xmm\ data}

\label{sec:xquan}

For all single clusters (17 systems) or obvious sub-components in double and triple systems (4 objects), the X-ray peak position was taken to be the (sub-)cluster centre. For these systems we undertook a more in-depth analysis assuming that a spherically symmetric approximation is appropriate.

Surface brightness profiles, centred on the X--ray peak, were extracted in the $[0.3$--$2]\,\keV$ band in $3\farcs3$ bins. Deprojected, PSF-corrected gas density profiles were then calculated using the method described  in \citet{cro08}. Global cluster parameters were then estimated self-consistently within $R_{500}$ via iteration about the $M_{500}$--$\YX$ relation of \citet[][see also \citealt{pra10}]{arn10}, viz.,

{\small
\begin{equation}
E(z)^{2/5}\M500 = 10^{14.567 \pm 0.010} \left[\frac{\YX}{2\times10^{14}\,{\msol}\,\keV}\right]^{0.561 \pm 0.018}\,{\rm h_{70}^{-1}\, \msol},\label{eqn:Yx}
\end{equation}
}

\noindent assuming the standard evolution predicted by the self-similar model purely based on gravitation. In addition, the X-ray luminosity in the $[0.1-2.4]$ keV band interior to $\Rv$, $\LX$ was calculated as described in \citet{pra09}. All resulting  X--ray properties are summarized in Table~\ref{tab:xray}.  Errors include only statistical uncertainties.  We did not attempt to include systematic errors due to redshift uncertainty or high background level; such estimates are beyond the scope of the paper.  The results for this sample are not used for quantitative statistical study (e.g derivation of scaling laws), which would require redshift confirmation (sources with $Q_{\rm z}<2$) and deeper  \xmm\ observations.

\begin{figure}[t]
\center
\includegraphics[scale=1.,angle=0,keepaspectratio,width=0.85\columnwidth]{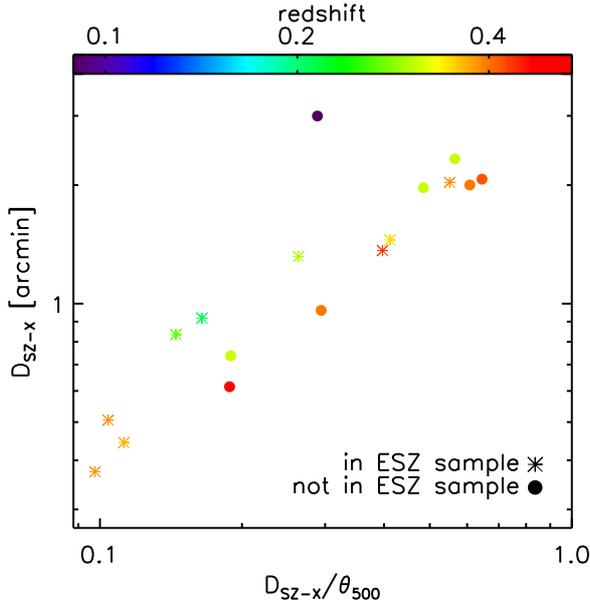}%
\caption{{\footnotesize Distance of blind SZ position to X-ray position, $D_{\rm SZ-X}$, as a function of $D_{\rm SZ-X}$, normalised to the cluster size $\theta_{500,X}$ for single confirmed systems. The clusters are colour-coded according to redshift. Note that the offset is typically less than $2\arcm$ and always less than $\theta_{500}$.}
\label{fig:dist}}
\end{figure}

The X--ray position for single systems is compared to the \planck\ position in Fig.~\ref{fig:dist}.  The offset behaviour is similar to that observed for known clusters in the ESZ sample \citep[see][for discussion]{planck2011-5.1a}.  Except for the outlier PLCK~G18.7+23.6, the positional offset is less than $2\arcm$ and is clearly dominated by the \planck\ reconstruction error which peaks at that value.  A physical offset is also expected, especially for  merging clusters. Such an offset would contribute less with increasing $z$ as it would be more and more poorly resolved. The small residual systematic variation of the offset with $z$, for $z>0.2$, suggests that physical offsets may indeed slightly contribute.  This is likely to be the case for PLCK~G18.7+23.6, a highly disturbed object  at $z=0.09$, the lowest $z$ of the sample, and which has an offset of $3\arcm$ corresponding to $0.3\Rv$. In all cases, the offset remains smaller than $\theta_{500}$ or the cluster extent (see  also Fig.~\ref{fig:gal}).

\subsection{\planck\  refined $Y_{SZ}$  estimate}
\label{sec:rsz}

The SZ signal extraction procedure is described in detail in \citet{planck2011-5.1a}. It consists of applying multi-frequency matched filters to the data that maximally enhance the signal-to-noise ratio of an SZ cluster source by optimally filtering the data. As shown in \citet{planck2011-5.1a}, SZ fluxes derived using this method can be significantly overestimated due to an over-estimation of the cluster size $\theta_{500}$.

\begin{figure}[t]
\center
\includegraphics[scale=1.,angle=0,keepaspectratio,width=0.87\columnwidth]{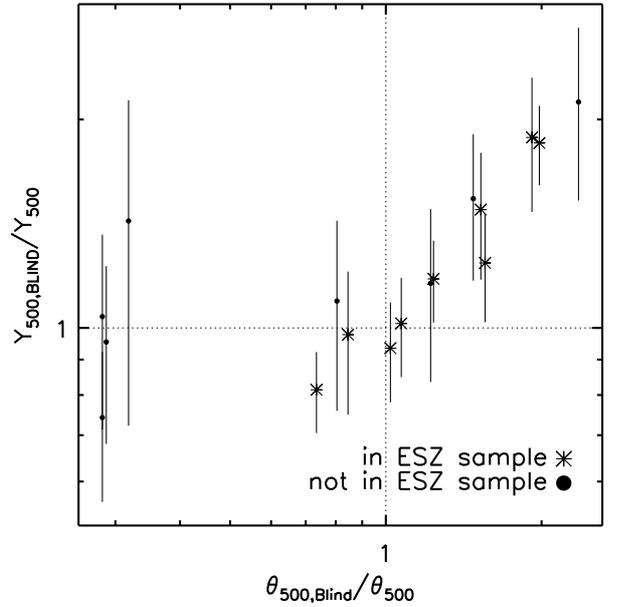}%
\caption{  {\footnotesize Comparison of the \planck\ blind and X-ray constrained $\YSZ$ measurements for single confirmed systems (see text, Sec.~\ref{sec:rsz}). The ratio is correlated with the ratio of the corresponding characteristic size, $\theta_{500}$. }\label{fig:yblindy}}
\end{figure} 

We can optimise the SZ photometry of the clusters presented here by using the X-ray estimate of the cluster position and  size  $\theta_{500}$, derived from $R_{500}$ measured using the \MYX\ relation as detailed in Sect.~\ref{sec:xquan}. For each cluster in the sample, we thus re--extract the SZ flux, calculating $\Yv$ with the X-ray position and size fixed to the refined values derived from the high-quality \xmm\ observation. The resulting $Y_{500}$ values are listed in Table~\ref{tab:xray}. In Fig.~\ref{fig:yblindy}, they are compared to the blind values as a function of the ratio of the \xmm\ and blind characteristic size $\theta_{500}$.  For most cases the values are consistent within the errors; however, there is a clear trend of SZ flux overestimation with size overestimation, which can reach as much as a factor of two (see detailed discussion of the cluster size--flux relation in \citealt{planck2011-5.1a} and \citealt{planck2011-5.2b}).

\section{X-ray and  SZ properties of newly detected clusters}

In this Section we consider the 17 systems confirmed as single-component clusters of galaxies, leaving aside the multiple systems which are discussed in the next Section.

\subsection{\rass\ properties}
\label{sec:rass}

\begin{figure}[t]
\center
\includegraphics[scale=1.,angle=0,keepaspectratio,width=0.95\columnwidth]{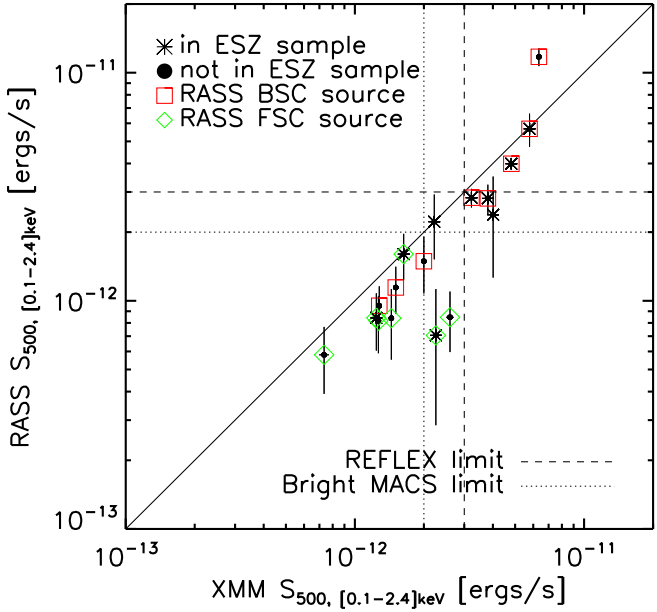}%
\caption{{\footnotesize  Unabsorbed \rass\ flux versus \xmm\  flux for  the 17 confirmed single-component clusters. The $[0.1$--$2.4]\,\keV$ flux is measured within an aperture of  $\theta_{500}$.  Clusters coincident with a \rass-BSC or a \rass-FSC source are marked with red squares and green diamonds,  respectively.  
{The most significant outlier at high flux is PLCK~G18.7+23.6 at $z=0.09$ (see Sec.~\ref{sec:rass} and Sec.~\ref{sec:lxz} for discussion)}}\label{fig:flux}}
\end{figure}

We  extracted $2\deg\times 2\deg$ count images   in the $[0.5$--$2.]\,\keV$ hard band from the \rass\  data  at the position of
  each cluster. We excised events associated with known   \rass-BSC and \rass-FSC  sources \citep{vog99,vog00}. We then carefully followed the methods   described in \citet{boe00} and \citet{rei02} to compute background corrected  growth curves and estimate an associated detection radius,   $R_{\rm d}$. The background was estimated from an outer  annulus with   $15\arcm<\theta<90\arcm$. When allowed by the quality of the growth curve,   the count rate within  the $\Rv$ aperture  was either  taken as the count rate within $R_{\rm d}$ when  $R_{\rm d}<\Rv$ or interpolated on the curve when $R_{\rm d}>\Rv$.  In the case of low quality growth curves, we computed a direct integrated  count rate  from the map within an aperture of $\Rv$.  Assuming the best fitting    \xmm\ spectral parameters for each cluster (i.e. $z$, temperature, abundance,   galactic $N_{\rm H}$) we derived the $[0.1$--$2.4]\,\keV$ band \rass\ flux.

The \rass\ values are compared to the \xmm\ values in Fig.~\ref{fig:flux}. There is a good agreement after  taking into account the \rass\ statistical errors. The slight offset ($<20\%$) is likely due to systematic errors linked to the \rass\ background estimate and/or calibration uncertainties.  The most significant outlier at high flux is PLCK~G18.7+23.6. A bright point source is present at the centre of this object (see Fig.~\ref{fig:gal}) that cannot be excised from the \rass\ data and which contaminates the signal. From the \xmm\ image (Fig.~\ref{fig:gal}), the known \rass-FSC or \rass-BSC sources within the \planck\ error box for 15 of the candidates can be clearly identified with the clusters.  Those are indicated in Fig.~\ref{fig:flux}. The two clusters with no \rass-FSC or \rass-BSC association, PLCK~G287.0+32.9 and PLCK~G292.5+22.0,  are in fact detected in \rass, but at low $\textrm{S/N}$ (2 and 3, respectively; see also Sec.~\ref{sec:scaling}). 
\begin{figure}[t]
\center
\includegraphics[scale=1.,angle=0,keepaspectratio,width=0.95\columnwidth]{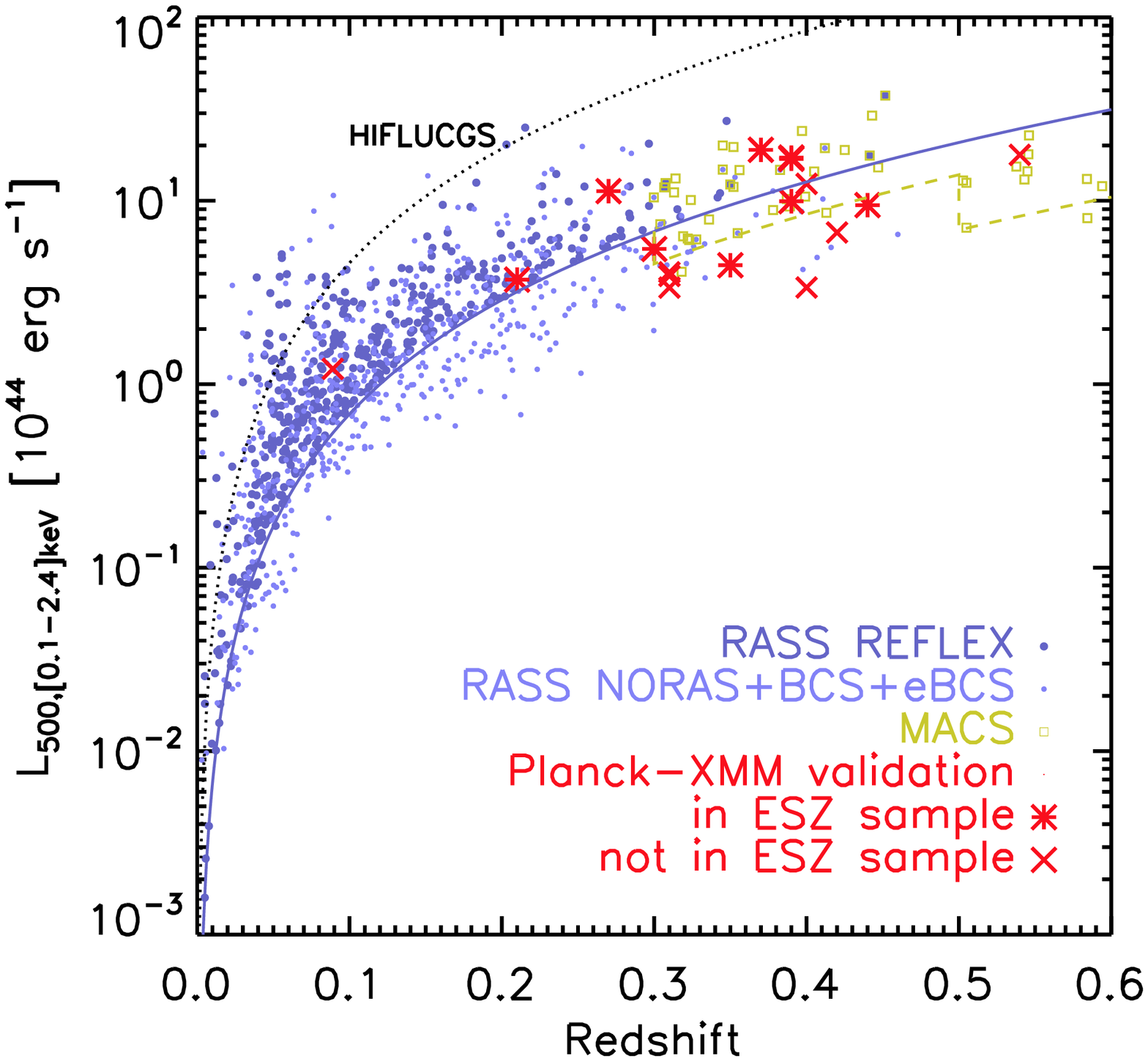}
\caption{ {\footnotesize 
The new SZ-discovered \planck\ clusters compared to clusters from the \rosat\  All-Sky Survey catalogues in the \Lxz\ plane.  The X--ray luminosity is that in the $[0.1$--$2.4]\,\keV$ band. Catalogues shown are the \reflex, \noras, \bcs, \ebcs\ and published  \macs\ catalogues. The solid line is the \reflex\ flux limit of $3 \times 10^{-12}\,\ergscm$, similar to that of the \bcs+\ebcs\ catalogues. The dotted line is the HIFLUCGS flux limit of $2 \times 10^{-11}\ergscm$ and the dashed line is from  the \macs\ flux limits.
See Sec.~\ref{sec:lxz} for references and details.} \label{fig:lxz}} 
\end{figure}
\begin{figure}[tp]
\center
\includegraphics[scale=1.,angle=0,keepaspectratio,width=0.95\columnwidth]{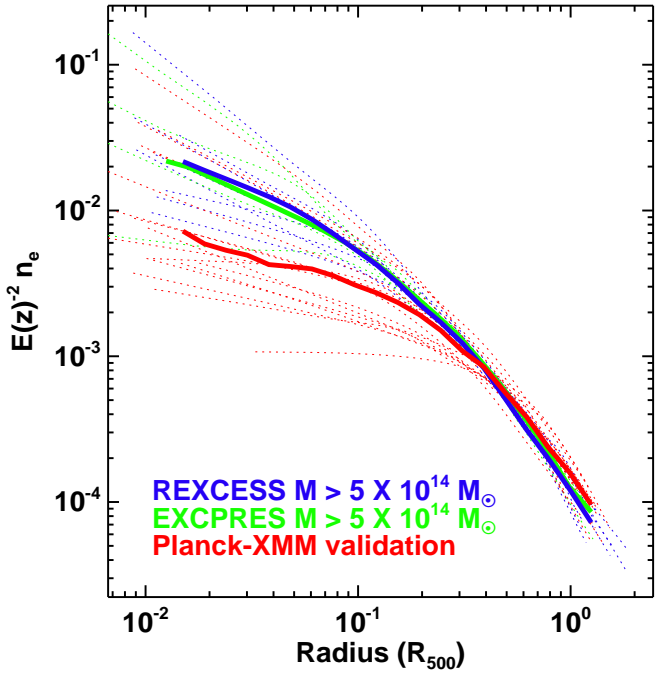}
\caption{{\footnotesize Scaled density profiles of the new \planck\ SZ clusters compared to those of similar mass systems from the representative X-ray samples \rexcess\ \citep{boe07} and \excpres\ (Arnaud et al., in prep.). $\Rv$ is estimated from the \MYX\ relation of \citet{arn10}. Thick lines show the mean profile of each sample. The density profiles of the \planck\ SZ-selected clusters are on average shallower than those of the X--ray selected clusters of the same mass.}}\label{fig:ne}
\end{figure}

\subsection{The \Lxz\  plane and comparison with \rass\ catalogues}
\label{sec:lxz}

In Fig.~\ref{fig:lxz}, the new clusters are shown in the \Lxz\  plane, plotted together with the clusters from large catalogues based on \rass\  data outside the Galactic Plane: \reflex\ \citep{boe04} in the Southern sky: \noras\ \citep{boe00}; \bcs\ \citep{ebe98}; and \ebcs\ \citep{ebe00} in the Northern Sky. The \noras\ is not flux limited. The \reflex\ flux limit of $ 3\times 10^{-12}\,\ergscm$  is shown. It is similar to that of  the \ebcs+\bcs\ limit of $2.8\times 10^{-12}\,\ergscm$.  Also shown are clusters from the published catalogues of the \macs\ survey with their corresponding flux limit. \macs\ is based on the \rass-BSC but in contrast to the above surveys, the X-ray extent of the \rass\ source is not a selection criterion, allowing more distant (but massive) clusters to be found \citep{ebe01}. Published \macs\ catalogues are the $z>0.5$ catalogue \citep[][]{ebe07} and the $0.3<z<0.5$ brightest cluster catalogue  \citep[][hereafter bright \macs]{ebe07}.   Luminosities plotted in Fig.~\ref{fig:lxz} are the homogenised values given in the MCXC \citep[Meta--Catalogue of X-ray detected Clusters of galaxies][]{pif10}.

The present sample of new \planck-detected systems spans a redshift range of $0.1\,\lesssim\,z \,\lesssim\,0.6$, with 15 out of 17 clusters above $z = 0.25$, a medium-distant redshift region of the \Lxz\  plane that is sparsely-populated by the \rass\ catalogues. As a consequence, our current sample has X--ray luminosities well below the flux limit of HIFLUCGS \citep{rei02} and REFLEX-DXL \citep{zha06}, two high-luminosity X--ray selected samples that  stand as the counterparts to our present high S/N SZ sample.  The closest   sample in X-ray luminosity and redshift to the new \planck\ clusters  are the \macs\ clusters, although the \planck\ clusters go to lower luminosity.

Most of the new \planck\ clusters naturally fall  below the \reflex\ flux limit or, equivalently, the \bcs+\ebcs\ limit in the North. However,  three clusters lie well above this limit:  PLCK~G18.7+23.6, PLCK~G171.9$-$40.7, PLCK~G271.2$-$31.0, in order of decreasing X--ray flux (Fig.~\ref{fig:flux} and Fig.~\ref{fig:lxz}).  As discussed above, PLCK~G18.7+23.6 at $z=0.09$ has a very bright central source and very diffuse ICM emission. It may have been misclassified as a point source in the \reflex\ survey.  {We also note that this cluster, although not included in the ESZ sample,  is the brightest X-ray cluster of the sample due to its low redshift $z=0.09$.} PLCK~G271.2$-$31.0 simply falls in the Large Magellanic Cloud LMC2 region, which was excluded in the \reflex\ survey \citep[see][Table 1] {boe01}.  However, PLCK~G171.9$-$40.7 at $z=0.27$ has a flux of $5.7\times 10^{-12}\ \ergscm$ (from fully consistent \rosat\ and \xmm\ measurements), and is a northern sky cluster that fulfills the \bcs\ flux and sky position criteria. Thus {\it a priori}, it should have been included in that survey.  Finally, six new clusters at $z\geq0.3$ are above the \macs\ flux limit. Of these, four are not associated with a \rass-BSC source and so could not be found in a \macs-like survey, and the other two are at lower declination than considered by \macs.

\subsection{Gas morphology and scaled density profiles}

Figure~\ref{fig:gal} shows $[0.3$--$2]\,\keV$ \xmm\  images of the newly-discovered clusters. Each image corresponds to the same physical size in units of $\Rv$ and is corrected for surface brightness dimming with redshift and divided by the emissivity in the $[0.3$--$2]\,\keV$ energy band. As detailed in \citet[][Sec.~ 3.2]{arn02}, the emissivity is computed from a redshifted thermal model convolved with the instrument response and taking into account Galactic absorption. The resulting image is proportional to the emission measure along the line of sight, which is then scaled by $E(z)^2\Rv$ according to the self-similar model. The colour table is the same for all clusters, so that the images would be identical if clusters obeyed strict self-similarity.  A first visual impression is that low surface brightness, morphologically-disturbed objects dominate the sample, which contains very few centrally-peaked, cool core-like objects.

The visual impression is confirmed and quantified when one looks  at the density profiles of the clusters shown in Fig.~\ref{fig:ne}. They are plotted together with the density profiles of similar mass clusters from the representative X-ray-selected samples \rexcess\ \citep[$z < 0.2$;][]{boe07} and \excpres\ ($0.4 < z < 0.6$; Arnaud et al, in prep.). For all three samples, the radii are scaled by $R_{500}$,  estimated from the \MYX\ relation (Eq.~\ref{eqn:Yx}). The thick lines show the mean profile. While the two X-ray-selected samples agree to a remarkable degree, the \planck-selected sample clearly consists of systems with much flatter density profiles, and the corresponding mean profile is significantly flatter than that of the X-ray selected samples.  This  shape is due to a number of very disturbed clusters with very flat profiles in the new \planck-discovered cluster sample. Let us consider the ten clusters with the flattest density profiles, flatter than the mean profile \planck\ cluster profile and flatter than all the \rexcess\ profiles. These ten objects include PLCK~G18.7+23.6 at $z=0.09$ discussed above,  PLCK~G286.6$-$31.3  at $z=0.21$ that is just at the \reflex\ flux limit and PLCK~G292.5+22.0 at $z=0.30$ that is just at the  \macs\ flux limit  (Fig.~\ref{fig:lxz}). The other seven clusters lie at medium redshift, $0.3 < z < 0.45$, and are all hot ($T_{\rm X}  \gtrsim 5\,\keV$)  and massive  ($\Mv \gtrsim 5\times\,10^{14}\,\msol$)  systems. They lie below the flux limit of both \reflex\ and bright \macs\ for a similar mass range \citep{ebe10}.  Thus \planck\ appears to have uncovered a population of massive, disturbed, low-luminosity systems.

\begin{figure*}[]
\begin{centering}
\begin{minipage}[t]{0.95\textwidth}
\resizebox{\hsize}{!} {
\includegraphics{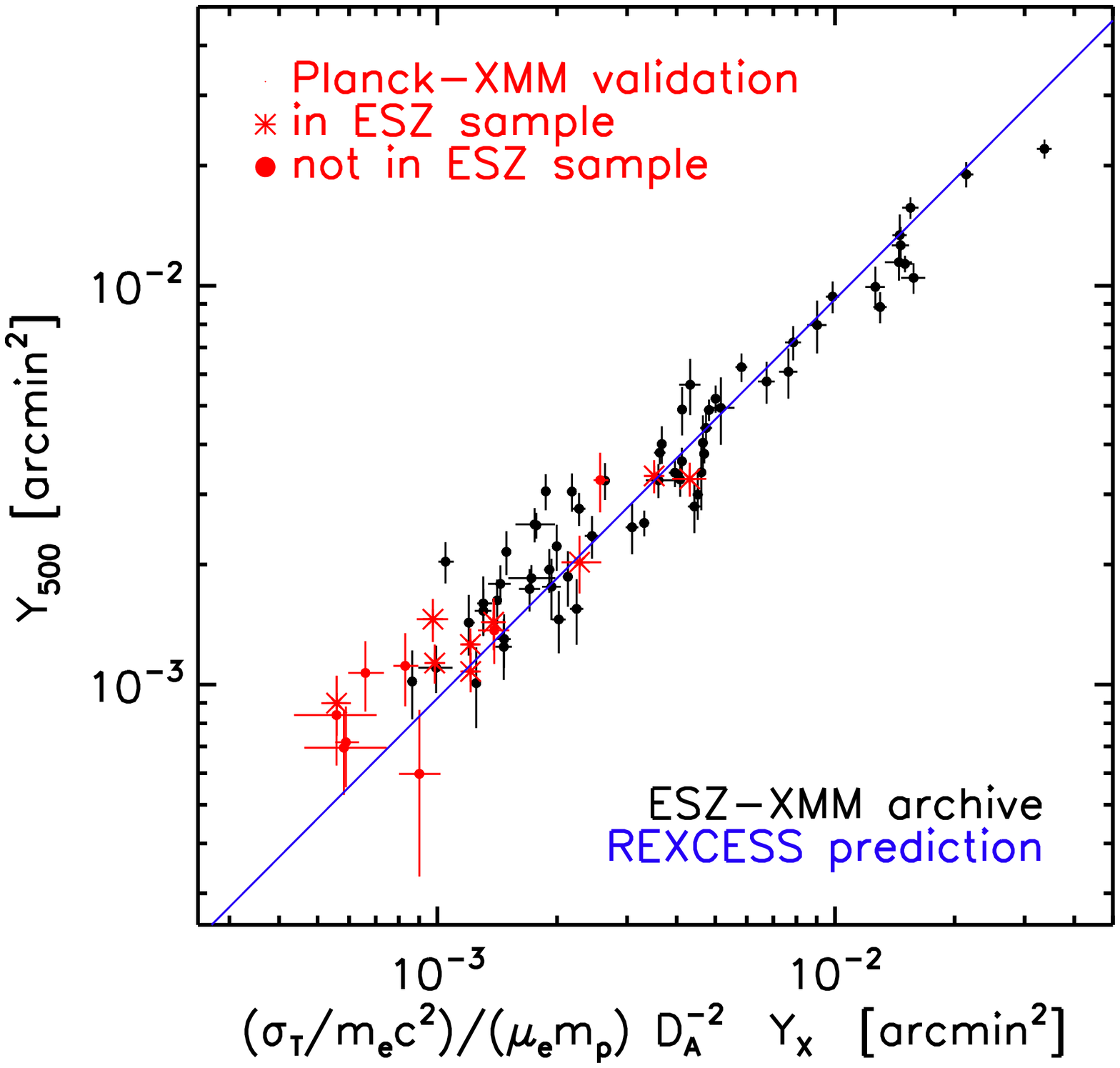}
\hspace{8mm}
\includegraphics{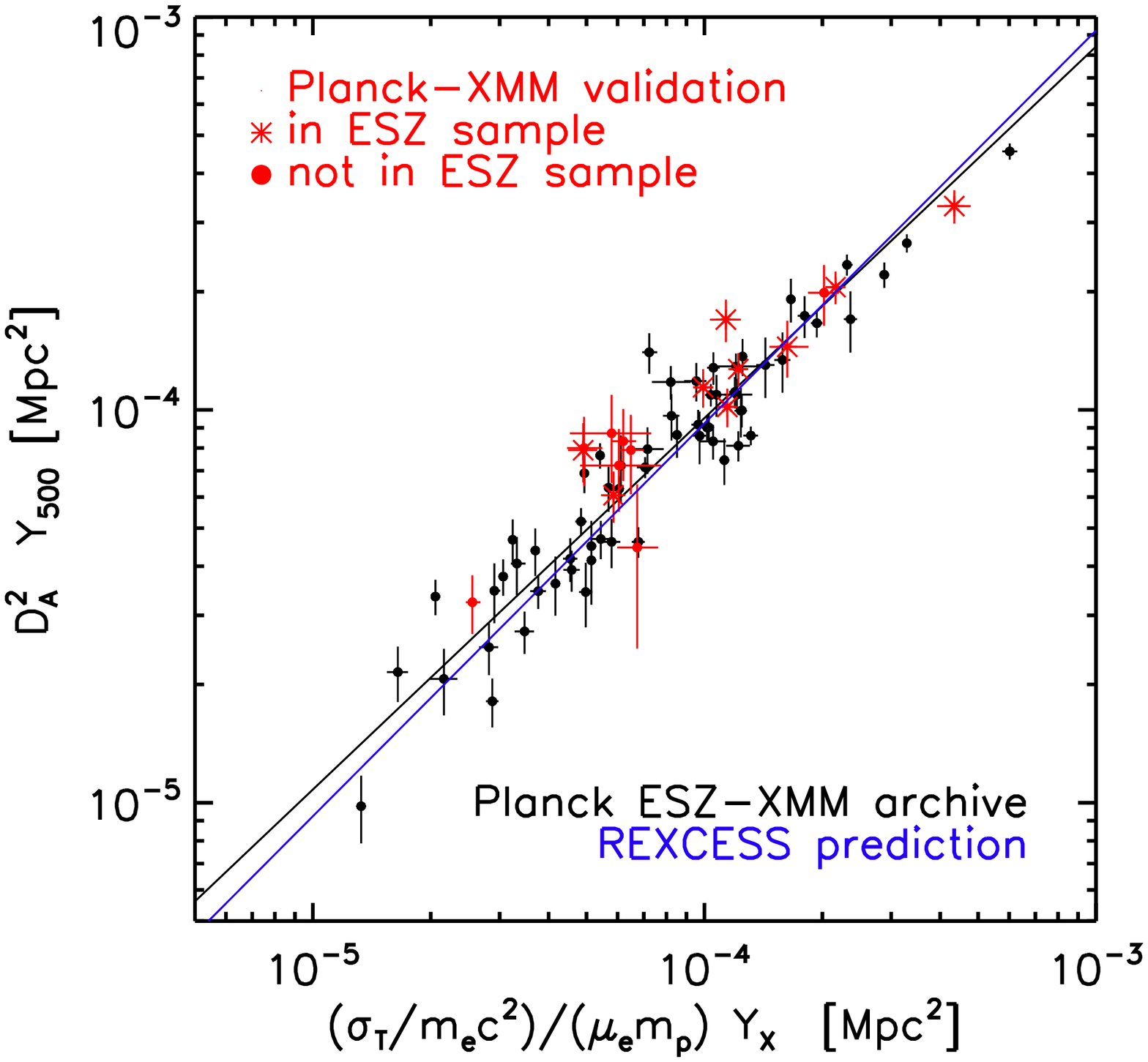}
} 
\vspace{8mm}
\end{minipage}
\begin{minipage}[t]{0.95\textwidth}
\resizebox{\hsize}{!} {
\includegraphics{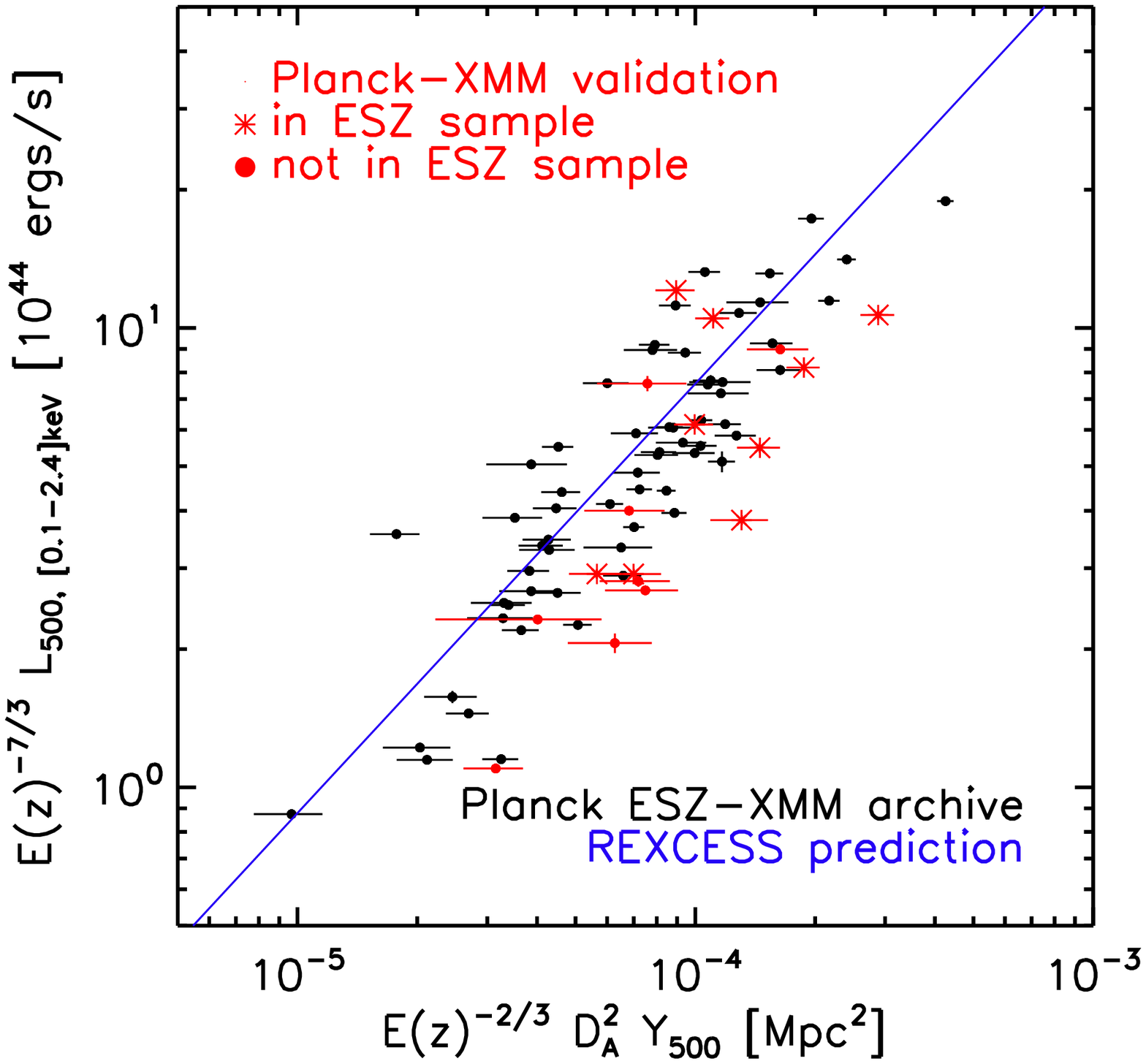}
\hspace{8mm}
\includegraphics{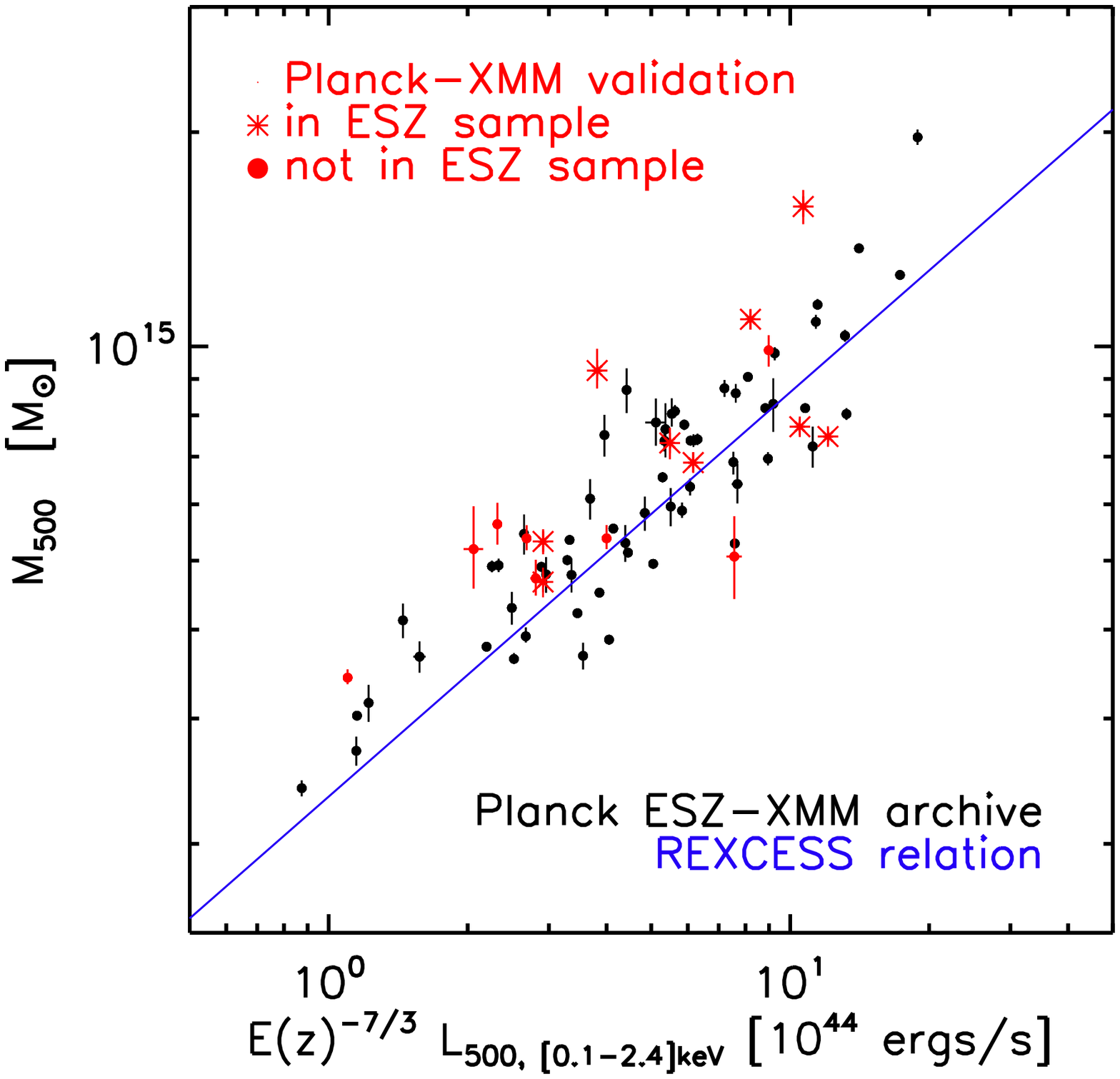}
} 
\end{minipage}
\end{centering}
\caption{{\footnotesize 
Scaling relations for the 17 new confirmed single-component clusters (red symbols). Black points show clusters in the  \planck\--ESZ sample with \xmm\ archival data as presented in \citet{planck2011-5.2b}. The solid black line denotes the corresponding scaling relation fits in each panel.  
The blue lines in the top and bottom right panels denote the predicted $\YSZ$ scaling relations  from the \rexcess\ X-ray observations \citep{arn10}. The blue line in the bottom left panel is the Malmquist bias corrected $M$--$L$ relation from the \rexcess\ sample \citep{pra09,arn10}. In all figures, $\Rv$ and $\Mv$ are estimated from the \MYX\ relation of \citet{arn10}.
Top row:  Relation between apparent SZ signal ($Y_{500}$, left) or intrinsic Compton parameter ($D_{\rm A}^2 Y_{500}$, right) and the corresponding normalised $\YX$ parameter. Bottom row:  Relation between X--ray luminosity and  $Y_{500}$  (left) and between mass and luminosity (right panel). The new clusters are on average less luminous at a given $Y_{500}$, or more massive at a given luminosity, than X-ray selected clusters. }
\label{fig:xsz}}
\end{figure*}

\subsection{SZ flux versus X-ray prediction and mass-proxy - mass relations}
\label{sec:scaling}

\citet{planck2011-5.2b} uses \xmm\ archival data to study the relations between the SZ signal and X-ray properties such as $\YX$ or the soft band luminosity $\LX$ on a  sub-sample of  clusters from the high signal-to-noise ratio ESZ sample.   This sample (hereafter the ESZ--{\it XMM}--archive sample) is SZ selected but by nature only comprises clusters from X--ray surveys.  As discussed extensively in \citet{planck2011-5.2b}, the analysis has demonstrated the excellent agreement between the observed scaling relations and the predictions based on \rexcess\ pressure profiles and numerical simulations  \citep{arn10}. In Fig.~\ref{fig:xsz}, we have placed  the new \planck\ \xmm\ confirmed clusters on the \YSZYX\ relations (top panels) and the  $\LX$--$\YSZ$ and $\Mv$--$\LX$ relations (bottom panels). 

\subsubsection{The \YSZYX\ relation}

The SZ signal, $Y_{500}$, is plotted as a function of the normalized $D_{\rm A}^{-2} \YX$ parameter in the left panel of Fig.~\ref{fig:xsz}. The new clusters follow the trend observed for the ESZ--{\it XMM}--archive and are consistent with the \rexcess\ prediction (blue line). However, a slight turnover is observed at low flux, with observed points systematically above the predicted relation. This excess is likely due to the Malmquist bias. Such a trend is also slightly apparent for the ESZ--{\it XMM}--archive sample but is less important \citep[see][for discussion of this effect]{planck2011-5.2b}. The low flux clusters span various $z$ values and are redistributed over the range of intrinsic Compton parameter. As a result there is slight  positive offset apparent in the $D_{\rm A}^{2}\YSZ$--$\YX$  relation for new clusters as compared to the ESZ--{\it XMM}--archive sample. 

This suggests that as far as the relation between $\YSZ$ and its X--ray equivalent $\YX$  is concerned, the new clusters are similar to X--ray selected clusters, although they are more dynamically disturbed.  This is expected if indeed the pressure is the quantity less affected by dynamical state and both $\YX$ and $\YSZ$ are low scatter mass proxies. However, independent mass estimates are required to check this point; they cannot be provided by X--ray measurements in view of the highly unrelaxed nature of the new clusters.

\subsubsection{The $\LX$--$\YSZ$ and $\Mv$--$\LX$ relations.}

As compared to X--ray selected clusters, the new
clusters fall on the low luminosity side of the $\LX$--$\YSZ$ relation (bottom left panel of fig.~\ref{fig:xsz}). In other words, they are under-luminous at given $\YSZ$. If the mass is indeed tightly related to  $\YSZ$ (or $\YX$) we then expect them to be underluminous at a given mass. This trend is consistently observed in the bottom--right panel, where $\Mv$ is estimated from $\YX$: the new clusters fall towards the high-mass, low-luminosity side of the $\Mv$--$\LX$ relation. However, confirmation requires independent mass estimates, e.g., from lensing data.

  As shown by \citet{pra09}, the underluminous region of the $L$--$M$ plane is populated by morphologically disturbed systems. This once again suggests that the majority of the new \planck-detected systems are disturbed, in agreement with the above discussion on the morphology and the scaled density profiles.

 The dispersion of the new clusters about the $\Mv$--$\LX$ relation also seems higher than that for X--ray selected objects. This suggests the existence of new extreme low-luminosity, high-mass objects that are being revealed by \planck. The two prominent outliers are PLCK~G287.0+32.9 ($z=0.39$) and PLCK~G292.5+22.0 ($z=0.3$), detected by \planck\ at high $\textrm{S/N}$ values of 10.6 and 6.9, respectively. They belong to the very hot ($T\gtrsim10\,\keV$) and very massive ($\Mv\gtrsim\,10^{15}\,\msol$) cluster category (Table~2) and are the only two clusters associated with neither a BSC nor an FSC source (Sec.~\ref{sec:rass} and Fig.~{\ref{fig:flux}). The flux of PLCK~G292.5+22.0 barely reaches the \macs\ limit for a mass of $\Mv\sim9.2\times10^{14}\,\msol$. It has a very disturbed morphology (Fig.~\ref{fig:gal}) and a flat density profile with a scaled central density of $4\times10^{-3}\,{\rm cm^{-3}}$ (Fig.~\ref{fig:ne}).

\begin{figure*}[t]
\begin{centering}
\begin{minipage}[t]{0.90\textwidth}
\resizebox{\hsize}{!} {
\includegraphics[scale=1.,angle=0,keepaspectratio]{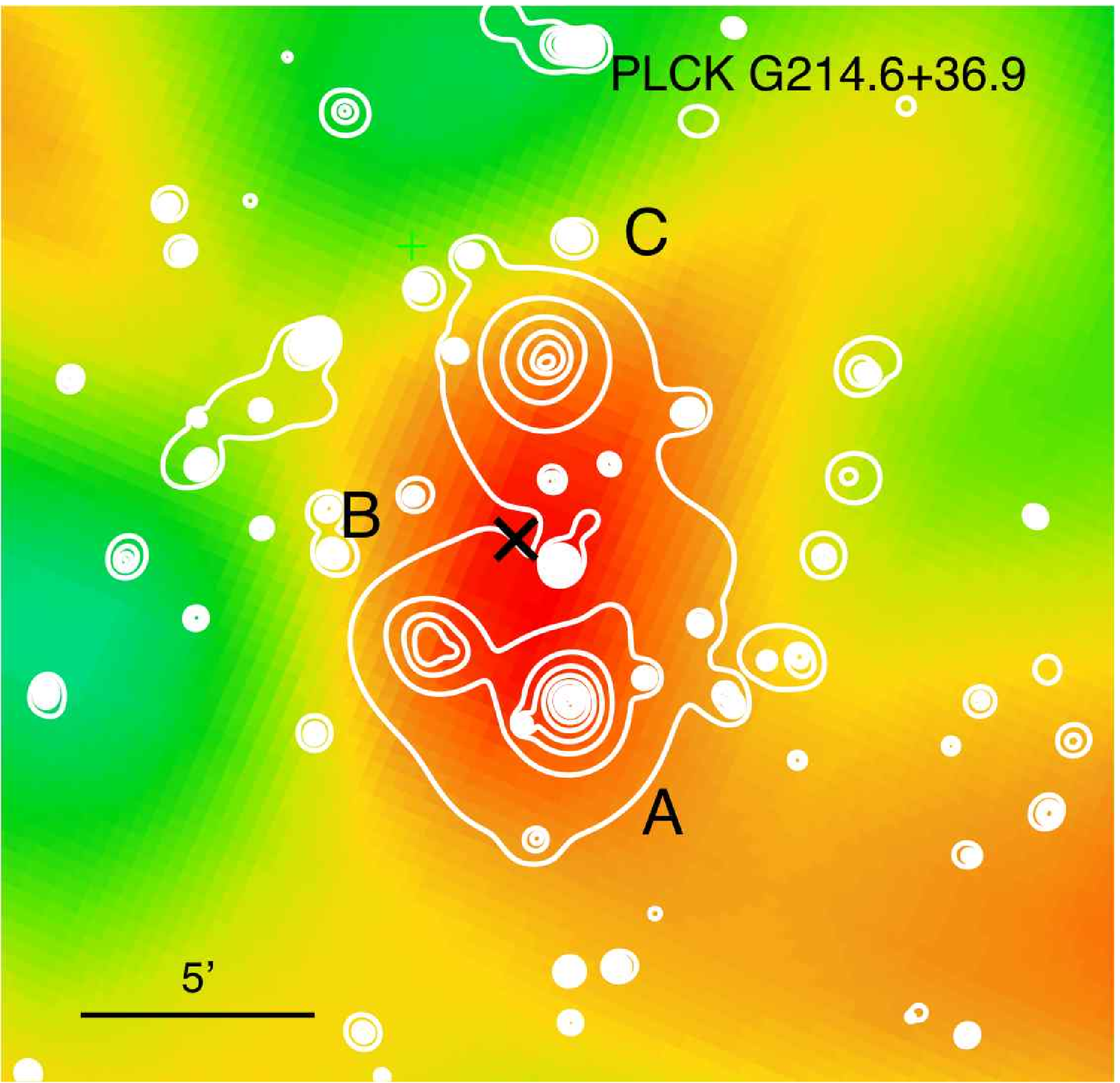}%
\hspace{2mm}
\includegraphics[scale=1.,angle=0,keepaspectratio]{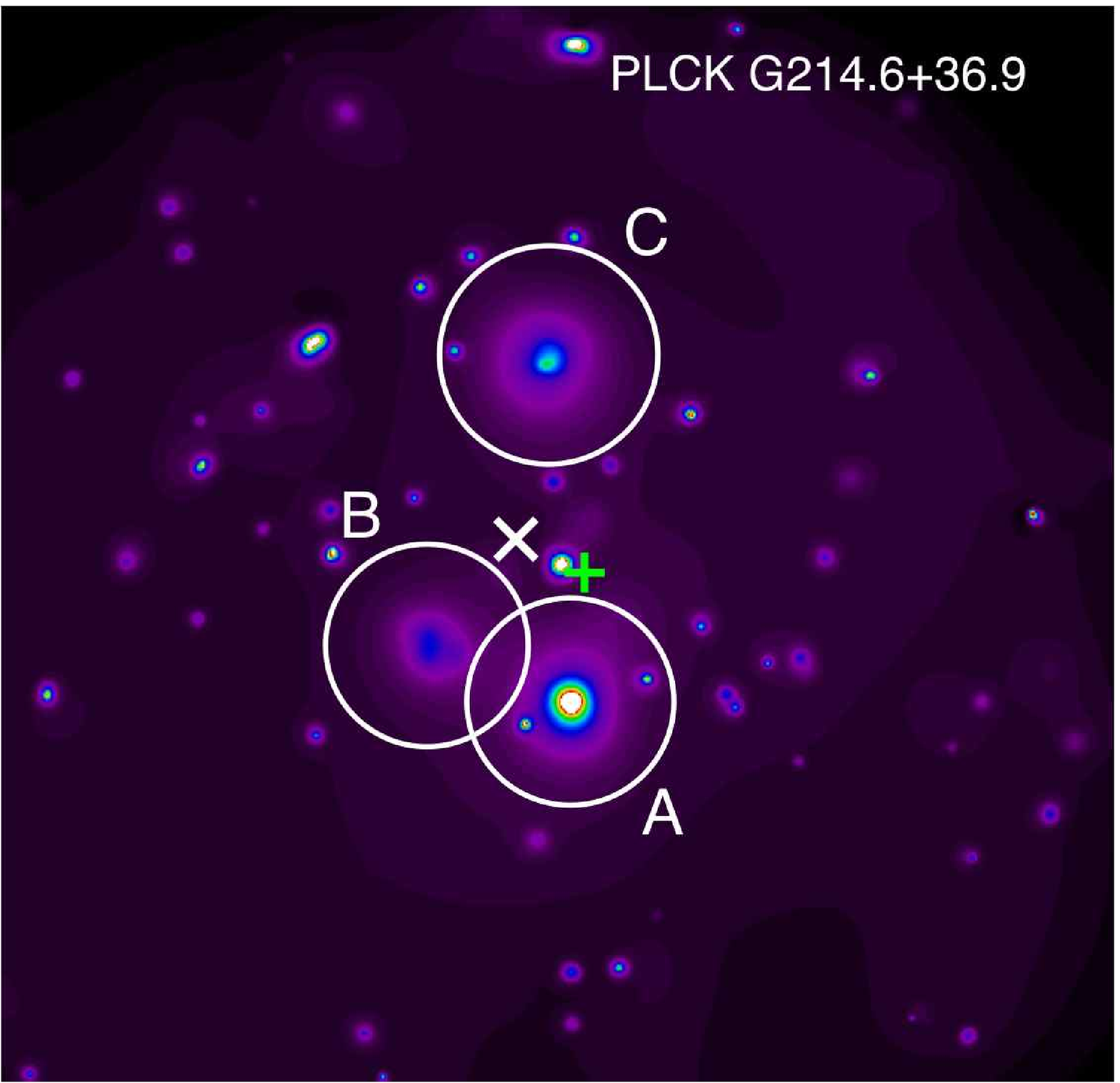}
\hspace{2mm}
\includegraphics[scale=1.,angle=0,keepaspectratio]{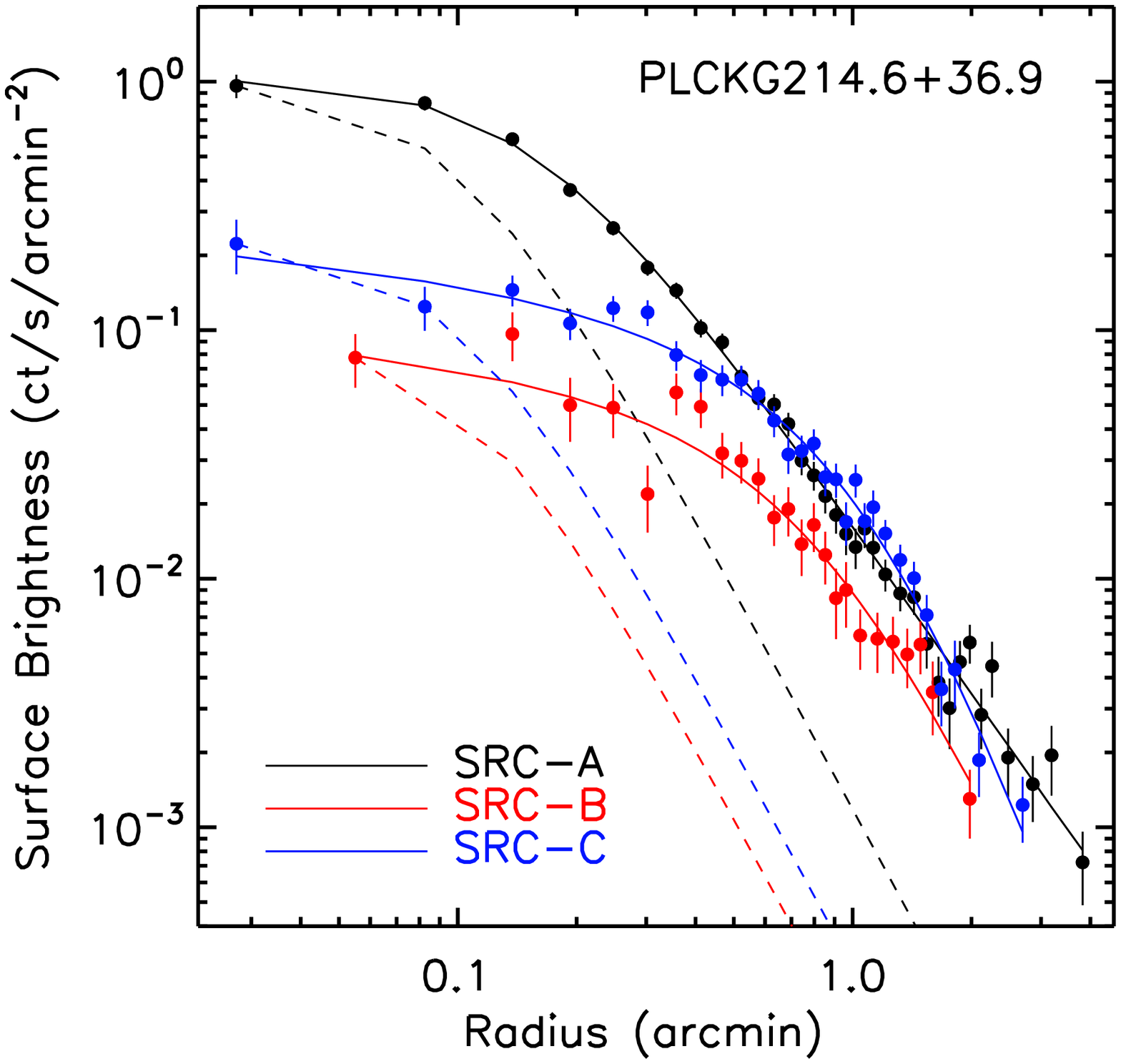}
}
\end{minipage}\\[2mm]
\begin{minipage}[t]{0.90\textwidth}
\resizebox{\hsize}{!} {
\includegraphics[scale=1.,angle=0,keepaspectratio]{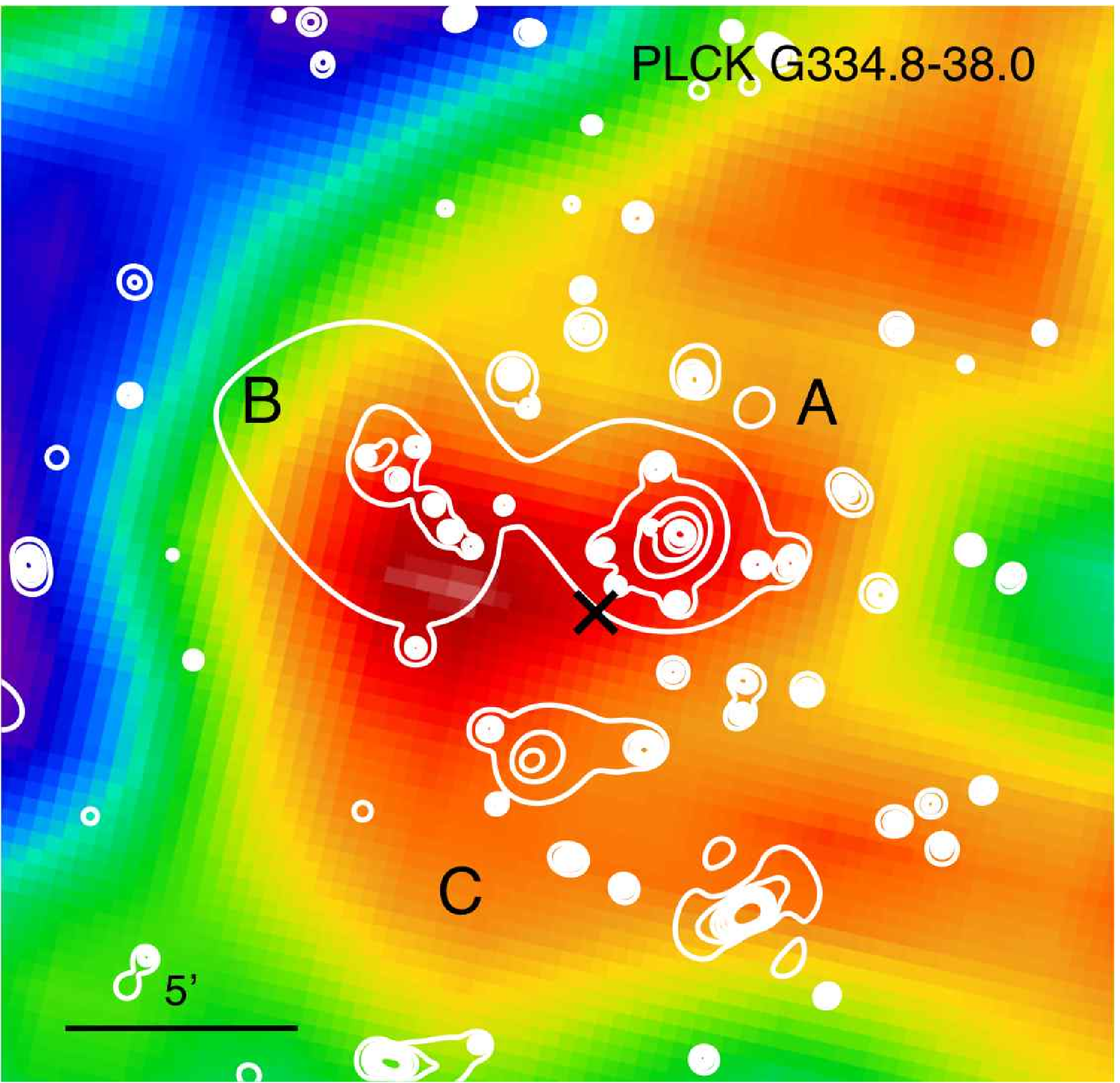}%
\hspace{2mm}
\includegraphics[scale=1.,angle=0,keepaspectratio]{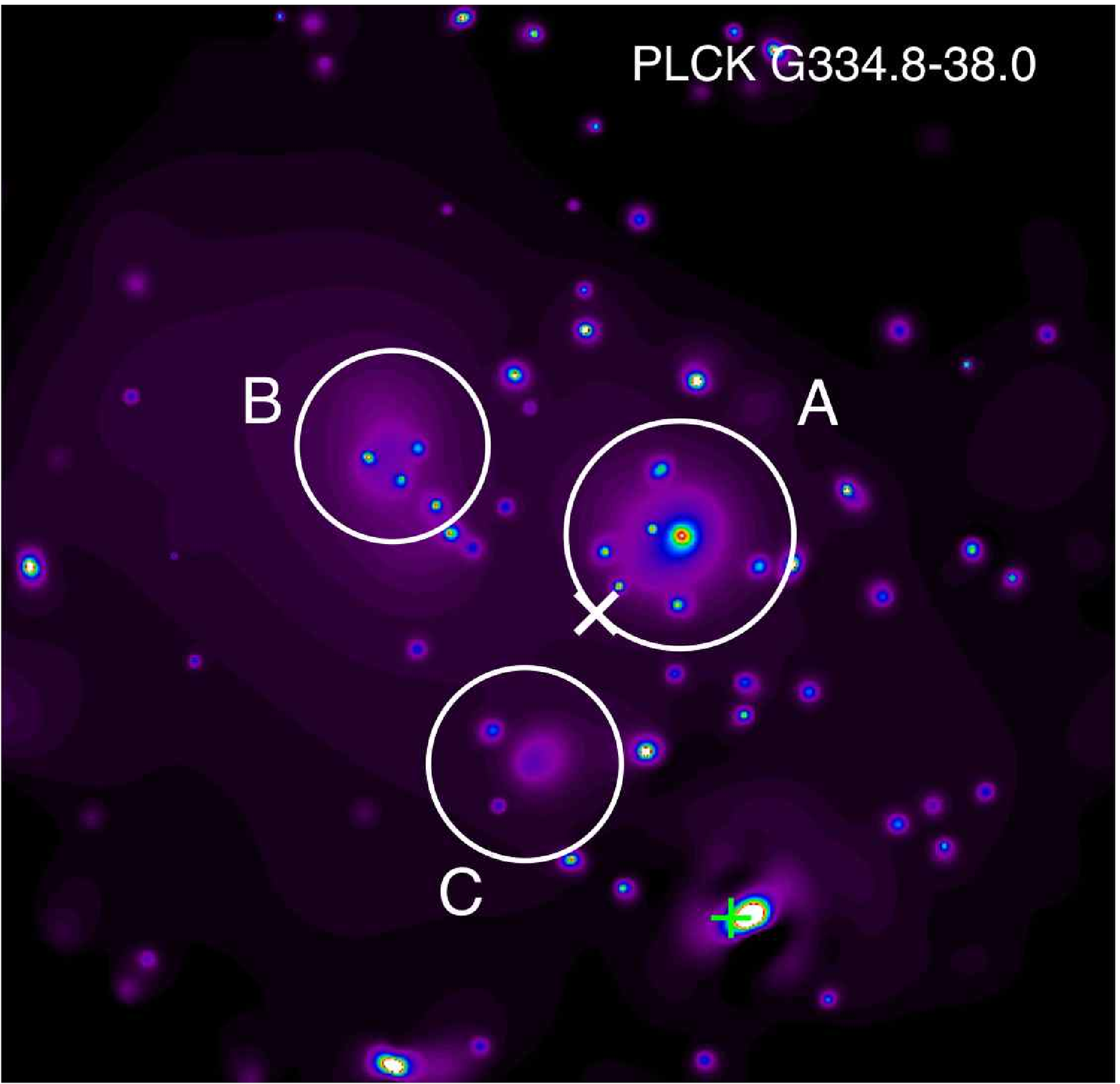}
\hspace{2mm}
\includegraphics[scale=1.,angle=0,keepaspectratio]{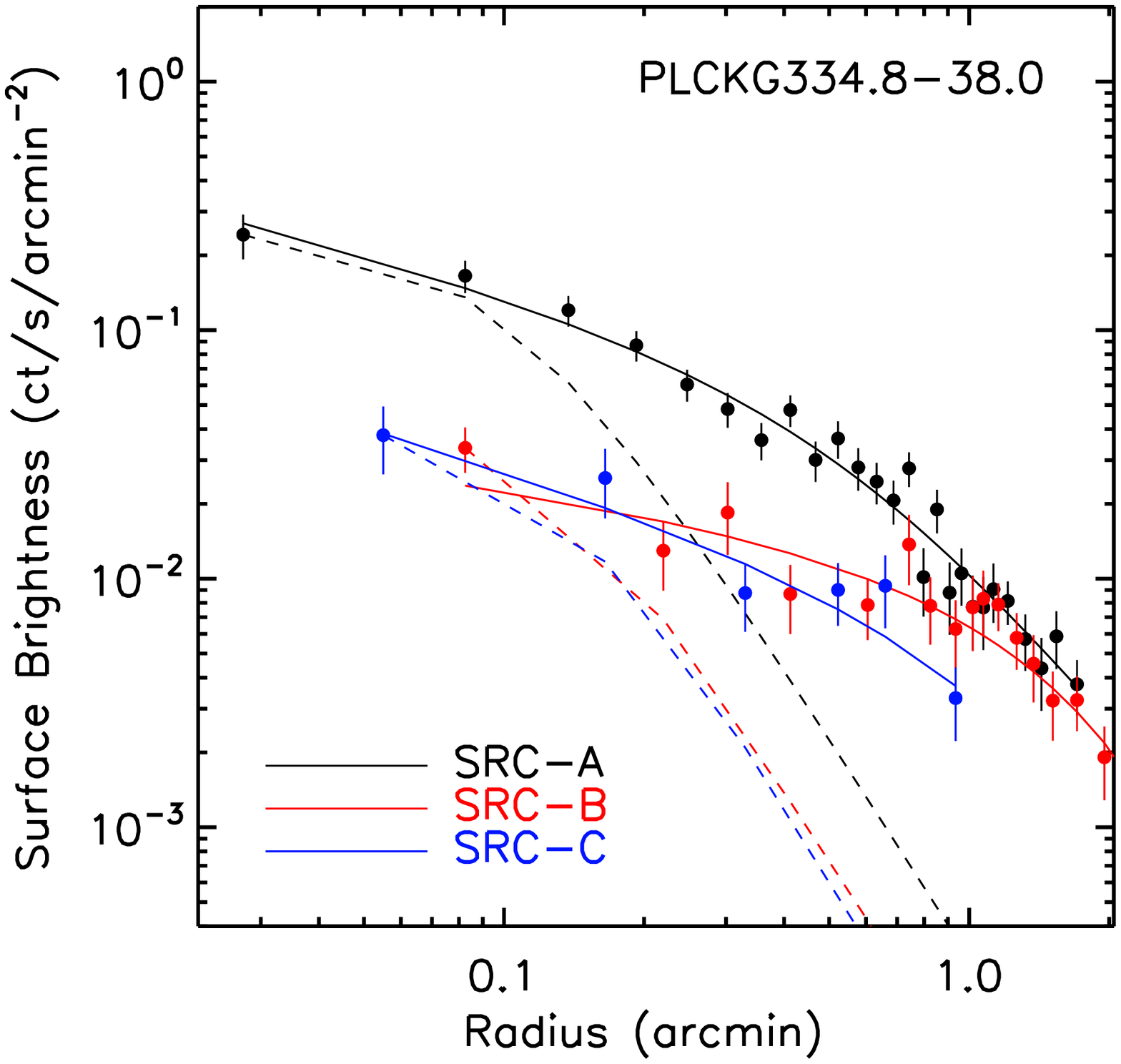}%
}
\end{minipage}\\
\end{centering}
\caption{{\footnotesize The triple systems PLCK~G214.6+37.0 (top) and
  PLCK~G334.8$-$38.0 (bottom). The left panels show the  \planck\ $Y_{SZ}$ map
  (derived from an Internal Linear Combination method) with contours from the \xmm\ wavelet filtered $[0.3$--$2]\,\keV$ image (middle panels) overlaid in white.  The cross
  marks the position of the re-extraction centre for flux re-analysis. Extended
  components found in the \xmm\ image are marked with letters (see
  text and Table~\ref{tab:xray}). The circles in each \xmm\ image denote the estimated $\Rv$ radius for each component. The right panels show the X--ray surface brightness profiles of the
  three components for each super cluster (points with uncertainties), and the best-fitting $\beta$-model (solid lines) compared to the profile of the Point Spread Function normalised at the central level (dashed lines). }}\label{fig:sc}
\end{figure*}

\section{Further analysis of multiple systems}
\label{sec:multi}

\subsection{Double systems}

Two of the new \planck\ sources (PLCK~G308.3$-$20.2 and PLCK~G337.1$-$26.0) were revealed by the \xmm\ validation observations to be double systems. X-ray images of these systems are included in the gallery in Fig.~\ref{fig:gal}.

\subsubsection{PLCK~G308.3$-$20.2}

Two clusters with quite regular morphology are clearly detected in the \xmm\ snapshot observation for this candidate (denoted A and B in Fig.~\ref{fig:gal}). The \planck\ position is very close ($1.5\arcm$) to that of the northern cluster A. This cluster is very hot ($\TX\sim10\,\keV$) and massive (Table~\ref{tab:xray}). From the X-ray spectroscopy, we estimated its redshift to be $z=0.48$. This estimate is robust, with a quality flag of 2 as reported in Table~\ref{tab:xray}. The second component, $B$, lies $7\arcm$ to the South-East of A. The lack of statistics prevents us from deriving a sufficiently reliable redshift estimate. Assuming it lies at the same redshift as A, its $\YX$ parameter is $6.0$ times less than that of A, and its derived mass is $2.7$ times less. Both clusters are seen as well-separated sources in \rass: A is associated with a \rass-BSC source; whereas B coincides with a \rass-FSC source.

\subsubsection{PLCK~G337.1$-$26.0}

The distance between components A and B (Fig.~\ref{fig:gal}) is $8.1\arcm$. Both have regular morphologies, and exhibit strong Fe~K lines,  allowing individual redshift estimation. They are found to lie at two clearly different redshifts: $z_{\rm Fe}=0.26$ for A; and $z_{\rm Fe}=0.12$ for B. A is the hotter of the two with $\TX=(6.2\pm 0.2)\,\keV$ and thus the more massive. The $\YX$ of cluster A is 15 times larger than that of $B$, making it the main contributor to the \planck\ SZ signal. 

The two clusters are seen as separate sources in \rass: cluster A as a \rass-BCS source; and B as a \rass-FCS source. The \xmm\ emission coincides perfectly with the \rass\  emission in each case. Additionally, the two clusters are also found $40\arcm$ off-axis in a PSPC pointed observation of a globular cluster, NGC\,6752 \citep{joh94},  where they are listed as sources within the globular cluster (sources 1 and 2 in \citeauthor{joh94}'s Table~8). Lacking spectroscopic information, \citet{joh94} could not specify the exact nature of the sources, which they assumed to be of Galactic origin. Note that it is not surprising that the sources were not  identified as extended sources, in view of the large PSPC PSF ($90\%$  encircled energy diameter of $\sim 6\arcm$)  at such off-axis angle. 

\subsection{Triple systems}

PLCK~G214.6+37.0 and PLCK~G334.8-38.0 were included in the \xmm\  pilot programme and are detected in the \planck\ survey with $\textrm{S/N}$ of 5.0 and 4.1, respectively. The wavelet-filtered X-ray surface brightness contours are
overlaid on the \planck\ maps in the left-hand panels of Fig.~\ref{fig:sc}.  For both sources, the \xmm\ observation revealed three extended X-ray components; their extended nature is evident in the surface brightness profiles shown in the right-hand panels.  

\subsubsection{PLCK~G214.6+37.0}\label{sec:scn} 

The Planck SZ source candidate position is located $\sim 5\arcm$ from the two southern components ($A$ and $B$).  A third component, $C$, lies approximately $7\arcm$ to the North (Fig.~\ref{fig:sc}, top panels).  X-ray spectral analysis of the Fe~K line indicates a redshift of $z_{\rm Fe}\sim~0.45$ for the brightest component. None of the sources is particularly hot, luminous, or massive (i.e. $M_{500}<2.5\times 10^{14}\,$M$_\odot$).  A  \rass-FSC source lies in the South-East and its counterpart is easily seen with \xmm. It is associated with a point source and is unassociated with the SZ emission. 

PLCKG214.6+37.0 falls in the SDSS area. We investigated the SDSS-DR7 database using refined positional information from the \xmm\ observation. We identified two bright galaxies with spectroscopic redshifts of $z=0.45$ whose  positions coincide with the peak of components $A$ and $C$, respectively. Furthermore, a bright galaxy with a photometric redshift of $0.46$ lies very close to the $B$ X-ray peak.  We also ran a dedicated algorithm \citep{fro11} to search  for an overdensity of SDSS galaxies at the location of the \planck\ SZ source. While we were unable to differentiate the three sub-structures, the analysis suggests the presence of a massive structure ($\sim10^{15}$~M$_\odot$) at $z\sim0.45$  around $A$ and $B$.   A further cross-correlation with SDSS-DR7 LRGs and the SCs catalogue from the SDSS-DR7 \citep{lii10} hints that this triple system is encompassed within a very large--scale structure located at $z\sim 0.45$, and whose centroid lies about $2\deg$ to the South (see Appendix~\ref{sec:sc_ap} for further details).

Thus there is good agreement between all redshift estimates, including the redshift of component $A$ estimated from the \xmm\ observation, the optical SDSS redshifts of the three components, and that of the larger-scale environment. This agreement strongly argues in favour of a real structure of (at least) three clusters, likely forming  the core of a larger-scale super cluster.

\subsubsection{PLCK~G334.8$-$38.0} 

Two extended X-ray components separated by $7\arcm$ are clearly visible in the \xmm\ image (denoted A and B in Fig.~\ref{fig:sc}, bottom panels). The  \planck\ SZ source candidate position lies between and slightly to the south of the components. A third fainter component, $C$, is seen $5\,\arcm$ to the South. The spectral analysis of component $A$ suggests a redshift of $z_{\rm Fe}\sim0.35$. Although, this estimate, based  on the Fe~L complex detection, has to be taken with caution, we adopted it as the redshift for all three X-ray components. Despite limited statistics, we derived temperatures of {$(2$--$3)\,\keV$  (with large uncertainties for cluster C)}, suggesting masses of $(0.5$--$1)\times 10^{14}$M$_\odot$. The only \rass\ source  found in the vicinity of the SZ source is clearly not associated with the three \xmm\ components, and coincides with an off--axis point source seen in the \xmm\ image.

\subsection{Comparison of X-ray and SZ properties}

As a first comparison of the X-ray and SZ properties, we simply  compared the $\YSZ$ \planck\ measurement with the predicted value from the summed contribution of the various components, derived from their estimated $\YX$ values.

For PLCKG308.3$-$20.2, the predicted summed contribution from A and B represents {$46\%$} of the total measured SZ signal (with  {$40\%$}   from A alone).
In the case of PLCK~G337.1$-$26.0, this amounts to  {$76\%$}  of the measured $\YSZ$ (with  {$62\%$}  coming from component A). The presence of component B marginally enhances the expected SZ signal. As the two clusters are not physically connected, no enhanced SZ emission is expected from their surrounding (i.e., due to mergers, shocks, etc.).  We recall that the reconstruction error in the SZ position for \planck\ blind SZ detections is $2\arcm$ on average \citep{mel11}}. The fact that the \planck\ SZ position lies almost in the middle of the two components (i.e. $3.3\arcm$ and $4.7\arcm$ from A and B, respectively) is probably coincidental.

The \planck\ $\YSZ$  values of PLCK~G214.6+37.0 and PLCK~G334.8$-$38.0 were recomputed at a fixed `barycentric' position of the three components (black cross in  Fig.~\ref{fig:sc}).
The sum over the three components of PLCK~G214.6+37.0 yields $Y_{500,{\rm   pred}}=3.2\times 10^{-4}$~arcmin$^2$, i.e. {$25\%$} of the measured value. It is  {$35\%$} of the $1\sigma$ lower limit of $\YSZ$, and  consistent within its $3\sigma$ error range.  

In the case of PLCK~G334.8$-$38.0, $Y_{500,{\rm pred}}=1.4\times 10^{-4}$~arcmin$^2$ accounts for only {$\sim21\%$}  of the measured SZ signal and {$29\%$}  of its $1\sigma$ lower limit.  However, the predicted value is consistent within the $3\sigma$ error range of the \planck\  value, that includes uncertainties on the structure size. 
We also note that  a fortuitous association between a spurious detection by \planck\ and such an association of extended X-ray sources is quite unlikely. Indeed, such a  configuration of multiple massive halos either physically connected or associated by projection effect is not usual, making this source even more puzzling.  The formal discrepancy between the SZ and X-ray signal  is likely partly due to lack of constraints on the structure size in the SZ measurement, even when the position is fixed to the X-ray position. It could also be the result of an under-estimate of the structure's redshift. Redshift measurements of various components are definitively required to assess the nature of this association and the \planck\ source.

For all systems, the cumulative contribution predicted by the $\YX$ measurements does not match the measured SZ signal, although it is compatible in all cases within the 3$\sigma$ uncertainty on $\YSZ$. However,  the SZ flux is estimated assuming a single component that follows the universal pressure profile, an inadequate approach for these systems.   Due to its moderate spatial resolution at SZ frequencies (i.e., $5$--$10\arcm$),  \planck\ cannot separate the emission of the two or three components contributing to the overall signal.  Nevertheless, a proper multi-component analysis can be carried out in the future. From the X--ray constraints on the system geometry, a spatial template can be built to improve extraction of the $\YSZ$ signal for each component. Indeed, such a detailed study might allow us to ascertain whether SZ or X--ray emission emanates from the  regions between the main system components. The current \xmm\ snapshot observations are not deep enough to build such an accurate template (i.e., measurement of the pressure profiles of the individual components). Together with accurate redshift measurements, deeper X-ray observations are needed to derive the pressure profile of individual clusters.

\section{Conclusion and perspectives}

In the framework of an \xmm\  DDT 
validation programme, the first 21 new SZ-detected clusters in the \planck\ survey have been confirmed. Six of these were confirmed in an initial Pilot programme, the results of which were used to improve the quality assessment and selection processes of cluster candidates.  The Pilot programme  also clearly demonstrated  the efficiency of \xmm\ for \planck\ candidate confirmation. Based on the detection of extended emission, snapshot exposures have been shown to be sufficient for unambiguous discrimination between clusters and false candidates. Importantly, for redshifts at least up to $z=1.5$, the spurious association of \planck\ candidates with faint extended sources in the position error box can be distinguished via a consistency check between the X--ray and SZ flux. A further 15 candidates were confirmed in a second programme focussed on high $\textrm{S/N}$ detections. The $100\%$  success rate above $\textrm{S/N}=5$ is the first illustration of the capability of the \planck\ survey to detect new clusters via their SZ signature. 

Except for two clusters, all confirmed single or double clusters are associated with \rass-BSC or FSC sources. The two non-associations are in fact detected in \rass, but at a low $\textrm{S/N}$ of $2$--$3$. The presence of significant \rass\ emission is thus a positive indicator of the validity of a \planck\ cluster candidate in the presently-covered $z$ range. However, association with a \rass\ source   within the position error box is not, by itself, sufficient for cluster candidate confirmation. Two of the false candidates in the Pilot programme, as well as one of the confirmed triple systems,  were each associated with a single \rass-FSC source that \xmm\ subsequently revealed to be a point source.  Furthermore such spurious association, and also the number of real candidates not detected in \rass, is expected to increase when probing higher $z$, i.e.,  at lower \planck\ $\textrm{S/N}$, or later in the mission.

The \xmm\ validation programme brings clear added value to simple candidate confirmation. The X-ray flux measurement and refined position is essential information for optimisation of deeper follow-up observations for detailed X-ray studies.  The refined position is also useful for  optical follow-up, such as for redshift measurements. Importantly, the determination of the exact cluster centre and extent allows a refined estimate of the SZ flux from \planck\ data. For the X-ray brightest objects, \xmm\ can directly provide the source redshift from the Fe~K line in the spectrum.  For the present sample of confirmed candidates,  17 of 27 individual clusters  (including those in multiple systems) have high quality redshift measurements. The new clusters span the redshift range $0.09<z<0.54$, with a median redshift of $z=0.37$.

In addition, the \xmm\ validation programme has provided a preview of the properties of the new clusters that \planck\ is discovering. Of the 21 confirmed candidates, 17 are single clusters, most of which are found to have highly irregular and/or disturbed morphologies (i.e. {$\sim 70\%$ from visual check}).  Two more confirmed candidates were revealed to be double systems, one of which is a projection of two physically independent clusters at different redshifts. More unexpected are two further newly-discovered triple systems that were not resolved by \planck. One of these is a true cluster association at $z\sim0.45$, as confirmed both from the \xmm\ data and in our subsequent analysis of SDSS data. It likely forms the core of a larger-scale supercluster, and is the first supercluster to be discovered via the SZ effect. Theoretically, the SZ signal from such a supercluster is expected to arise from the sum of the signal from the individual clusters, plus  a possible additional contribution from a filamentary inter-cluster gas structure, the existence of which has not yet been observationally proven. This \planck-\xmm\ discovery may open the way to  constrain  the existence and   properties of such filamentary matter, via deeper combined \planck\ SZ and X--ray studies. The current \xmm\ snapshot observations do not allow conclusive comparison between the SZ and X--ray signals. Deeper observations are needed, sufficient to determine the pressure profile of individual subclusters.

The \planck\ SZ survey has already started to complement existing X--ray surveys, particularly above $z\sim0.3$. Notably, it is finding new clusters below the flux limit of catalogues based on extended \rass\ source detection, such as the \reflex\ survey, and new clusters brighter than the flux limit of  the \macs\ survey above $z=0.3$.  Such discoveries are due to a combination of larger effective sky coverage and the intrinsic limitations of a \rass--based cluster survey.  In practice, surveys considering extended \rass\ sources, such as  \reflex,  have a higher flux limit than that corresponding to \planck's sensitivity. By considering \rass-BSC sources without extent criteria, the \macs\ survey reaches a lower flux limit, at the price of extensive optical confirmation follow-up that does not cover the whole sky. Furthermore, the \rass-BSC detection algorithm was primarily designed for point source detection and can miss very diffuse sources similar to the clusters with flat morphology that \planck\ is revealing. Four of our confirmed clusters are above the \macs\ limit but are not associated with a \rass-BSC source.

For the single-component clusters, we have been able to derive the first estimates of their physical properties such as $\LX$, $\YX$ (with $\Mv$ estimated using $\YX$ as a mass proxy), and density profiles. These properties  suggest that the new clusters are massive, dynamically-complex, objects. These SZ-selected objects have, on average, lower luminosities, flatter density profiles, and a more disturbed morphology than their X-ray selected counterparts. As a result, the dispersion around the $M$--$L$ relation may be larger than previously thought, with new clusters like PLCK~G292.5+22.0 at $z=0.3$ barely reaching the \macs\ flux limit for an estimated mass of $\Mv\sim10^{15}\,\msol$. This suggests that there is a non-negligible population of massive, dynamically perturbed (merging) clusters that do not appear in all-sky X--ray surveys. Furthermore, as  the bulk of cluster cosmology is currently undertaken using X-ray-selected samples, the lack of these clusters may have implications for measures of the cosmologically-sensitive exponential end of the mass function. 

The above preview of newly-detected \planck\ cluster properties must be confirmed with deeper, multi-wavelength, follow-up observations. Such observations include optical redshift confirmation (see the ENO observations presented here), detailed pressure profiles from deeper \xmm\ observations and mass estimates. The latter require the combination of lensing, optical and X--ray data,  in view of the highly unrelaxed nature of the objects.  

Continuation of the confirmation of \planck\ candidates and the characterisation of the \planck\ selection function constitutes a major effort, and requires a good understanding of the properties of the newly-discovered clusters. As we have shown in this paper, 
\xmm\ can play a major role in this process. The \xmm\ validation programme is  presently ongoing. It is currently focussed on \planck\ detections both in the $\textrm{S/N}>5$ range and at lower $\textrm{S/N}$, thus potentially leading to the discovery of more distant clusters.

\begin{acknowledgements}
  The \planck\ Collaboration thanks Norbert Schartel for his
  support to the validation process and granting discretionary time for the observation of \planck\ cluster candidates.  
  The present work is based:
  on observations obtained with \xmm, an ESA science mission with
  instruments and contributions directly funded by ESA Member States
  and the USA (NASA);  and on observations made with the IAC80
  telescope operated on the island of Tenerife by the Instituto de
  Astrof\'isica de Canarias (IAC) in the Spanish Observatorio del
  Teide.  This research has made use of the following databases: SIMBAD, operated at CDS, Strasbourg, France; the NED database, which is operated by the Jet Propulsion Laboratory, California Institute of Technology, under contract with the National Aeronautics and Space Administration; BAX, which is operated by the Laboratoire dÕAstrophysique de Tarbes-Toulouse (LATT), under contract with the Centre National dÕEtudes Spatiales (CNES); and the   SZ repository operated by
IAS Data and Operation Center (IDOC) under contract with CNES.
A description of the Planck Collaboration and a
list of its members, indicating which technical or scientific
activities they have been involved in, can be found at  http://www.rssd.esa.int/Planck
\_Collaboration.
The Planck Collaboration acknowledges the support of: ESA; CNES and CNRS/INSU-IN2P3-INP (France); ASI, CNR, and INAF (Italy); NASA and DoE (USA); STFC and UKSA (UK); CSIC, MICINN and JA (Spain); Tekes, AoF and CSC (Finland); DLR and MPG (Germany); CSA (Canada); DTU Space (Denmark); SER/SSO (Switzerland); RCN (Norway); SFI (Ireland); FCT/MCTES (Portugal); and DEISA (EU)

\end{acknowledgements}

\bibliographystyle{aa}
\bibliography{Planck2011-5.1b.bib,Planck_bib.bib}

\appendix

\section{Redshift determination based on optical counterparts}
\begin{figure}[t]
 \centering
\includegraphics[angle=0,keepaspectratio,width=\columnwidth]{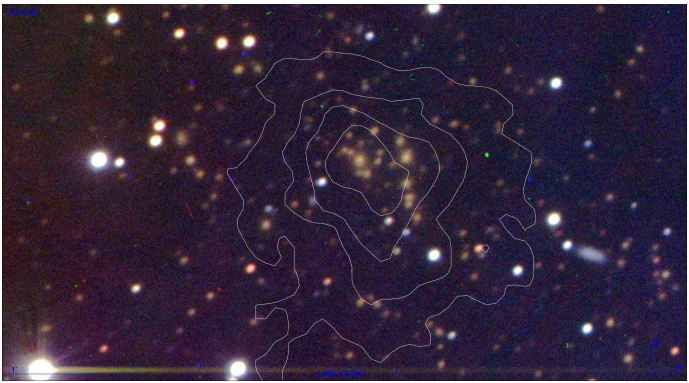}
 \caption{{\footnotesize RGB colour-composite image of the optical counterpart of cluster
   PLCK~G100.2-30.4, as obtained with the IAC80 telescope. R,G,B colours are $g,r$
   and $i$--band Sloan filters, respectively. Although the IAC80 field of view is
  $7\arcmin$, this image is trimmed to $4.8\arcmin \times 2.7\arcmin$, and  is
   centred at the \xmm\  counterpart (white iso-contours). Red galaxies
   defining the red-sequence of the cluster are clearly seen at the centre of
   the image. }}
\label{fig:image-PLCK100-30}
\end{figure}

\subsection{PLCK~G100.2$-$30.4}
\label{sec:apa}

We observed PLCK~G100.2$-$30.4 with the CAMELOT camera on the 0.82-m IAC80 telescope at the Observatorio del Teide (Tenerife, Spain), as part of a validation campaign for newly detected \planck\ clusters that started in semester 2010\,A.
We obtained four images in the Sloan $g,r,i$ and $z$--bands, all centred at the location of the \planck\ cluster candidate, with a field of view of
$7\arcm$ and a pixel scale of $0\farcs304$. The integration time achieved in each filter was approximately $3\,$ksec, yielding a limiting magnitude of
22.9, 21.7, 20.1 and 20.2 for $g,r,i$ and $z$, respectively, for a $6\sigma$ detection.

Image data reduction was undertaken using standard {\sc iraf}
routines. The source detection, catalogue extraction and photometry measurements on the processed images were performed using {\sc SExtractor} \citep{sextractor}. Sources were identified independently in the four bands using a $1.5\sigma$ {\sc SExtractor} detection threshold in the filtered maps (i.e. equivalent to $\textrm{S/N}$$\sim 3$).
The colour-composite image of the $g,r$ and $i$ filters (see
Figure~\ref{fig:image-PLCK100-30}) clearly shows an excess of red galaxies at
the location of the X-ray detection.

We have obtained photometric redshifts for all galaxies in the field
using the {\sc bpz} code \citep{bpz00}. We use the photometry information
from the $g, r$ and $i$--bands, as they provide more reliable redshift estimates.
From the final galaxy catalogue, we identify eight galaxies located within a radius of $1.5\arcm$ from the peak of the X-ray emission which all have a photometric redshift estimate of about 0.38.  Based on this information, we estimate the photometric redshift of the cluster to be $z_{\rm phot}
= 0.38 \pm 0.04$.

\begin{figure}
 \begin{center}
\includegraphics[width=0.8\columnwidth]{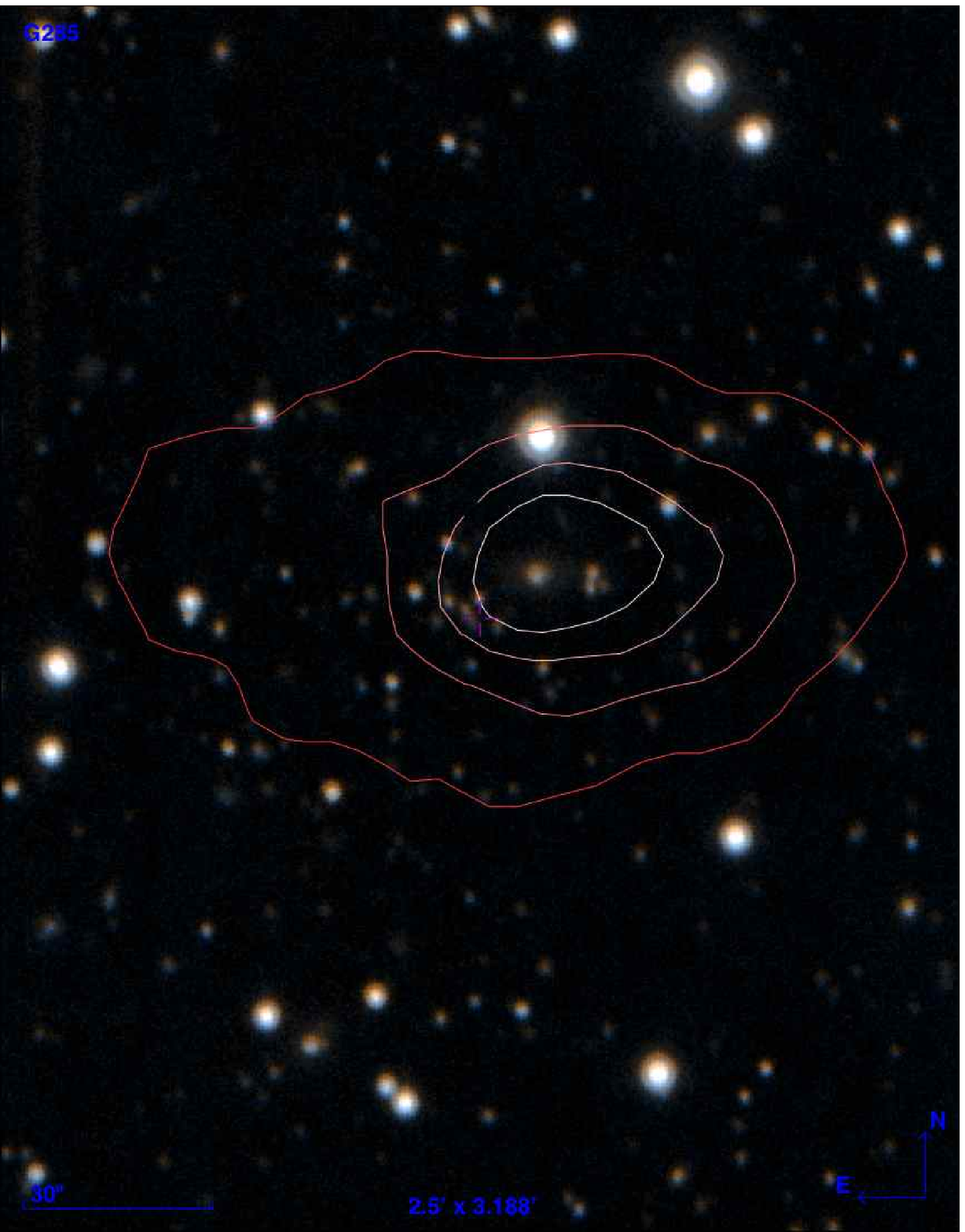}
 \end{center}
 \vspace{-15pt}
 \begin{center}
 \includegraphics[width=0.8\columnwidth]{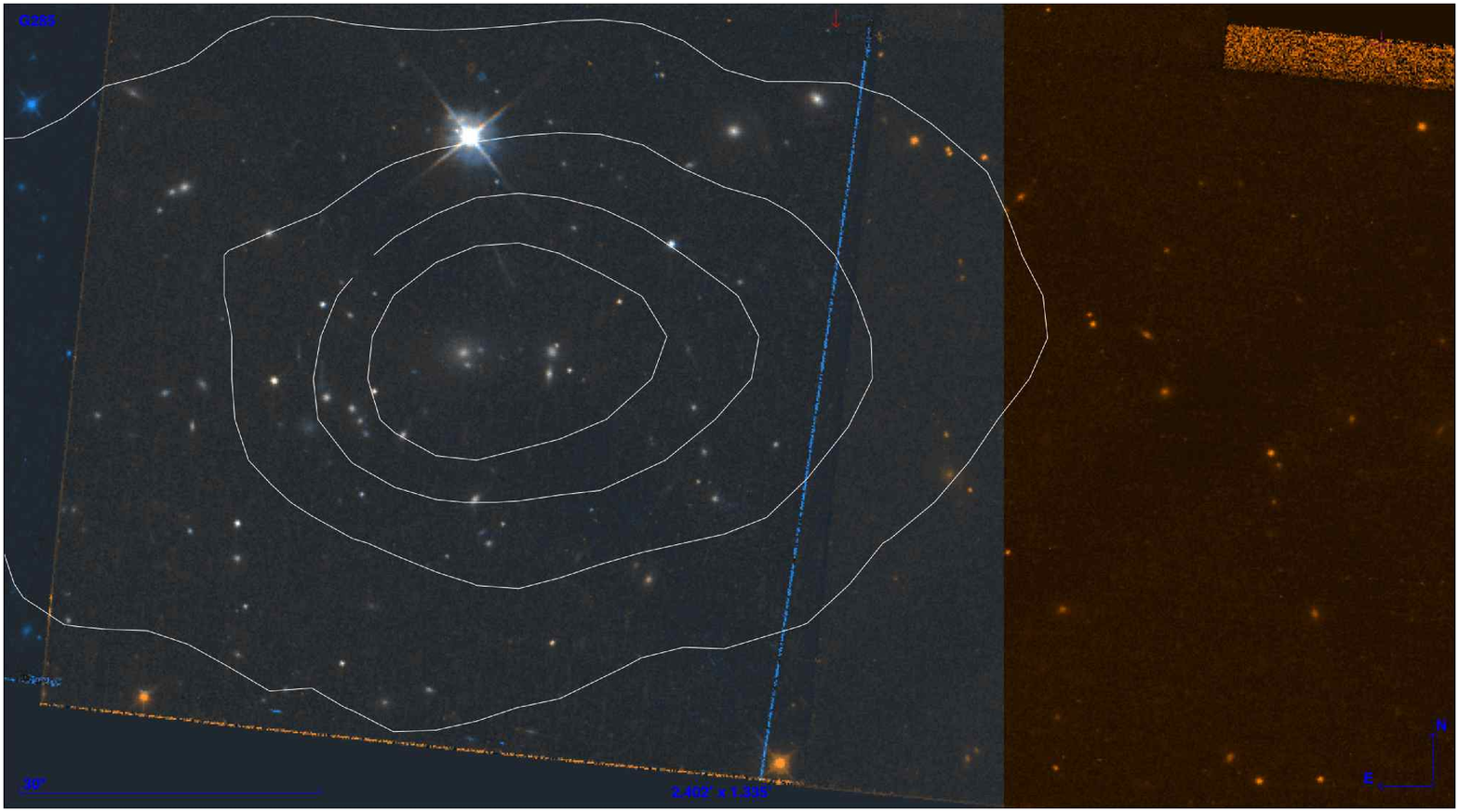}
 \end{center}
 \caption{{\footnotesize Reduced images for PLCK~G285.0$-$23.7, based on archival data. Top:
   Combined V and R--band images taken from ESO archives (SUSI2 data). Bottom:
   Combined F814W and F606W images taken from WFPC2 archival data.  The \xmm\ iso-contours are overlaid.}}
\label{fig:image-PLCK285.0-23.7}
\end{figure}

\subsection{PLCK~G285.0$-$23.7}

After a detailed search in the ESO archive\footnote{http://archive.eso.org} of all existing observations within $5\arcm$ around the location of {PLCK~G285.0$-$23.7}, we found ten images taken with SUSI2 and three spectra taken with EMMI,  all obtained at the ESO-NTT 3.5m telescope \footnote{http://www.eso.org/sci/facilities/lasilla/telescopes/ntt/}.
The first panel in Figure~\ref{fig:image-PLCK285.0-23.7} shows the resulting colour composite of the central region of this cluster, based on V (1600s) and R band (1680s) images with the SUSI2 instrument at NTT.

Publicly--available WFPC2 images for this region also exist in the {\it HST} archive\footnote{http://archive.eso.org/wdb/wdb/hst/science/query}.  The second panel in Figure~\ref{fig:image-PLCK285.0-23.7} shows the reduced colour composite based on the F814W and F606W filter images. Both images have an integration time of 1200~s. From the colour--redshift relation for those images, we derived  a redshift for this cluster of $z\sim0.37$.

\subsection{PLCK~G286.3$-$38.4}

After a detailed search in the ESO archive of all existing observations within
$5\arcmin$ around the location of PLCK~G286.3$-$38.4, nine images (NTT/SUSI2),
and three spectra (NTT/EMMI) were found. The spectroscopic data from NTT/EMMI are associated with a proposal to characterise the optical counterpart of a potential galaxy cluster associated with the X-ray source RX\,J0359.1$-$7205.

\begin{figure}
 \centering
 \includegraphics[width=8cm]{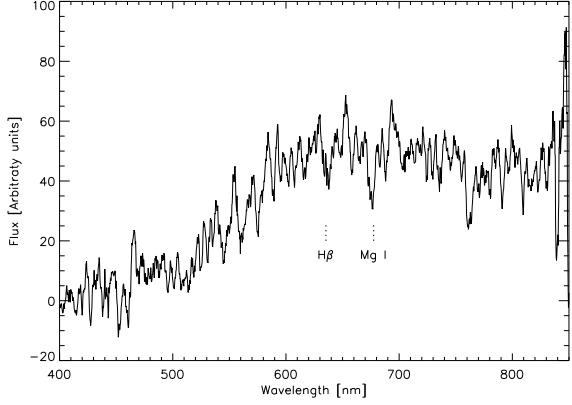}
 \caption{{\footnotesize Reduced spectrum for PLCK286.0-38.4, based on the EMMI data obtained
   from ESO archives. A cross-correlation analysis yields a redshift
   of $z=0.307$, which is consistent with the preliminary estimate of the two
   absorption features H$\beta$ and \ion{Mg}{I} which are also indicated in the figure. }}
\label{fig:spectrum-PLCK286.0-38.4}
\end{figure}

We undertook data reduction of the three EMMI spectra using standard {\sc  iraf} routines. Fig.~\ref{fig:spectrum-PLCK286.0-38.4} shows the combined final spectrum. Although this spectrum has a very low signal-to-noise ratio, a preliminary redshift estimate could be obtained from cross-correlating the reduced spectrum with a reference template spectrum for an early-type galaxy (taken from  http://www.arcetri.astro.it/\verb!~!k20/), using the {\sc iraf} routine {\sc fxcor}. The derived redshift estimate is $z=0.307 \pm 0.003$, although the significance of the cross-correlation peak is very low.
Nevertheless, this redshift estimate is apparently compatible with the preliminary identification of two absorption features (H$\beta$ and Mg-I) in the reduced spectrum (see Fig.~\ref{fig:spectrum-PLCK286.0-38.4}), which gives us more confidence in the result.

\section{Search for large scale structure in the field of PLCK~G214.6+36.9 using SDSS}
\label{sec:sc_ap}

We searched for superclusters in the direction of PLCK~G214.6+36.9  by calculating the luminosity density field of the spectroscopic sample of luminous red galaxies (LRG) from the SDSS DR7. To correct for the finite width of the survey magnitude window,  galaxy luminosities were  weighted. Superclusters are delineated by an appropriate luminosity density level.  For the LRG superclusters, we set this level at 3.0 times the mean density. This level was obtained  by  comparing the SDSS main galaxy sample superclusters with those in the LRG sample in the volume where these samples overlap. The procedure is explained in detail in \citet{lii10}. 

The best candidate is a supercluster containing 10 LRGs with a mass centre  at RA=137\fdg5, Dec=13\fdg6, lying at $z=0.45$. Since each LRG likely indicates the presence of a galaxy cluster (like the two LRGs that lie in the observed X--ray clusters) the supercluster is likely to contain 10 clusters. The estimated total luminosity of the supercluster is  $3\times 10^{12}\,{\rm L_{\sun}}$, the maximum extent about $70 {\rm h^{-1}}$~Mpc. The co-moving distance along the line-of-sight between the two LRGs hosted by the X-ray clusters is about $4.1 h^{-1}$~Mpc. Using a $M/L$ value of 200 (in solar units), the total  mass of the supercluster is about $10^{15}\,\msol$.  This supercluster is typical among other LRG superclusters at that distance. Fig.~\ref{fig:sc} shows the projected luminosity density contours of the candidate supercluster, together with the location of the PLCK~G214.6+36.9 and the X-ray clusters.

\begin{figure}
  \centering
  \includegraphics[width=\columnwidth]{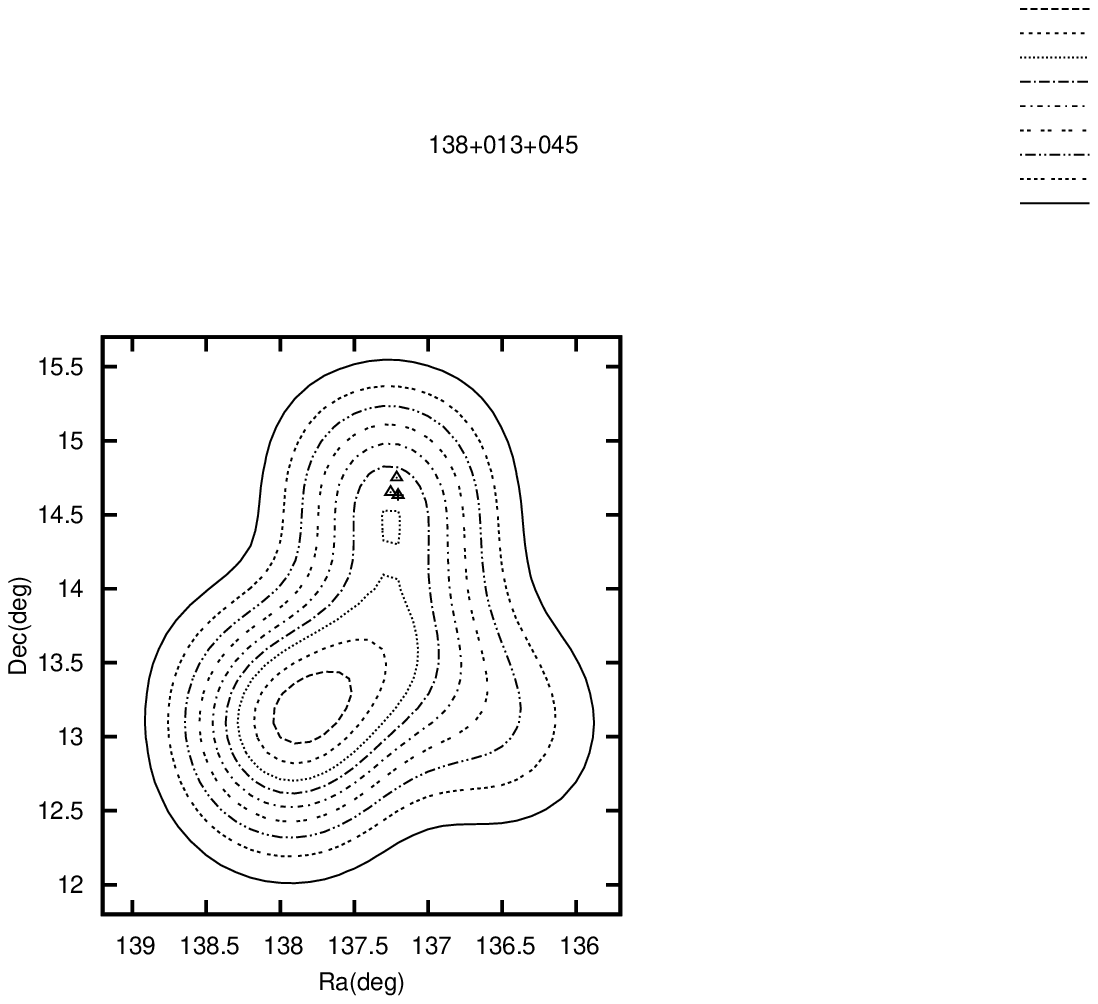}
  \caption{{\footnotesize Projected luminosity density map for the best supercluster candidate. The X-ray sources are marked by triangles, the SZ source by the cross. The minimum density level and the level step are $10^{11}{\rm L_{\sun}}$ per square degree.}}
\label{fig:scc}
\end{figure}

\end{document}

%% file: Proj_Ref_5_1b_authors_and_institutes.tex
\author{\small
Planck Collaboration:
N.~Aghanim\inst{48}
\and
M.~Arnaud\inst{59}
\and
M.~Ashdown\inst{57, 4}
\and
J.~Aumont\inst{48}
\and
C.~Baccigalupi\inst{70}
\and
A.~Balbi\inst{30}
\and
A.~J.~Banday\inst{79, 7, 64}
\and
R.~B.~Barreiro\inst{54}
\and
M.~Bartelmann\inst{78, 64}
\and
J.~G.~Bartlett\inst{3, 55}
% ...
}
\institute{\small
Aalto University Mets\"{a}hovi Radio Observatory, Mets\"{a}hovintie 114, FIN-02540 Kylm\"{a}l\"{a}, Finland\\
\and
Agenzia Spaziale Italiana Science Data Center, c/o ESRIN, via Galileo Galilei, Frascati, Italy\\
\and
Astroparticule et Cosmologie, CNRS (UMR7164), Universit\'{e} Denis Diderot Paris 7, B\^{a}timent Condorcet, 10 rue A. Domon et L\'{e}onie Duquet, Paris, France\\
\and
Astrophysics Group, Cavendish Laboratory, University of Cambridge, J J Thomson Avenue, Cambridge CB3 0HE, U.K.\\
\and
Atacama Large Millimeter/submillimeter Array, ALMA Santiago Central Offices, Alonso de Cordova 3107, Vitacura, Casilla 763 0355, Santiago, Chile\\
\and
CITA, University of Toronto, 60 St. George St., Toronto, ON M5S 3H8, Canada\\
\and
CNRS, IRAP, 9 Av. colonel Roche, BP 44346, F-31028 Toulouse cedex 4, France\\
\and
California Institute of Technology, Pasadena, California, U.S.A.\\
\and
Centre of Mathematics for Applications, University of Oslo, Blindern, Oslo, Norway\\
\and
Centro de Astrof\'{\i}sica, Universidade do Porto, Rua das Estrelas, 4150-762 Porto, Portugal\\
\and
DAMTP, University of Cambridge, Centre for Mathematical Sciences, Wilberforce Road, Cambridge CB3 0WA, U.K.\\
\and
DSM/Irfu/SPP, CEA-Saclay, F-91191 Gif-sur-Yvette Cedex, France\\
\and
DTU Space, National Space Institute, Juliane Mariesvej 30, Copenhagen, Denmark\\
\and
Departamento de F\'{\i}sica, Universidad de Oviedo, Avda. Calvo Sotelo s/n, Oviedo, Spain\\
\and
Department of Astronomy and Astrophysics, University of Toronto, 50 Saint George Street, Toronto, Ontario, Canada\\
\and
Department of Physics \& Astronomy, University of British Columbia, 6224 Agricultural Road, Vancouver, British Columbia, Canada\\
\and
Department of Physics and Astronomy, University of Southern California, Los Angeles, California, U.S.A.\\
\and
Department of Physics and Astronomy, University of Sussex, Brighton BN1 9QH, U.K.\\
\and
Department of Physics, Gustaf H\"{a}llstr\"{o}min katu 2a, University of Helsinki, Helsinki, Finland\\
\and
Department of Physics, Purdue University, 525 Northwestern Avenue, West Lafayette, Indiana, U.S.A.\\
\and
Department of Physics, University of California, Berkeley, California, U.S.A.\\
\and
Department of Physics, University of California, One Shields Avenue, Davis, California, U.S.A.\\
\and
Department of Physics, University of California, Santa Barbara, California, U.S.A.\\
\and
Department of Physics, University of Illinois at Urbana-Champaign, 1110 West Green Street, Urbana, Illinois, U.S.A.\\
\and
Dipartimento di Fisica G. Galilei, Universit\`{a} degli Studi di Padova, via Marzolo 8, 35131 Padova, Italy\\
\and
Dipartimento di Fisica, Universit\`{a} La Sapienza, P. le A. Moro 2, Roma, Italy\\
\and
Dipartimento di Fisica, Universit\`{a} degli Studi di Milano, Via Celoria, 16, Milano, Italy\\
\and
Dipartimento di Fisica, Universit\`{a} degli Studi di Trieste, via A. Valerio 2, Trieste, Italy\\
\and
Dipartimento di Fisica, Universit\`{a} di Ferrara, Via Saragat 1, 44122 Ferrara, Italy\\
\and
Dipartimento di Fisica, Universit\`{a} di Roma Tor Vergata, Via della Ricerca Scientifica, 1, Roma, Italy\\
\and
Discovery Center, Niels Bohr Institute, Blegdamsvej 17, Copenhagen, Denmark\\
\and
Dpto. Astrof\'{i}sica, Universidad de La Laguna (ULL), E-38206 La Laguna, Tenerife, Spain\\
\and
European Southern Observatory, ESO Vitacura, Alonso de Cordova 3107, Vitacura, Casilla 19001, Santiago, Chile\\
\and
European Space Agency, ESAC, Camino bajo del Castillo, s/n, Urbanizaci\'{o}n Villafranca del Castillo, Villanueva de la Ca\~{n}ada, Madrid, Spain\\
\and
European Space Agency, ESAC, Planck Science Office, Camino bajo del Castillo, s/n, Urbanizaci\'{o}n Villafranca del Castillo, Villanueva de la Ca\~{n}ada, Madrid, Spain\\
\and
European Space Agency, ESTEC, Keplerlaan 1, 2201 AZ Noordwijk, The Netherlands\\
\and
Helsinki Institute of Physics, Gustaf H\"{a}llstr\"{o}min katu 2, University of Helsinki, Helsinki, Finland\\
\and
INAF - Osservatorio Astronomico di Padova, Vicolo dell'Osservatorio 5, Padova, Italy\\
\and
INAF - Osservatorio Astronomico di Roma, via di Frascati 33, Monte Porzio Catone, Italy\\
\and
INAF - Osservatorio Astronomico di Trieste, Via G.B. Tiepolo 11, Trieste, Italy\\
\and
INAF/IASF Bologna, Via Gobetti 101, Bologna, Italy\\
\and
INAF/IASF Milano, Via E. Bassini 15, Milano, Italy\\
\and
INRIA, Laboratoire de Recherche en Informatique, Universit\'{e} Paris-Sud 11, B\^{a}timent 490, 91405 Orsay Cedex, France\\
\and
IPAG: Institut de Plan\'{e}tologie et d'Astrophysique de Grenoble, Universit\'{e} Joseph Fourier, Grenoble 1 / CNRS-INSU, UMR 5274, Grenoble, F-38041, France\\
\and
Imperial College London, Astrophysics group, Blackett Laboratory, Prince Consort Road, London, SW7 2AZ, U.K.\\
\and
Infrared Processing and Analysis Center, California Institute of Technology, Pasadena, CA 91125, U.S.A.\\
\and
Institut N\'{e}el, CNRS, Universit\'{e} Joseph Fourier Grenoble I, 25 rue des Martyrs, Grenoble, France\\
\and
Institut d'Astrophysique Spatiale, CNRS (UMR8617) Universit\'{e} Paris-Sud 11, B\^{a}timent 121, Orsay, France\\
\and
Institut d'Astrophysique de Paris, CNRS UMR7095, Universit\'{e} Pierre \& Marie Curie, 98 bis boulevard Arago, Paris, France\\
\and
Institute of Astronomy and Astrophysics, Academia Sinica, Taipei, Taiwan\\
\and
Institute of Astronomy, University of Cambridge, Madingley Road, Cambridge CB3 0HA, U.K.\\
\and
Institute of Theoretical Astrophysics, University of Oslo, Blindern, Oslo, Norway\\
\and
Instituto de Astrof\'{\i}sica de Canarias, C/V\'{\i}a L\'{a}ctea s/n, La Laguna, Tenerife, Spain\\
\and
Instituto de F\'{\i}sica de Cantabria (CSIC-Universidad de Cantabria), Avda. de los Castros s/n, Santander, Spain\\
\and
Jet Propulsion Laboratory, California Institute of Technology, 4800 Oak Grove Drive, Pasadena, California, U.S.A.\\
\and
Jodrell Bank Centre for Astrophysics, Alan Turing Building, School of Physics and Astronomy, The University of Manchester, Oxford Road, Manchester, M13 9PL, U.K.\\
\and
Kavli Institute for Cosmology Cambridge, Madingley Road, Cambridge, CB3 0HA, U.K.\\
\and
LERMA, CNRS, Observatoire de Paris, 61 Avenue de l'Observatoire, Paris, France\\
\and
Laboratoire AIM, IRFU/Service d'Astrophysique - CEA/DSM - CNRS - Universit\'{e} Paris Diderot, B\^{a}t. 709, CEA-Saclay, F-91191 Gif-sur-Yvette Cedex, France\\
\and
Laboratoire Traitement et Communication de l'Information, CNRS (UMR 5141) and T\'{e}l\'{e}com ParisTech, 46 rue Barrault F-75634 Paris Cedex 13, France\\
\and
Laboratoire de Physique Subatomique et de Cosmologie, CNRS/IN2P3, Universit\'{e} Joseph Fourier Grenoble I, Institut National Polytechnique de Grenoble, 53 rue des Martyrs, 38026 Grenoble cedex, France\\
\and
Laboratoire de l'Acc\'{e}l\'{e}rateur Lin\'{e}aire, Universit\'{e} Paris-Sud 11, CNRS/IN2P3, Orsay, France\\
\and
Lawrence Berkeley National Laboratory, Berkeley, California, U.S.A.\\
\and
Max-Planck-Institut f\"{u}r Astrophysik, Karl-Schwarzschild-Str. 1, 85741 Garching, Germany\\
\and
Max-Planck-Institut f\"{u}r Extraterrestrische Physik, Giessenbachstra{\ss}e, 85748 Garching, Germany\\
\and
MilliLab, VTT Technical Research Centre of Finland, Tietotie 3, Espoo, Finland\\
\and
National University of Ireland, Department of Experimental Physics, Maynooth, Co. Kildare, Ireland\\
\and
Niels Bohr Institute, Blegdamsvej 17, Copenhagen, Denmark\\
\and
Observational Cosmology, Mail Stop 367-17, California Institute of Technology, Pasadena, CA, 91125, U.S.A.\\
\and
SISSA, Astrophysics Sector, via Bonomea 265, 34136, Trieste, Italy\\
\and
SUPA, Institute for Astronomy, University of Edinburgh, Royal Observatory, Blackford Hill, Edinburgh EH9 3HJ, U.K.\\
\and
School of Physics and Astronomy, Cardiff University, Queens Buildings, The Parade, Cardiff, CF24 3AA, U.K.\\
\and
Space Research Institute (IKI), Russian Academy of Sciences, Profsoyuznaya Str, 84/32, Moscow, 117997, Russia\\
\and
Space Sciences Laboratory, University of California, Berkeley, California, U.S.A.\\
\and
Stanford University, Dept of Physics, Varian Physics Bldg, 382 Via Pueblo Mall, Stanford, California, U.S.A.\\
\and
Tartu Observatory, Toravere, Tartumaa, 61602, Estonia\\
\and
Tuorla Observatory, Department of Physics and Astronomy, University of Turku, V\"ais\"al\"antie 20, FIN-21500, Piikki\"o, Finland\\
\and
Universit\"{a}t Heidelberg, Institut f\"{u}r Theoretische Astrophysik, Albert-\"{U}berle-Str. 2, 69120, Heidelberg, Germany\\
\and
Universit\'{e} de Toulouse, UPS-OMP, IRAP, F-31028 Toulouse cedex 4, France\\
% ...
}

%% file: Planck.tex
\def\setsymbol#1#2{\expandafter\def\csname #1\endcsname{#2}}
\def\getsymbol#1{\csname #1\endcsname}

\def\Planck{{\it Planck\/}}

\def\HeJT{$^4$He-JT}

\def\allearlypapers{\nocite{planck2011-1.1, planck2011-1.3, planck2011-1.4, planck2011-1.5, planck2011-1.6, planck2011-1.7, planck2011-1.10, planck2011-1.10sup, planck2011-5.1a, planck2011-5.1b, planck2011-5.2a, planck2011-5.2b, planck2011-5.2c, planck2011-6.1, planck2011-6.2, planck2011-6.3a, planck2011-6.4a, planck2011-6.4b, planck2011-6.6, planck2011-7.0, planck2011-7.2, planck2011-7.3, planck2011-7.7a, planck2011-7.7b, planck2011-7.12, planck2011-7.13}}

\newbox\tablebox    \newdimen\tablewidth
\def\leaderfil{\leaders\hbox to 5pt{\hss.\hss}\hfil}
%
%
\def\endPlancktable{\tablewidth=\columnwidth 
    $$\hss\copy\tablebox\hss$$
    \vskip-\lastskip\vskip -2pt}
\def\endPlancktablewide{\tablewidth=\textwidth 
    $$\hss\copy\tablebox\hss$$
    \vskip-\lastskip\vskip -2pt}
\def\tablenote#1 #2\par{\begingroup \parindent=0.8em
    \abovedisplayshortskip=0pt\belowdisplayshortskip=0pt
    \noindent
    $$\hss\vbox{\hsize\tablewidth \hangindent=\parindent \hangafter=1 \noindent
    \hbox to \parindent{\sup{\rm #1}\hss}\strut#2\strut\par}\hss$$
    \endgroup}
\def\doubleline{\vskip 3pt\hrule \vskip 1.5pt \hrule \vskip 5pt}

%
\def\L2{\ifmmode L_2\else $L_2$\fi}
\def\dtt{\Delta T/T}
\def\DeltaT{\ifmmode \Delta T\else $\Delta T$\fi}
\def\deltat{\ifmmode \Delta t\else $\Delta t$\fi}
\def\fknee{\ifmmode f_{\rm knee}\else $f_{\rm knee}$\fi}
\def\Fmax{\ifmmode F_{\rm max}\else $F_{\rm max}$\fi}
\def\solar{\ifmmode{\rm M}_{\mathord\odot}\else${\rm M}_{\mathord\odot}$\fi}
\def\sol{\solar}
\def\mag{\sup{m}}
\def\inv{\ifmmode^{-1}\else$^{-1}$\fi}
\def\mo{\ifmmode^{-1}\else$^{-1}$\fi}
\def\sup#1{\ifmmode ^{\rm #1}\else $^{\rm #1}$\fi}
\def\expo#1{\ifmmode \times 10^{#1}\else $\times 10^{#1}$\fi}
\def\,{\thinspace}
\def\lsim{\mathrel{\raise .4ex\hbox{\rlap{$<$}\lower 1.2ex\hbox{$\sim$}}}}
\def\gsim{\mathrel{\raise .4ex\hbox{\rlap{$>$}\lower 1.2ex\hbox{$\sim$}}}}
\let\lea=\lsim
\let\gea=\gsim
\def\simprop{\mathrel{\raise .4ex\hbox{\rlap{$\propto$}\lower 1.2ex\hbox{$\sim$}}}}
\def\deg{\ifmmode^\circ\else$^\circ$\fi}
\def\pdeg{\ifmmode $\setbox0=\hbox{$^{\circ}$}\rlap{\hskip.11\wd0 .}$^{\circ}
          \else \setbox0=\hbox{$^{\circ}$}\rlap{\hskip.11\wd0 .}$^{\circ}$\fi}
\def\arcs{\ifmmode {^{\scriptstyle\prime\prime}}
          \else $^{\scriptstyle\prime\prime}$\fi}
\def\arcm{\ifmmode {^{\scriptstyle\prime}}
          \else $^{\scriptstyle\prime}$\fi}
\newdimen\sa  \newdimen\sb
\def\parcs{\sa=.07em \sb=.03em
     \ifmmode \hbox{\rlap{.}}^{\scriptstyle\prime\kern -\sb\prime}\hbox{\kern -\sa}
     \else \rlap{.}$^{\scriptstyle\prime\kern -\sb\prime}$\kern -\sa\fi}
\def\parcm{\sa=.08em \sb=.03em
     \ifmmode \hbox{\rlap{.}\kern\sa}^{\scriptstyle\prime}\hbox{\kern-\sb}
     \else \rlap{.}\kern\sa$^{\scriptstyle\prime}$\kern-\sb\fi}
\def\ra[#1 #2 #3.#4]{#1\sup{h}#2\sup{m}#3\sup{s}\llap.#4}
\def\dec[#1 #2 #3.#4]{#1\deg#2\arcm#3\arcs\llap.#4}
\def\deco[#1 #2 #3]{#1\deg#2\arcm#3\arcs}
\def\rra[#1 #2]{#1\sup{h}#2\sup{m}}
\def\page{\vfill\eject}
\def\dots{\relax\ifmmode \ldots\else $\ldots$\fi}
%
%
\def\WHzsr{\ifmmode $W\,Hz\mo\,sr\mo$\else W\,Hz\mo\,sr\mo\fi}
\def\mHz{\ifmmode $\,mHz$\else \,mHz\fi}
\def\GHz{\ifmmode $\,GHz$\else \,GHz\fi}
\def\mKs{\ifmmode $\,mK\,s$^{1/2}\else \,mK\,s$^{1/2}$\fi}
\def\muKs{\ifmmode \,\mu$K\,s$^{1/2}\else \,$\mu$K\,s$^{1/2}$\fi}
\def\muKRJs{\ifmmode \,\mu$K$_{\rm RJ}$\,s$^{1/2}\else \,$\mu$K$_{\rm RJ}$\,s$^{1/2}$\fi}
\def\muKHz{\ifmmode \,\mu$K\,Hz$^{-1/2}\else \,$\mu$K\,Hz$^{-1/2}$\fi}
\def\MJysr{\ifmmode \,$MJy\,sr\mo$\else \,MJy\,sr\mo\fi}
\def\MJysrmK{\ifmmode \,$MJy\,sr\mo$\,mK$_{\rm CMB}\mo\else \,MJy\,sr\mo\,mK$_{\rm CMB}\mo$\fi}
\def\microns{\ifmmode \,\mu$m$\else \,$\mu$m\fi}
\def\micron{\microns}
\def\muK{\ifmmode \,\mu$K$\else \,$\mu$\hbox{K}\fi}
\def\microK{\ifmmode \,\mu$K$\else \,$\mu$\hbox{K}\fi}
\def\muW{\ifmmode \,\mu$W$\else \,$\mu$\hbox{W}\fi}
\def\kms{\ifmmode $\,km\,s$^{-1}\else \,km\,s$^{-1}$\fi}
\def\kmsMpc{\ifmmode $\,\kms\,Mpc\mo$\else \,\kms\,Mpc\mo\fi}
%
%


\setsymbol{LFI:center:frequency:70GHz:units}{70.3\,GHz}
\setsymbol{LFI:center:frequency:44GHz:units}{44.1\,GHz}
\setsymbol{LFI:center:frequency:30GHz:units}{28.5\,GHz}

\setsymbol{LFI:center:frequency:70GHz}{70.3}
\setsymbol{LFI:center:frequency:44GHz}{44.1}
\setsymbol{LFI:center:frequency:30GHz}{28.5}

\setsymbol{LFI:center:frequency:LFI18:Rad:M:units}{71.7\GHz}
\setsymbol{LFI:center:frequency:LFI19:Rad:M:units}{67.5\GHz}
\setsymbol{LFI:center:frequency:LFI20:Rad:M:units}{69.2\GHz}
\setsymbol{LFI:center:frequency:LFI21:Rad:M:units}{70.4\GHz}
\setsymbol{LFI:center:frequency:LFI22:Rad:M:units}{71.5\GHz}
\setsymbol{LFI:center:frequency:LFI23:Rad:M:units}{70.8\GHz}
\setsymbol{LFI:center:frequency:LFI24:Rad:M:units}{44.4\GHz}
\setsymbol{LFI:center:frequency:LFI25:Rad:M:units}{44.0\GHz}
\setsymbol{LFI:center:frequency:LFI26:Rad:M:units}{43.9\GHz}
\setsymbol{LFI:center:frequency:LFI27:Rad:M:units}{28.3\GHz}
\setsymbol{LFI:center:frequency:LFI28:Rad:M:units}{28.8\GHz}
\setsymbol{LFI:center:frequency:LFI18:Rad:S:units}{70.1\GHz}
\setsymbol{LFI:center:frequency:LFI19:Rad:S:units}{69.6\GHz}
\setsymbol{LFI:center:frequency:LFI20:Rad:S:units}{69.5\GHz}
\setsymbol{LFI:center:frequency:LFI21:Rad:S:units}{69.5\GHz}
\setsymbol{LFI:center:frequency:LFI22:Rad:S:units}{72.8\GHz}
\setsymbol{LFI:center:frequency:LFI23:Rad:S:units}{71.3\GHz}
\setsymbol{LFI:center:frequency:LFI24:Rad:S:units}{44.1\GHz}
\setsymbol{LFI:center:frequency:LFI25:Rad:S:units}{44.1\GHz}
\setsymbol{LFI:center:frequency:LFI26:Rad:S:units}{44.1\GHz}
\setsymbol{LFI:center:frequency:LFI27:Rad:S:units}{28.5\GHz}
\setsymbol{LFI:center:frequency:LFI28:Rad:S:units}{28.2\GHz}

\setsymbol{LFI:center:frequency:LFI18:Rad:M}{71.7}
\setsymbol{LFI:center:frequency:LFI19:Rad:M}{67.5}
\setsymbol{LFI:center:frequency:LFI20:Rad:M}{69.2}
\setsymbol{LFI:center:frequency:LFI21:Rad:M}{70.4}
\setsymbol{LFI:center:frequency:LFI22:Rad:M}{71.5}
\setsymbol{LFI:center:frequency:LFI23:Rad:M}{70.8}
\setsymbol{LFI:center:frequency:LFI24:Rad:M}{44.4}
\setsymbol{LFI:center:frequency:LFI25:Rad:M}{44.0}
\setsymbol{LFI:center:frequency:LFI26:Rad:M}{43.9}
\setsymbol{LFI:center:frequency:LFI27:Rad:M}{28.3}
\setsymbol{LFI:center:frequency:LFI28:Rad:M}{28.8}
\setsymbol{LFI:center:frequency:LFI18:Rad:S}{70.1}
\setsymbol{LFI:center:frequency:LFI19:Rad:S}{69.6}
\setsymbol{LFI:center:frequency:LFI20:Rad:S}{69.5}
\setsymbol{LFI:center:frequency:LFI21:Rad:S}{69.5}
\setsymbol{LFI:center:frequency:LFI22:Rad:S}{72.8}
\setsymbol{LFI:center:frequency:LFI23:Rad:S}{71.3}
\setsymbol{LFI:center:frequency:LFI24:Rad:S}{44.1}
\setsymbol{LFI:center:frequency:LFI25:Rad:S}{44.1}
\setsymbol{LFI:center:frequency:LFI26:Rad:S}{44.1}
\setsymbol{LFI:center:frequency:LFI27:Rad:S}{28.5}
\setsymbol{LFI:center:frequency:LFI28:Rad:S}{28.2}


\setsymbol{LFI:white:noise:sensitivity:70GHz:units}{152.6\muKs}
\setsymbol{LFI:white:noise:sensitivity:44GHz:units}{173.1\muKs}
\setsymbol{LFI:white:noise:sensitivity:30GHz:units}{146.8\muKs}

\setsymbol{LFI:white:noise:sensitivity:70GHz}{152.6}
\setsymbol{LFI:white:noise:sensitivity:44GHz}{173.1}
\setsymbol{LFI:white:noise:sensitivity:30GHz}{146.8}

\setsymbol{LFI:white:noise:sensitivity:LFI18:Rad:M:units}{512.0\muKs}
\setsymbol{LFI:white:noise:sensitivity:LFI19:Rad:M:units}{581.4\muKs}
\setsymbol{LFI:white:noise:sensitivity:LFI20:Rad:M:units}{590.8\muKs}
\setsymbol{LFI:white:noise:sensitivity:LFI21:Rad:M:units}{455.2\muKs}
\setsymbol{LFI:white:noise:sensitivity:LFI22:Rad:M:units}{492.0\muKs}
\setsymbol{LFI:white:noise:sensitivity:LFI23:Rad:M:units}{507.7\muKs}
\setsymbol{LFI:white:noise:sensitivity:LFI24:Rad:M:units}{462.2\muKs}
\setsymbol{LFI:white:noise:sensitivity:LFI25:Rad:M:units}{413.6\muKs}
\setsymbol{LFI:white:noise:sensitivity:LFI26:Rad:M:units}{478.6\muKs}
\setsymbol{LFI:white:noise:sensitivity:LFI27:Rad:M:units}{277.7\muKs}
\setsymbol{LFI:white:noise:sensitivity:LFI28:Rad:M:units}{312.3\muKs}
\setsymbol{LFI:white:noise:sensitivity:LFI18:Rad:S:units}{465.7\muKs}
\setsymbol{LFI:white:noise:sensitivity:LFI19:Rad:S:units}{555.6\muKs}
\setsymbol{LFI:white:noise:sensitivity:LFI20:Rad:S:units}{623.2\muKs}
\setsymbol{LFI:white:noise:sensitivity:LFI21:Rad:S:units}{564.1\muKs}
\setsymbol{LFI:white:noise:sensitivity:LFI22:Rad:S:units}{534.4\muKs}
\setsymbol{LFI:white:noise:sensitivity:LFI23:Rad:S:units}{542.4\muKs}
\setsymbol{LFI:white:noise:sensitivity:LFI24:Rad:S:units}{399.2\muKs}
\setsymbol{LFI:white:noise:sensitivity:LFI25:Rad:S:units}{392.6\muKs}
\setsymbol{LFI:white:noise:sensitivity:LFI26:Rad:S:units}{418.6\muKs}
\setsymbol{LFI:white:noise:sensitivity:LFI27:Rad:S:units}{302.9\muKs}
\setsymbol{LFI:white:noise:sensitivity:LFI28:Rad:S:units}{285.3\muKs}

\setsymbol{LFI:white:noise:sensitivity:LFI18:Rad:M}{512.0}
\setsymbol{LFI:white:noise:sensitivity:LFI19:Rad:M}{581.4}
\setsymbol{LFI:white:noise:sensitivity:LFI20:Rad:M}{590.8}
\setsymbol{LFI:white:noise:sensitivity:LFI21:Rad:M}{455.2}
\setsymbol{LFI:white:noise:sensitivity:LFI22:Rad:M}{492.0}
\setsymbol{LFI:white:noise:sensitivity:LFI23:Rad:M}{507.7}
\setsymbol{LFI:white:noise:sensitivity:LFI24:Rad:M}{462.2}
\setsymbol{LFI:white:noise:sensitivity:LFI25:Rad:M}{413.6}
\setsymbol{LFI:white:noise:sensitivity:LFI26:Rad:M}{478.6}
\setsymbol{LFI:white:noise:sensitivity:LFI27:Rad:M}{277.7}
\setsymbol{LFI:white:noise:sensitivity:LFI28:Rad:M}{312.3}
\setsymbol{LFI:white:noise:sensitivity:LFI18:Rad:S}{465.7}
\setsymbol{LFI:white:noise:sensitivity:LFI19:Rad:S}{555.6}
\setsymbol{LFI:white:noise:sensitivity:LFI20:Rad:S}{623.2}
\setsymbol{LFI:white:noise:sensitivity:LFI21:Rad:S}{564.1}
\setsymbol{LFI:white:noise:sensitivity:LFI22:Rad:S}{534.4}
\setsymbol{LFI:white:noise:sensitivity:LFI23:Rad:S}{542.4}
\setsymbol{LFI:white:noise:sensitivity:LFI24:Rad:S}{399.2}
\setsymbol{LFI:white:noise:sensitivity:LFI25:Rad:S}{392.6}
\setsymbol{LFI:white:noise:sensitivity:LFI26:Rad:S}{418.6}
\setsymbol{LFI:white:noise:sensitivity:LFI27:Rad:S}{302.9}
\setsymbol{LFI:white:noise:sensitivity:LFI28:Rad:S}{285.3}


\setsymbol{LFI:knee:frequency:70GHz:units}{29.5\mHz}
\setsymbol{LFI:knee:frequency:44GHz:units}{56.2\mHz}
\setsymbol{LFI:knee:frequency:30GHz:units}{113.7\mHz}

\setsymbol{LFI:knee:frequency:70GHz}{29.5}
\setsymbol{LFI:knee:frequency:44GHz}{56.2}
\setsymbol{LFI:knee:frequency:30GHz}{113.7}

\setsymbol{LFI:knee:frequency:LFI18:Rad:M:units}{16.3\mHz}
\setsymbol{LFI:knee:frequency:LFI19:Rad:M:units}{15.1\mHz}
\setsymbol{LFI:knee:frequency:LFI20:Rad:M:units}{18.7\mHz}
\setsymbol{LFI:knee:frequency:LFI21:Rad:M:units}{37.2\mHz}
\setsymbol{LFI:knee:frequency:LFI22:Rad:M:units}{12.7\mHz}
\setsymbol{LFI:knee:frequency:LFI23:Rad:M:units}{34.6\mHz}
\setsymbol{LFI:knee:frequency:LFI24:Rad:M:units}{46.2\mHz}
\setsymbol{LFI:knee:frequency:LFI25:Rad:M:units}{24.9\mHz}
\setsymbol{LFI:knee:frequency:LFI26:Rad:M:units}{67.6\mHz}
\setsymbol{LFI:knee:frequency:LFI27:Rad:M:units}{187.4\mHz}
\setsymbol{LFI:knee:frequency:LFI28:Rad:M:units}{122.2\mHz}
\setsymbol{LFI:knee:frequency:LFI18:Rad:S:units}{17.7\mHz}
\setsymbol{LFI:knee:frequency:LFI19:Rad:S:units}{22.0\mHz}
\setsymbol{LFI:knee:frequency:LFI20:Rad:S:units}{8.7\mHz}
\setsymbol{LFI:knee:frequency:LFI21:Rad:S:units}{25.9\mHz}
\setsymbol{LFI:knee:frequency:LFI22:Rad:S:units}{15.8\mHz}
\setsymbol{LFI:knee:frequency:LFI23:Rad:S:units}{129.8\mHz}
\setsymbol{LFI:knee:frequency:LFI24:Rad:S:units}{100.9\mHz}
\setsymbol{LFI:knee:frequency:LFI25:Rad:S:units}{38.9\mHz}
\setsymbol{LFI:knee:frequency:LFI26:Rad:S:units}{58.9\mHz}
\setsymbol{LFI:knee:frequency:LFI27:Rad:S:units}{104.4\mHz}
\setsymbol{LFI:knee:frequency:LFI28:Rad:S:units}{40.7\mHz}

\setsymbol{LFI:knee:frequency:LFI18:Rad:M}{16.3}
\setsymbol{LFI:knee:frequency:LFI19:Rad:M}{15.1}
\setsymbol{LFI:knee:frequency:LFI20:Rad:M}{18.7}
\setsymbol{LFI:knee:frequency:LFI21:Rad:M}{37.2}
\setsymbol{LFI:knee:frequency:LFI22:Rad:M}{12.7}
\setsymbol{LFI:knee:frequency:LFI23:Rad:M}{34.6}
\setsymbol{LFI:knee:frequency:LFI24:Rad:M}{46.2}
\setsymbol{LFI:knee:frequency:LFI25:Rad:M}{24.9}
\setsymbol{LFI:knee:frequency:LFI26:Rad:M}{67.6}
\setsymbol{LFI:knee:frequency:LFI27:Rad:M}{187.4}
\setsymbol{LFI:knee:frequency:LFI28:Rad:M}{122.2}
\setsymbol{LFI:knee:frequency:LFI18:Rad:S}{17.7}
\setsymbol{LFI:knee:frequency:LFI19:Rad:S}{22.0}
\setsymbol{LFI:knee:frequency:LFI20:Rad:S}{8.7}
\setsymbol{LFI:knee:frequency:LFI21:Rad:S}{25.9}
\setsymbol{LFI:knee:frequency:LFI22:Rad:S}{15.8}
\setsymbol{LFI:knee:frequency:LFI23:Rad:S}{129.8}
\setsymbol{LFI:knee:frequency:LFI24:Rad:S}{100.9}
\setsymbol{LFI:knee:frequency:LFI25:Rad:S}{38.9}
\setsymbol{LFI:knee:frequency:LFI26:Rad:S}{58.9}
\setsymbol{LFI:knee:frequency:LFI27:Rad:S}{104.4}
\setsymbol{LFI:knee:frequency:LFI28:Rad:S}{40.7}


\setsymbol{LFI:slope:70GHz:units}{$-1.03$\mHz}
\setsymbol{LFI:slope:44GHz:units}{$-0.89$\mHz}
\setsymbol{LFI:slope:30GHz:units}{$-0.87$\mHz}

\setsymbol{LFI:slope:70GHz}{$-1.03$}
\setsymbol{LFI:slope:44GHz}{$-0.89$}
\setsymbol{LFI:slope:30GHz}{$-0.87$}

\setsymbol{LFI:slope:LFI18:Rad:M:units}{$-1.04$\mHz}
\setsymbol{LFI:slope:LFI19:Rad:M:units}{$-1.09$\mHz}
\setsymbol{LFI:slope:LFI20:Rad:M:units}{$-0.69$\mHz}
\setsymbol{LFI:slope:LFI21:Rad:M:units}{$-1.56$\mHz}
\setsymbol{LFI:slope:LFI22:Rad:M:units}{$-1.01$\mHz}
\setsymbol{LFI:slope:LFI23:Rad:M:units}{$-0.96$\mHz}
\setsymbol{LFI:slope:LFI24:Rad:M:units}{$-0.83$\mHz}
\setsymbol{LFI:slope:LFI25:Rad:M:units}{$-0.91$\mHz}
\setsymbol{LFI:slope:LFI26:Rad:M:units}{$-0.95$\mHz}
\setsymbol{LFI:slope:LFI27:Rad:M:units}{$-0.87$\mHz}
\setsymbol{LFI:slope:LFI28:Rad:M:units}{$-0.88$\mHz}
\setsymbol{LFI:slope:LFI18:Rad:S:units}{$-1.15$\mHz}
\setsymbol{LFI:slope:LFI19:Rad:S:units}{$-1.00$\mHz}
\setsymbol{LFI:slope:LFI20:Rad:S:units}{$-0.95$\mHz}
\setsymbol{LFI:slope:LFI21:Rad:S:units}{$-0.92$\mHz}
\setsymbol{LFI:slope:LFI22:Rad:S:units}{$-1.01$\mHz}
\setsymbol{LFI:slope:LFI23:Rad:S:units}{$-0.95$\mHz}
\setsymbol{LFI:slope:LFI24:Rad:S:units}{$-0.73$\mHz}
\setsymbol{LFI:slope:LFI25:Rad:S:units}{$-1.16$\mHz}
\setsymbol{LFI:slope:LFI26:Rad:S:units}{$-0.79$\mHz}
\setsymbol{LFI:slope:LFI27:Rad:S:units}{$-0.82$\mHz}
\setsymbol{LFI:slope:LFI28:Rad:S:units}{$-0.91$\mHz}

\setsymbol{LFI:slope:LFI18:Rad:M}{$-1.04$}
\setsymbol{LFI:slope:LFI19:Rad:M}{$-1.09$}
\setsymbol{LFI:slope:LFI20:Rad:M}{$-0.69$}
\setsymbol{LFI:slope:LFI21:Rad:M}{$-1.56$}
\setsymbol{LFI:slope:LFI22:Rad:M}{$-1.01$}
\setsymbol{LFI:slope:LFI23:Rad:M}{$-0.96$}
\setsymbol{LFI:slope:LFI24:Rad:M}{$-0.83$}
\setsymbol{LFI:slope:LFI25:Rad:M}{$-0.91$}
\setsymbol{LFI:slope:LFI26:Rad:M}{$-0.95$}
\setsymbol{LFI:slope:LFI27:Rad:M}{$-0.87$}
\setsymbol{LFI:slope:LFI28:Rad:M}{$-0.88$}
\setsymbol{LFI:slope:LFI18:Rad:S}{$-1.15$}
\setsymbol{LFI:slope:LFI19:Rad:S}{$-1.00$}
\setsymbol{LFI:slope:LFI20:Rad:S}{$-0.95$}
\setsymbol{LFI:slope:LFI21:Rad:S}{$-0.92$}
\setsymbol{LFI:slope:LFI22:Rad:S}{$-1.01$}
\setsymbol{LFI:slope:LFI23:Rad:S}{$-0.95$}
\setsymbol{LFI:slope:LFI24:Rad:S}{$-0.73$}
\setsymbol{LFI:slope:LFI25:Rad:S}{$-1.16$}
\setsymbol{LFI:slope:LFI26:Rad:S}{$-0.79$}
\setsymbol{LFI:slope:LFI27:Rad:S}{$-0.82$}
\setsymbol{LFI:slope:LFI28:Rad:S}{$-0.91$}


\setsymbol{LFI:FWHM:70GHz:units}{13\parcm01}
\setsymbol{LFI:FWHM:44GHz:units}{27\parcm92}
\setsymbol{LFI:FWHM:30GHz:units}{32\parcm65}

\setsymbol{LFI:FWHM:70GHz}{13.01}
\setsymbol{LFI:FWHM:44GHz}{27.92}
\setsymbol{LFI:FWHM:30GHz}{32.65}

\setsymbol{LFI:FWHM:LFI18:units}{13\parcm39}
\setsymbol{LFI:FWHM:LFI19:units}{13\parcm01}
\setsymbol{LFI:FWHM:LFI20:units}{12\parcm75}
\setsymbol{LFI:FWHM:LFI21:units}{12\parcm74}
\setsymbol{LFI:FWHM:LFI22:units}{12\parcm87}
\setsymbol{LFI:FWHM:LFI23:units}{13\parcm27}
\setsymbol{LFI:FWHM:LFI24:units}{22\parcm98}
\setsymbol{LFI:FWHM:LFI25:units}{30\parcm46}
\setsymbol{LFI:FWHM:LFI26:units}{30\parcm31}
\setsymbol{LFI:FWHM:LFI27:units}{32\parcm65}
\setsymbol{LFI:FWHM:LFI28:units}{32\parcm66}

\setsymbol{LFI:FWHM:LFI18}{13.39}
\setsymbol{LFI:FWHM:LFI19}{13.01}
\setsymbol{LFI:FWHM:LFI20}{12.75}
\setsymbol{LFI:FWHM:LFI21}{12.74}
\setsymbol{LFI:FWHM:LFI22}{12.87}
\setsymbol{LFI:FWHM:LFI23}{13.27}
\setsymbol{LFI:FWHM:LFI24}{22.98}
\setsymbol{LFI:FWHM:LFI25}{30.46}
\setsymbol{LFI:FWHM:LFI26}{30.31}
\setsymbol{LFI:FWHM:LFI27}{32.65}
\setsymbol{LFI:FWHM:LFI28}{32.66}



\setsymbol{LFI:FWHM:uncertainty:LFI18:units}{0.170\arcm}
\setsymbol{LFI:FWHM:uncertainty:LFI19:units}{0.174\arcm}
\setsymbol{LFI:FWHM:uncertainty:LFI20:units}{0.170\arcm}
\setsymbol{LFI:FWHM:uncertainty:LFI21:units}{0.156\arcm}
\setsymbol{LFI:FWHM:uncertainty:LFI22:units}{0.164\arcm}
\setsymbol{LFI:FWHM:uncertainty:LFI23:units}{0.171\arcm}
\setsymbol{LFI:FWHM:uncertainty:LFI24:units}{0.652\arcm}
\setsymbol{LFI:FWHM:uncertainty:LFI25:units}{1.075\arcm}
\setsymbol{LFI:FWHM:uncertainty:LFI26:units}{1.131\arcm}
\setsymbol{LFI:FWHM:uncertainty:LFI27:units}{1.266\arcm}
\setsymbol{LFI:FWHM:uncertainty:LFI28:units}{1.287\arcm}

\setsymbol{LFI:FWHM:uncertainty:LFI18}{0.170}
\setsymbol{LFI:FWHM:uncertainty:LFI19}{0.174}
\setsymbol{LFI:FWHM:uncertainty:LFI20}{0.170}
\setsymbol{LFI:FWHM:uncertainty:LFI21}{0.156}
\setsymbol{LFI:FWHM:uncertainty:LFI22}{0.164}
\setsymbol{LFI:FWHM:uncertainty:LFI23}{0.171}
\setsymbol{LFI:FWHM:uncertainty:LFI24}{0.652}
\setsymbol{LFI:FWHM:uncertainty:LFI25}{1.075}
\setsymbol{LFI:FWHM:uncertainty:LFI26}{1.131}
\setsymbol{LFI:FWHM:uncertainty:LFI27}{1.266}
\setsymbol{LFI:FWHM:uncertainty:LFI28}{1.287}


\setsymbol{HFI:center:frequency:100GHz:units}{100\,GHz}
\setsymbol{HFI:center:frequency:143GHz:units}{143\,GHz}
\setsymbol{HFI:center:frequency:217GHz:units}{217\,GHz}
\setsymbol{HFI:center:frequency:353GHz:units}{353\,GHz}
\setsymbol{HFI:center:frequency:545GHz:units}{545\,GHz}
\setsymbol{HFI:center:frequency:857GHz:units}{857\,GHz}

\setsymbol{HFI:center:frequency:100GHz}{100}
\setsymbol{HFI:center:frequency:143GHz}{143}
\setsymbol{HFI:center:frequency:217GHz}{217}
\setsymbol{HFI:center:frequency:353GHz}{353}
\setsymbol{HFI:center:frequency:545GHz}{545}
\setsymbol{HFI:center:frequency:857GHz}{857}


\setsymbol{HFI:Ndetectors:100GHz}{8}
\setsymbol{HFI:Ndetectors:143GHz}{11}
\setsymbol{HFI:Ndetectors:217GHz}{12}
\setsymbol{HFI:Ndetectors:353GHz}{12}
\setsymbol{HFI:Ndetectors:545GHz}{3}
\setsymbol{HFI:Ndetectors:857GHz}{4}


\setsymbol{HFI:FWHM:Maps:100GHz:units}{9\parcm88}
\setsymbol{HFI:FWHM:Maps:143GHz:units}{7\parcm18}
\setsymbol{HFI:FWHM:Maps:217GHz:units}{4\parcm87}
\setsymbol{HFI:FWHM:Maps:353GHz:units}{4\parcm65}
\setsymbol{HFI:FWHM:Maps:545GHz:units}{4\parcm72}
\setsymbol{HFI:FWHM:Maps:857GHz:units}{4\parcm39}
\setsymbol{HFI:FWHM:Maps:100GHz}{9.88}
\setsymbol{HFI:FWHM:Maps:143GHz}{7.18}
\setsymbol{HFI:FWHM:Maps:217GHz}{4.87}
\setsymbol{HFI:FWHM:Maps:353GHz}{4.65}
\setsymbol{HFI:FWHM:Maps:545GHz}{4.72}
\setsymbol{HFI:FWHM:Maps:857GHz}{4.39}


\setsymbol{HFI:beam:ellipticity:Maps:100GHz}{1.15}
\setsymbol{HFI:beam:ellipticity:Maps:143GHz}{1.01}
\setsymbol{HFI:beam:ellipticity:Maps:217GHz}{1.06}
\setsymbol{HFI:beam:ellipticity:Maps:353GHz}{1.05}
\setsymbol{HFI:beam:ellipticity:Maps:545GHz}{1.14}
\setsymbol{HFI:beam:ellipticity:Maps:857GHz}{1.19}


\setsymbol{HFI:FWHM:Mars:100GHz:units}{9\parcm37}
\setsymbol{HFI:FWHM:Mars:143GHz:units}{7\parcm04}
\setsymbol{HFI:FWHM:Mars:217GHz:units}{4\parcm68}
\setsymbol{HFI:FWHM:Mars:353GHz:units}{4\parcm43}
\setsymbol{HFI:FWHM:Mars:545GHz:units}{3\parcm80}
\setsymbol{HFI:FWHM:Mars:857GHz:units}{3\parcm67}

\setsymbol{HFI:FWHM:Mars:100GHz}{9.37}
\setsymbol{HFI:FWHM:Mars:143GHz}{7.04}
\setsymbol{HFI:FWHM:Mars:217GHz}{4.68}
\setsymbol{HFI:FWHM:Mars:353GHz}{4.43}
\setsymbol{HFI:FWHM:Mars:545GHz}{3.80}
\setsymbol{HFI:FWHM:Mars:857GHz}{3.67}


\setsymbol{HFI:beam:ellipticity:Mars:100GHz}{1.18}
\setsymbol{HFI:beam:ellipticity:Mars:143GHz}{1.03}
\setsymbol{HFI:beam:ellipticity:Mars:217GHz}{1.14}
\setsymbol{HFI:beam:ellipticity:Mars:353GHz}{1.09}
\setsymbol{HFI:beam:ellipticity:Mars:545GHz}{1.25}
\setsymbol{HFI:beam:ellipticity:Mars:857GHz}{1.03}


\setsymbol{HFI:CMB:relative:calibration:100GHz}{$\lsim 1\%$}
\setsymbol{HFI:CMB:relative:calibration:143GHz}{$\lsim 1\%$}
\setsymbol{HFI:CMB:relative:calibration:217GHz}{$\lsim 1\%$}
\setsymbol{HFI:CMB:relative:calibration:353GHz}{$\lsim 1\%$}
\setsymbol{HFI:CMB:relative:calibration:545GHz}{}
\setsymbol{HFI:CMB:relative:calibration:857GHz}{}


\setsymbol{HFI:CMB:absolute:calibration:100GHz}{$\lsim 2\%$}
\setsymbol{HFI:CMB:absolute:calibration:143GHz}{$\lsim 2\%$}
\setsymbol{HFI:CMB:absolute:calibration:217GHz}{$\lsim 2\%$}
\setsymbol{HFI:CMB:absolute:calibration:353GHz}{$\lsim 2\%$}
\setsymbol{HFI:CMB:absolute:calibration:545GHz}{}
\setsymbol{HFI:CMB:absolute:calibration:857GHz}{}


\setsymbol{HFI:FIRAS:gain:calibration:accuracy:statistical:100GHz}{}
\setsymbol{HFI:FIRAS:gain:calibration:accuracy:statistical:143GHz}{}
\setsymbol{HFI:FIRAS:gain:calibration:accuracy:statistical:217GHz}{}
\setsymbol{HFI:FIRAS:gain:calibration:accuracy:statistical:353GHz}{2.5\%}
\setsymbol{HFI:FIRAS:gain:calibration:accuracy:statistical:545GHz}{1\%}
\setsymbol{HFI:FIRAS:gain:calibration:accuracy:statistical:857GHz}{0.5\%}


\setsymbol{HFI:FIRAS:gain:calibration:accuracy:systematic:100GHz}{}
\setsymbol{HFI:FIRAS:gain:calibration:accuracy:systematic:143GHz}{}
\setsymbol{HFI:FIRAS:gain:calibration:accuracy:systematic:217GHz}{}
\setsymbol{HFI:FIRAS:gain:calibration:accuracy:systematic:353GHz}{}
\setsymbol{HFI:FIRAS:gain:calibration:accuracy:systematic:545GHz}{7\%}
\setsymbol{HFI:FIRAS:gain:calibration:accuracy:systematic:857GHz}{7\%}


\setsymbol{HFI:FIRAS:zero:point:accuracy:100GHz:units}{0.8\MJysr}
\setsymbol{HFI:FIRAS:zero:point:accuracy:143GHz:units}{}
\setsymbol{HFI:FIRAS:zero:point:accuracy:217GHz:units}{}
\setsymbol{HFI:FIRAS:zero:point:accuracy:353GHz:units}{1.4\MJysr}
\setsymbol{HFI:FIRAS:zero:point:accuracy:545GHz:units}{2.2\MJysr}
\setsymbol{HFI:FIRAS:zero:point:accuracy:857GHz:units}{1.7\MJysr}

\setsymbol{HFI:FIRAS:zero:point:accuracy:100GHz}{0.8}
\setsymbol{HFI:FIRAS:zero:point:accuracy:143GHz}{}
\setsymbol{HFI:FIRAS:zero:point:accuracy:217GHz}{}
\setsymbol{HFI:FIRAS:zero:point:accuracy:353GHz}{1.4}
\setsymbol{HFI:FIRAS:zero:point:accuracy:545GHz}{2.2}
\setsymbol{HFI:FIRAS:zero:point:accuracy:857GHz}{1.7}


\setsymbol{HFI:unit:conversion:100GHz:units}{0.2415\MJysrmK}
\setsymbol{HFI:unit:conversion:143GHz:units}{0.3694\MJysrmK}
\setsymbol{HFI:unit:conversion:217GHz:units}{0.4811\MJysrmK}
\setsymbol{HFI:unit:conversion:353GHz:units}{0.2883\MJysrmK}
\setsymbol{HFI:unit:conversion:545GHz:units}{0.05826\MJysrmK}
\setsymbol{HFI:unit:conversion:857GHz:units}{0.002238\MJysrmK}

\setsymbol{HFI:unit:conversion:100GHz}{0.2415}
\setsymbol{HFI:unit:conversion:143GHz}{0.3694}
\setsymbol{HFI:unit:conversion:217GHz}{0.4811}
\setsymbol{HFI:unit:conversion:353GHz}{0.2883}
\setsymbol{HFI:unit:conversion:545GHz}{0.05826}
\setsymbol{HFI:unit:conversion:857GHz}{0.002238}


\setsymbol{HFI:colour:correction:alpha=-2:V1.01:100GHz}{0.9893}
\setsymbol{HFI:colour:correction:alpha=-2:V1.01:143GHz}{0.9759}
\setsymbol{HFI:colour:correction:alpha=-2:V1.01:217GHz}{1.0007}
\setsymbol{HFI:colour:correction:alpha=-2:V1.01:353GHz}{1.0028}
\setsymbol{HFI:colour:correction:alpha=-2:V1.01:545GHz}{1.0019}
\setsymbol{HFI:colour:correction:alpha=-2:V1.01:857GHz}{0.9889}


\setsymbol{HFI:colour:correction:alpha=0:V1.01:100GHz}{1.0008}
\setsymbol{HFI:colour:correction:alpha=0:V1.01:143GHz}{1.0148}
\setsymbol{HFI:colour:correction:alpha=0:V1.01:217GHz}{0.9909}
\setsymbol{HFI:colour:correction:alpha=0:V1.01:353GHz}{0.9888}
\setsymbol{HFI:colour:correction:alpha=0:V1.01:545GHz}{0.9878}
\setsymbol{HFI:colour:correction:alpha=0:V1.01:857GHz}{1.0014}